\documentclass{article}
\usepackage{amssymb}
\usepackage{graphicx}
\usepackage{amsmath}

\begin{document}

\title{\textsc{An introduction to the tomographic picture of quantum
mechanics}}
\date{}
\author{A. Ibort$^a$, V.I. Man'ko$^b$, G. Marmo$^c$, A. Simoni$^c$, F.
Ventriglia$^c$ \\
%EndAName
{\footnotesize \textit{$^a$Departamento de Matem\`{a}ticas, Universidad
Carlos III de Madrid, }}\\
{\footnotesize \textit{Av.da de la Universidad 30, 28911 Legan\'{e}s,
Madrid, Spain }}\\
{\footnotesize {(e-mail: \texttt{albertoi@math.uc3m.es})}}\\
{\footnotesize \textit{$^b$P.N.Lebedev Physical Institute, Leninskii
Prospect 53, Moscow 119991, Russia}}\\
{\footnotesize {(e-mail: \texttt{manko@na.infn.it})}}\\
\textsl{{\footnotesize {$^c$Dipartimento di Scienze Fisiche dell' Universit%
\`{a} ``Federico II" e Sezione INFN di Napoli,}}}\\
\textsl{{\footnotesize {Complesso Universitario di Monte S. Angelo, via
Cintia, 80126 Naples, Italy}}}\\
{\footnotesize {(e-mail: \texttt{marmo@na.infn.it, simoni@na.infn.it,
ventriglia@na.infn.it})}}}
\maketitle

\begin{abstract}
Starting from the famous Pauli problem on the possibility to associate
quantum states with probabilities, the formulation of quantum mechanics in
which quantum states are described by fair probability distributions
(tomograms, i.e. tomographic probabilities) is reviewed in a pedagogical
style. The relation between the quantum state description and the classical
state description is elucidated. The difference of those sets of tomograms
is described by inequalities equivalent to a complete set of uncertainty
relations for the quantum domain and to nonnegativity of probability density
on phase space in the classical domain. Intersection of such sets is
studied. The mathematical mechanism which allows to construct different
kinds of tomographic probabilities like symplectic tomograms, spin
tomograms, photon number tomograms, etc., is clarified and a connection with
abstract Hilbert space properties is established. Superposition rule and
uncertainty relations in terms of probabilities as well as quantum basic
equation like quantum evolution and energy spectra equations are given in
explicit form. A method to check experimentally uncertainty relations is
suggested using optical tomograms. Entanglement phenomena and the connection
with semigroups acting on simplexes are studied in detail for spin states in
the case of two qubits. The star-product formalism is associated with the
tomographic probability formulation of quantum mechanics. \newline
\end{abstract}

\tableofcontents

\newpage

\section{Introduction}

Pure quantum states are usually associated with wave functions \cite%
{Schrod26} or vectors in a Hilbert space \cite{Dirac}. Mixed quantum states
are associated with density matrices \cite{Landau27} or density states \cite%
{VonNeumann27}. Pauli \cite{Pauli, Pauli1} posed the problem whether it was
possible to associate quantum states with probability distributions as it
happens in classical statistical mechanics. Pauli problem was more concrete,
namely, is it possible to reconstruct the quantum state (i.e., the wave
function) from the knowledge of the probability distribution for the
position and the probability distribution for the momentum? The answer to
this particular question is negative (see the discussion in Reichenbach's
book \cite{R} and a recent example in \cite{Ventriglia1}). But the general
idea of Pauli to associate quantum states with probability distributions was
implemented by introducing the tomographic probability representation of
quantum states \cite{Mancini}. This representation is based on the Radon
transform \cite{Radon1917} of Wigner function \cite{Wig32}, suggested in
\cite{BerBer, VogelRis} to connect the measurable optical tomographic
probability \cite{Raymer, Mlynek} to reconstruct the Wigner function of a
photon quantum state. The mathematical nature of the tomographic probability
representation was clarified in \cite{Ventriglia2, Ventriglia3} (but see
also, e.g., \cite{QSO1996,Agarwal,WeigAmiet, Weigert,Cassinelli,Paini}). In
quantum mechanics we have the conventional Heisenberg and Schr\"{o}dinger
representations. The tomographic probability representation is another one.
The physical properties of quantum systems can be studied in the tomographic
probability representation as well as in the Heisenberg and Schr\"{o}dinger
representations or in the Feynman representation \cite{Feyn48} based on the
use of path integral as main ingredient of the quantum picture. Since the
tomographic picture deals with probabilities which describe the quantum
states, we will show how important quantum aspects as uncertainty relations
and superposition principle can be described in terms of tomographic
probabilities. We suggest a method of checking the Heisenberg uncertainty
relations using quantum state tomograms. It is worthy to mention that all
the available representations of quantum mechanics are equivalent (see,
e.g., the review \cite{AmJPhys}). One cannot say that some representation is
better or worse than the others. Nevertheless, each representation has
peculiar properties due to which some quantum aspects become clearer and
simpler than in other representations. For sure, the superposition principle
can be formulated in the easiest and clearest form using the natural linear
structure of a Hilbert space whose vectors are realized by complex wave
functions. The tomographic probability picture is very natural for problems
of quantum information and quantum entanglement, so each picture has its own
merits.

We shall try to present in a pedagogical style the construction of the
tomographic probability representation both for discrete (spin, qubit)
variables, studied in \cite{DodPL,OlgaJetph}, and for continuous variables
like position and momentum following \cite{Ventriglia1, Ventriglia2,
Ventriglia3, BeppeOlga, Cosmo, Cosimo}. The tomographic probability can be
used also in classical statistical mechanics \cite{manman, manmen2}. The
tomogram of a classical state is the Radon transform of the standard
probability density on the classical phase space. In this setting both
classical and quantum states can be described by tomographic probability
distributions. The difference between classical and quantum states in the
tomographic description is related to the different physical constrains the
state tomograms have to satisfy in order to be either in the classical or in
the quantum domain. In the tomographic representation the quantum-classical
relation is formulated in terms of properties of the tomographic probability
densities. It will be obvious that the ambient space of Radon transforms of
functions defined on the phase space contains the subset of quantum
tomograms (tomograms admissible in the quantum domain), the subset of
classical tomograms (those admissible in the classical domain) and the
subset of those which are not admissible neither in quantum nor in classical
domain. The subsets of quantum and classical tomograms have a not empty
intersection. In this work we study the properties of tomograms providing a
characterization of classical and quantum domains as well as of their
intersection.

The paper is organized as follows. After a preliminary section 2 in which
the Pauli problem is considered and its tomographic solution is introduced,
there are two main parts. Part one is devoted to discuss quantum mechanics
formulations on phase space and their relations to tomography. It contains
nine sections. Weyl systems are discussed in section 3.1. Wigner functions
are considered in section 3.2, and their transformation properties under the
action of the automorphisms of the Weyl-Heisenberg group are discussed in
section 3.3. Tomograms and Radon transform are studied in section 3.4.
Classical and quantum probability distributions are considered in section
3.5. State reconstruction procedure is studied in section 3.6. Tomographic
families of rank-one projectors are introduced in section 3.7, while a
general and abstract setting of tomographic maps is presented in section
3.8. Finally, a unified approach to construct the most commonly used
tomographic families of observables is given in section 3.9. In the second
part we describe quantum mechanics in the tomographic picture. It contains
six sections. In the sections 4.1 and 4.2, the superposition rule and the
uncertainty relations respectively are discussed in the tomographic
probability representation. In the section 4.3 some examples of classical or
quantum distributions are presented. In section 4.4 we study basic equations
(time evolution and eigenvalue equations) in tomographic representation. An
application to entanglement and separability on examples of two qubit states
is given in section 4.5. Bell inequalities and association with a semigroup
structure are considered in section 4.6. In section 5 some conclusions and
perspectives are finally drawn.

\section{The Pauli problem: the original formulation and the tomographic
solution}

\label{Pauli_problem}

In his book \cite{Pauli1}, after having introduced the wave function in the
position representation by means of $\psi \left( \vec{x}\right) $ and in
momentum representation by means of $\varphi \left( \vec{p}\right) ,$ along
with the probability densities
\begin{equation}
W\left( \vec{x}\right) =\left\vert \psi \left( \vec{x}\right) \right\vert
^{2}=\psi \left( \vec{x}\right) ^{\ast }\psi \left( \vec{x}\right) ;\quad
W\left( \vec{p}\right) =\varphi \left( \vec{p}\right) ^{\ast }\varphi \left(
\vec{p}\right) ,
\end{equation}
Pauli claims: These functions $\psi \left( \vec{x}\right) $ and $\varphi
\left( \vec{p}\right) $, which are usually called `probability amplitudes'
are not, however, directly observable with regard to their phases; this
holds only for the probability densities $W\left( \vec{x}\right) $ and $%
W\left( \vec{p}\right) $. In a footnote Pauli states: The mathematical
problem as to whether for given functions $W\left( \vec{x}\right) $ and $%
W\left( \vec{p}\right) $, the wave function $\psi ,$ if such a function
exists, is always uniquely determined (i.e. if $W\left( \vec{x}\right) $ and
$W\left( \vec{p}\right) $ are physically compatible) has still not being
investigated in all its generality. \cite{Pauli1}

As it was shown in Reichenbach's book \cite{R}, the negative answer to the
original Pauli problem may be given by means of counterexamples. For
instance \cite{Ventriglia1}, consider the two squeezed states with wave
functions in the position representation ($\hbar =1$):
\begin{eqnarray}
\psi _{1}(q) &=&N\exp (-\alpha q^{2}+i\beta q), \\
\psi _{2}(q) &=&N\exp (-\alpha ^{\ast }q^{2}+i\beta q),  \notag \\
\mathrm{Re}\,{\alpha }\geq 0 &;&\beta =\beta ^{\ast }~;~N=\sqrt[4]{\frac{%
\alpha +\alpha ^{\ast }}{\pi }}  \notag
\end{eqnarray}
and in the momentum representation
\begin{eqnarray}
\psi _{1}(p) &=&\frac{N}{\sqrt{2\alpha }}\exp \left[ -\frac{\left( \beta
+p\right) ^{2}}{4\alpha }\right] , \\
\psi _{2}(p) &=&\frac{N}{\sqrt{2\alpha ^{\ast }}}\exp \left[ -\frac{\left(
\beta +p\right) ^{2}}{4\alpha ^{\ast }}\right] .  \notag
\end{eqnarray}
One then has
\begin{eqnarray}
\left\vert \psi _{1}(q)\right\vert ^{2} &=&\left\vert \psi
_{2}(q)\right\vert ^{2}=\left\vert N\right\vert ^{2}\exp \left[ -(\alpha
+\alpha ^{\ast })q^{2}\right] , \\
\left\vert \psi _{1}(p)\right\vert ^{2} &=&\left\vert \psi
_{2}(p)\right\vert ^{2}=\frac{\left\vert N\right\vert ^{2}}{2\left\vert
\alpha \right\vert }\exp \left[ -\left( \beta +p\right) ^{2}\left( \frac{%
\alpha +\alpha ^{\ast }}{4\alpha \alpha ^{\ast }}\right) \right] .  \notag
\end{eqnarray}
The fidelity $f=\left\vert \left\langle \psi _{1}|\psi _{2}\right\rangle
\right\vert ^{2}$ is
\begin{equation*}
f=\frac{\alpha +\alpha ^{\ast }}{2\sqrt{\alpha \alpha ^{\ast }}}.
\end{equation*}
One can see that $f\neq 1,$ which means that the states are different.

To understand the reasons why the knowledge of the two marginal
distributions of position and momentum are not sufficient for reconstructing
a state, consider the family of dimensionless observables, depending on two
real parameters $\mu ,\nu :$
\begin{equation}
X(\mu ,\nu )=\mu Q+\nu P\ ,
\end{equation}
where, restoring the Planck constant $\hbar ,$ $Q$ and $P$ generate the
Weyl--Heisenberg algebra $[Q,P]=i\hbar I$. The spectrum of $X(\mu ,\nu )$ is
the real line, which we parametrize by $X$, with corresponding improper
eigenvector $\left\vert X\mu \nu \right\rangle $. In the position
representation,
\begin{equation}
X(\mu ,\nu )=-i\hbar \nu \frac{d}{dq}+\mu q,
\end{equation}
and its (improper) eigenfunctions may be chosen as
\begin{equation}
\varphi _{X\mu \nu }(q)=\left\langle q|X\mu \nu \right\rangle =Ne^{-i\frac{%
\mu }{2\hbar \nu }q^{2}+i\frac{X}{\hbar \nu }q},\quad N=\frac{1}{\sqrt{2\pi
\hbar |\nu |}}
\end{equation}
The normalization is such that $\left\langle X^{\prime }|{X}\right\rangle
=\delta (X-X^{\prime }).$ Analogously, in the momentum representation, the
eigenfunctions may be chosen as
\begin{equation}
\tilde{\varphi}_{X\mu \nu }(p)=\left\langle p|X\mu \nu \right\rangle =\frac{1%
}{\sqrt{2\pi \hbar |\mu |}}e^{i\frac{\nu }{2\hbar \mu }p^{2}-i\frac{X}{\hbar
\mu }p}
\end{equation}
Now, define the symplectic tomogram of a (normalized) pure state $\left\vert
\psi \right\rangle $ with respect to the family $X(\mu ,\nu )$ as (see,
e.g., \cite{manmenPhys lett})
\begin{eqnarray}
\mathcal{T}_{\psi }(X,\mu ,\nu ) &=&|\left\langle {X\mu \nu }|\psi
\right\rangle |^{2}=\frac{1}{2\pi \hbar \left\vert \nu \right\vert }%
\left\vert \int \psi (q)e^{i\frac{\mu }{2\hbar \nu }q^{2}-i\frac{X}{\hbar
\nu }q}dq\right\vert ^{2},\quad \nu \neq 0  \notag \\
&=&\frac{1}{2\pi \hbar \left\vert \mu \right\vert }\left\vert \int \tilde{%
\psi}(p)e^{-i\frac{\nu }{2\hbar \mu }p^{2}+i\frac{X}{\hbar \mu }%
p}dp\right\vert ^{2},\quad \mu \neq 0
\end{eqnarray}
where the Fourier transform of the wave function has been introduced
\begin{equation}
\tilde{\psi}(p)=\frac{1}{\sqrt{2\pi \hbar }}\int \psi (y)e^{-i\frac{py}{%
\hbar }}dy.
\end{equation}
In other words, $\mathcal{T}_{\psi }(X,\mu ,\nu )dX$ is the marginal
probability such that a measure in the given state $\left\vert \psi
\right\rangle $ of the observable $X(\mu ,\nu ),$ with fixed $\mu ,\nu ,$
has values in $(X,X+dX).$ Of course, as $\left\vert \psi \right\rangle $ is
normalized,
\begin{equation}
\int \mathcal{T}_{\psi }(X,\mu ,\nu )dX=1
\end{equation}
independently of the chosen values of $\mu ,\nu .$

Then, it is apparent that the Pauli problem amounts to reconstruct the given
state $\left\vert \psi \right\rangle $ from the knowledge of the two
marginal probability distributions of position, $\mathcal{T}_{\psi }(X,1,0),$
and momentum, $\mathcal{T}_{\psi }(X,0,1).$ As a matter of fact, the
following reconstruction formula holds (see, e.g., \cite{Found Phys 97}) for
the density matrix $\rho (q,q^{\prime })=\psi (q)\psi ^{\ast }(q^{\prime }):$%
\begin{eqnarray}
\rho (q,q^{\prime }) &=&\frac{1}{2\pi }\int \mathcal{T}_{\psi }(X,\mu ,\frac{%
q-q^{\prime }}{\hbar })e^{i\left[ X-\frac{\mu }{2}(q+q^{\prime })\right]
}dXd\mu  \label{symprec} \\
&=&\frac{1}{(2\pi )^{2}}\int \mathcal{T}_{\psi }(X,\mu ,\nu )e^{\frac{i}{%
\hbar }p(q-q^{\prime })}\exp \left[ i(X-\mu q-\nu p)\right] dXd\mu d\nu dp
\notag
\end{eqnarray}
Thus, the answer to the Pauli problem is negative because the reconstruction
requires the knowledge of many different marginal probability distributions,
corresponding to many different observables of the family $X(\mu ,\nu ).$ A
minimal set of such observables is called a \textit{quorum} \cite{Fano}, and
the characterization of a tomographic quorum will be discussed in the next
sections.

\newpage

\section{Quantum mechanics on phase space and tomography}

\subsection{Weyl systems}

The notion of Weyl system is useful as a tool to formulate a quantum version
of classical hamiltonian mechanics by using the symplectic form on a linear
classical phase space. We will use it in the following as a suitable setting
to discuss the quantum and classical tomographic maps and their relations.
Here we briefly recall the definition of Weyl system, mainly to fix
notations. Hereafter $\hbar =1$.

Given a symplectic vector space $(V,\omega )$, where $V$ has even dimension $%
2n$ and $\omega$ is a nondegenerate skew--symmetric bilinear form on it, a
Weyl system is a strongly continuous map from $V$ to the group of unitary
operators on some Hilbert space $\mathcal{H}$:
\begin{equation}
W\,:\,V\rightarrow \mathcal{U}(\mathcal{H})
\end{equation}
satisfying the condition
\begin{equation}  \label{comm}
W(v_{1})W(v_{2})W^{\dagger }(v_{1}+v_{2})=I e^{\frac{i}{2}\omega
(e_{1},e_{2})}\,.
\end{equation}
It is a projective unitary representation of the Abelian vector group
associated with $V.$

A theorem due to von Neumann \cite{Von} establishes that such a map exists
for any finite dimensional symplectic vector space. Indeed, the Hilbert
space $\mathcal{H}$ can be realized as the space of square integrable
functions on any Lagrangian subspace of $V$. By using a Lagrangian subspace $%
L$, this is a subspace $L$ of dimension half of the dimension of the space $%
V $ and such that $\omega (x,y) = 0$ for all $x,y \in L$, and its dual $%
L^{\ast }$ it is possible to decompose $V$ into $V\cong L\oplus L^{\ast
}=T^{\ast }L.$ Because $L$ is Lagrangian, Eq. (\ref{comm}) implies that the
unitary operators corresponding to these vectors will commute. We consider
the Lebesgue measure on $L$ and we construct the following specific
realization of $W$. The action of the restrictions $W|_{L},W|_{L^{\ast }}$
of the map $W$ to the subspaces $L$ and $L^*$, on ${\mathcal{L}}%
^{2}(L,d^{n}x)$ is given by:
\begin{equation}
(W|_{L}(y)\psi )(x)=\psi (x+y)\,,\quad (W|_{L^{\ast }}(\alpha )\psi
)(x)=e^{i\alpha (x)}\psi (x)\,,
\end{equation}
for $x,y\in L$, $\alpha \in L^{\ast }$, $\psi \in {\mathcal{L}}%
^{2}(L,d^{n}x) .$

The strong continuity requirement in the definition of $W$ allows to use
Stone theorem to get
\begin{equation}
W(tv)=e^{itR(v)}\,,\quad \forall \,v\in V\,,
\end{equation}
with $R(v)$ the infinitesimal generator of the one parameter unitary group $%
W(tv)$, $t\in \mathbb{R}$, depending linearly on $v$.

In the following we shall be mainly concerned with the simple case $V=%
\mathbb{R}^{2},$ where, for simplicity, we introduce $(q,p)$ coordinates.
Then, by denoting
\begin{equation}
P=R(1,0);\ Q=R(0,1)
\end{equation}
we may also represent the Weyl map in the familiar form $v=(q,p)\rightarrow
W(v)=D(q,p)$, where the displacement operator $D(q,p)$ is given by the
formula:
\begin{equation}  \label{displacement}
D(q,p) =\exp \left[ i\left( pQ - qP \right) \right] ,
\end{equation}
which turns out to be an irreducible representation of the Weyl--Heisenberg
group (see section \ref{weyl_aut} for more details on this topic).

By introducing complex coordinates $z=\frac{1}{\sqrt{2}}(q+ip)$ for $V,$ we
may write the displacement operator, eq. (\ref{displacement}), in terms of
creation and annihilation operators as:
\begin{equation}  \label{displacement_complex}
D(z)=\exp \left( za^{\dagger }-z^{\ast} a\right) .
\end{equation}

Generalizations of previous formulae to the case $V=\mathbb{R}^{2n}$ are
obvious, just by considering $n-$dimensional vector operators.

Thus, a Weyl system provides, for any choice of a Lagrangian subspace,
canonical pairs of operators and the displacements operators, i. e. a
projective irreducible representation of the translation group. It is now
possible to associate an operator with any function $f$ on $V$ admitting a
(symplectic) Fourier transform $\tilde{f}.$ Consider
\begin{equation}
f(q,p)=\int d^{n}\alpha d^{n}x\tilde{f}(\alpha ,x)\exp [i(\alpha q-xp)].
\end{equation}
The operator $W(f)$ associated to $f$ is obtained by a clever use of the
Fourier transform, substituting the (symplectic) Fourier kernel $\exp
[i(\alpha q-xp)]$ with the Weyl displacement operator $D(\alpha ,x)=\exp %
\left[ i\left( \alpha Q-xP\right) \right] :$
\begin{equation}
W(f)=\int d^{n}\alpha d^{n}x\tilde{f}(\alpha ,x)\exp \left[ i\left( \alpha
Q-xP\right) \right] .
\end{equation}
The above formula defines a unitary isomorphism between the space of the
Hilbert-Schmidt operators on ${\mathcal{L}}^{2}(L,d^{n}x)$ and the square
integrable functions of ${\mathcal{L}}^{2}(L\oplus L^{\ast
},d^{n}xd^{n}\alpha ).$

\subsection{Wigner functions}

\label{Wigner_functions}

We have considered a Weyl system to be a projective unitary representation
of an Abelian vector group $V$ of even dimension. Another useful
interpretation of a Weyl system comes from the following considerations.

Consider a fiducial vector $\left\vert \psi _{0}\right\rangle $ in the
Hilbert space $\mathcal{H}$ carrying the projective unitary representation
of $V.$ We may consider an immersion of $V$ into $\mathcal{H}$ by means of
the map
\begin{equation}
V\ni v\longmapsto \left\vert v\right\rangle =W(v)\left\vert \psi
_{0}\right\rangle \ .
\end{equation}
We denote this map by $\mathcal{W}_{0}:V\rightarrow \mathcal{H}$. The image
of $\mathcal{W}_{0}$ is a submanifold of $\mathcal{H}$, it is not a
subspace. If we consider a fiducial operator $A_{0}\in \mathrm{End}(\mathcal{%
H}\mathbb{)}$ , we may in a similar way immerse $V$ into the space of
operators acting on $\mathcal{H}$ by setting
\begin{equation*}
\mathcal{W}_{0}:V\rightarrow \mathrm{End}(\mathcal{H}\mathbb{)}\ ,\
v\longmapsto W^{\dagger }(v)A_{0}W(v)=:A(v)\ .
\end{equation*}
As it happens with any immersion of a manifold $\mathcal{M}$ into a manifold
$\mathcal{N}$, we can consider the pull-back to $V$ of the covariant tensor
fields on $\mathcal{H}$. As a matter of fact, on account of the
probabilistic interpretation of quantum mechanics, the immersion of $V$
should be considered to take place into the manifold of rays, i.e. $\mathcal{%
R}(\mathcal{H})$ the manifold of rays in $\mathcal{H}$ or the complex
projective space associated with $\mathcal{H}.$

With any immersion $\phi :\mathcal{M}\rightarrow \mathcal{N}$ we have a map $%
\phi ^{\ast }:\mathcal{F}\left( \mathcal{N}\right) \rightarrow \mathcal{F}(%
\mathcal{M})$ which, however, need not be surjective. Moreover, dealing with
infinite dimensional manifolds ($\mathcal{H}$ or $\mathrm{End}(\mathcal{H}%
\mathbb{)}$) the properties of the pulled-back tensors depend on the
specific immersion we deal with. In particular we may require the map $v
\longmapsto \left| v\right\rangle $ , or $v \longmapsto A(v),$ to satisfy
appropriate measurability, continuity or differentiability properties.
Depending on the use of the pulled-back tensors we may prefer one immersion
over the other and, for instance, prefer the Schr\"{o}dinger picture over
the Heisenberg picture.

By using the decomposition of $V$ into different Lagrangian subspaces, say $%
L\times L^{\prime }$ (we use the cartesian product notation to stress we are
considering it as a manifold rather than as a vector space), we may consider
`eigenvectors' of the position operators, say $\left\vert q\right\rangle $\
or $\left\vert q^{\prime }\right\rangle $\ . In this way we may `pull-back'
any vector $\left\vert \psi \right\rangle $ to a wave function $\psi
(q)=\left\langle q|\psi \right\rangle $ or $\psi (q^{\prime })=\left\langle
q^{\prime }|\psi \right\rangle ;$ similarly for an observable $A$ we have
the `matrix coefficient' $\left\langle q\left\vert A\right\vert q^{\prime
}\right\rangle =f_{A}(q,q^{\prime }).$

Had we chosen the decomposition $V=L\times L^{\ast }$, i.e.\ using
`eigenvectors'\ $\left\vert q\right\rangle $\ and $\left\vert p\right\rangle
$, we would have had $\left\langle q\left\vert A\right\vert p\right\rangle $
or $\left\langle p\left\vert A\right\vert q\right\rangle ,$ classically this
corresponds to the use of boundary values, $q$ and $q^{\prime },$ or initial
Cauchy data, $q$ and $p.$ These various matrix coefficient functions are
connected by means of the completeness relations
\begin{equation}
\int \left\vert q\right\rangle dq\left\langle q\right\vert =I=\int
\left\vert p\right\rangle dp\left\langle p\right\vert .
\end{equation}

This general idea of considering the representation as an immersion of a
manifold $V$ into the Hilbert space $\mathcal{H}$ \cite{Georg}, allows to
consider the pull-back of the algebra structure on the operators. We may
define a star-product by setting
\begin{equation}
\phi ^{\ast }(A)\ast \phi ^{\ast }(B):=\phi ^{\ast }(AB)\ .
\end{equation}
In this way, on the subspace of functions we obtain by means of the
pull-back of the operators a nonlocal and noncommutative product.

In the Schr\"{o}dinger picture we would have
\begin{equation}
f_{A}(v)=\frac{\left\langle v|Av\right\rangle }{\left\langle
v|v\right\rangle }\ ,
\end{equation}
we could also consider
\begin{equation}
\tilde{f}_{A}(v)=\left\langle \psi _{0}|W^{\dagger }(v)A_{0}W(v)|\psi
_{0}\right\rangle ,
\end{equation}
and so on. In general, these functions are called symbols of the
corresponding operators and carry a specific qualification to keep track of
the specific immersion one is considering (we have Weyl symbols, Berezin
symbols, H\"{o}rmander symbols, and so on).

At this point it should be clear that if we deal also with unbounded
operators on $\mathcal{H}$ we end up with a variety of situations already at
the level of topologies we are willing to consider on $\mathrm{End}(\mathcal{%
H}\mathbb{)}.$ Additional problems will arise from the specific choice of
the fiducial vector we start with (if we need to take derivatives, we better
deal with smooth or analytic vectors \cite{nelson}). Usually, it is better
to leave some of these choices unsettled and take them into full account
only in each specific problem.

Having clarified some aspects connected with the pull-back of observables,
let us turn now to the pull-back of states. We recall that states are
usually considered to be normalized positive functionals on the space of
observables. States are not a vector space but we may consider convex
combinations. Pure states are those that cannot be written as convex
combination. To avoid some pathologies, very often states are also required
to be normal.

Thanks to Gleason's theorem, states are also called density operators,
however this may be misleading because it may give the impression that they
should be considered with the same topologies as the operator algebra.
However the star-product we have considered allows to distinguish the $%
\mathcal{L}^{1}-$algebra, associated with operators, which acts on the $%
\mathcal{L}^{2}-$space of states. Thus, while the pull-back of states or
observables always provides us with functions on $V,$ the subsets to which
they belong enjoy quite different properties and therefore it is advisable
to avoid considering them as mathematical entities of the same kind.

The distinction will reappear to be crucial when we would like to compare
with the analogous situation in classical mechanics described on the same
phase space $V.$ Again here states and observables are quite different, pure
classical states end up being distributions, while observables, due to the
possibility of taking Poisson brackets, are usually required to be smooth
functions. This distinction plays a very relevant role when we consider
their associated Radon transforms to represent tomographic states or
tomographic observables.

We may now define a Wigner function the way it was considered by Wigner.
Given a state $\left\vert \psi \right\rangle $ we form the rank-one
projector
\begin{equation}
\rho =\frac{\left\vert \psi \right\rangle \left\langle \psi \right\vert }{%
\left\langle \psi |\psi \right\rangle }
\end{equation}
which defines a normalized positive functional on the space of
observables.We may consider, as we have stressed, either the density state
\begin{equation}
\rho (q,q^{\prime })=\frac{\left\langle q\left\vert \psi \right\rangle
\left\langle \psi \right\vert q^{\prime }\right\rangle }{\left\langle \psi
|\psi \right\rangle }
\end{equation}
or the function
\begin{equation}
\mathcal{\tilde{W}}(q,p):=\frac{\left\langle q\left\vert \psi \right\rangle
\left\langle \psi \right\vert p\right\rangle }{\left\langle \psi |\psi
\right\rangle },
\end{equation}
i.e. the matrix coefficients of the rank-one projector in the (position,
position) representation or in the (position, momentum) representation. By
using the completeness relation we have
\begin{equation}
\left\langle \psi |q^{\prime }\right\rangle =\int \left\langle \psi
\left\vert p\right\rangle dp\left\langle p\right\vert q^{\prime
}\right\rangle
\end{equation}
which provides us with the transformation from one representation to another.

It is now feasible to use convex combinations of pure states to define
generic states and their associated Wigner representations. The Wigner
function $\mathcal{W}(p,q)$ of a density state $\rho (q,q^{\prime })$ is
defined, restoring the Planck constant $\hbar ,$ as:
\begin{equation}
\mathcal{W}(q,p):=\int \rho (q+\frac{x}{2},q-\frac{x}{2})\exp (-\frac{i}{%
\hbar }px)dx\ ,
\end{equation}
and it results
\begin{equation}
\mathcal{W}(q,p)=2\exp (\frac{i}{\hbar }2pq)\int \mathcal{\tilde{W}}%
(q^{\prime },p^{\prime })\exp \left[ \frac{i}{\hbar }\left( q^{\prime
}p^{\prime }-2pq^{\prime }-2qp^{\prime }\right) \right] \frac{dq^{\prime
}dp^{\prime }}{\sqrt{2\pi \hbar }} \ .
\end{equation}
Now a simple manipulation, changing the variable $x/2=s$ and extracting from
the integral the bra $\left\langle \psi \right| $ and ket $\left| \psi
\right\rangle $, we get:
\begin{equation}
\mathcal{W}(q,p)=2\left\langle \psi \right| \left[ \int e^{-2ips/\hbar
}\left| q-s\rangle \langle q+s\right| ds\right] \left| \psi \right\rangle .
\label{moyal_rep}
\end{equation}

\subsection{Transformation properties of Wigner functions}

\label{weyl_aut}

We would like to understand the transformation properties of Wigner
functions under linear symplectic maps and dilations on phase space. The
exploration of these issues will suggest the possibility of extending the
definition of Wigner's function to the space of irreducible representations
of the Weyl--Heisenberg group and to discern their homogeneity dependence on
Planck's constant $\hbar$. We will make these comments precise in what
follows.

The Weyl map allows to associate automorphisms $\nu_\phi$ on the space of
unitary operators with elements $\phi$ of the symplectic linear group $Sp
(V,\omega )$ of $V$, according to the following diagram
\begin{equation}
\begin{tabular}{ccc}
$V$ & $\overset{W}{\longrightarrow }$ & $\mathcal{U} (\mathcal{H})$ \\
$\phi \downarrow $ &  & $\downarrow \nu _\phi$ \\
$V$ & $\overset{W}{\longrightarrow }$ & $\mathcal{U}(\mathcal{H})$%
\end{tabular}%
\end{equation}
for any $\phi\in Sp(V,\omega)$, by setting
\begin{equation}  \label{inner_aut}
\nu _\phi (W(v))=W(\phi(v))=U_{\phi}^{\dagger }W(v)U_{\phi}, \quad \forall v
\in V\ .
\end{equation}
Recall that an isomorphism $\phi \colon V \to V$ is symplectic if $\omega
(\phi(u), \phi(v)) = \omega (u,v)$ for all $u,v\in V$. Moreover, the group
of all symplectic isomorphisms of $V$ can be identified with the matrix
symplectic group $Sp(n)$ by choosing a symplectic basis on $V$. In other
words, the automorphism $\nu _{\phi},$ corresponding to the symplectic
linear transformation $\phi$ of $V,$ is a inner automorphism of the group of
unitary operators, that is there exists a unitary operator $U_\phi$ such
that $\nu_\phi (V) = U_\phi^\dagger V U_\phi$ for all $V\in \mathcal{U}(%
\mathcal{H})$, because it belongs to the connected component of the identity
of the automorphism group \cite{Kadison}. At the level of the infinitesimal
generators of the unitary group, we have
\begin{equation}
U_{\phi}^{\dagger }R(v)U_{\phi}=R(\phi(v))\,.
\end{equation}

Further insight on the physical meaning of the Wigner function of a density
state $\rho $, was obtained from its representation as the expectation value
of the shifted parity operator $\mathcal{P}(q,p)$ (see, e.g., Royer \cite%
{Royer}). In fact, we can write the expression given by Eq. (\ref{moyal_rep}%
) of the Wigner function $\mathcal{W}(q,p)$ corresponding to the state $%
\left\vert \psi \right\rangle $, as:
\begin{equation}
\mathcal{W}(q,p)=2\left\langle \psi \right\vert \mathcal{P}(q,p)\left\vert
\psi \right\rangle ,
\end{equation}
with $\mathcal{P}(q,p)$ being the shifted parity operator:
\begin{equation}
\mathcal{P}(q,p)=\int e^{-2ips/\hbar }\left\vert q-s\rangle \langle
q+s\right\vert ds.
\end{equation}
Notice that $\mathcal{P}(0,0)=\int \left\vert -s\rangle \langle s\right\vert
ds$ is just the parity operator defined as $(\mathcal{P}\psi )(q)=\psi (-q)$
or, equivalently the unitary operator $\mathcal{P}$ satisfying:
\begin{equation}
\mathcal{P}Q\mathcal{P}=-Q,\quad \mathcal{P}P\mathcal{P}=-P.  \label{parity}
\end{equation}
Now, we get immediately that:
\begin{equation}
\mathcal{P}(q,p)=D(q,p)\mathcal{P}D(q,p)^{\dagger },
\end{equation}
where the displacement operators $D(q,p)$ have the usual form given by Eq. (%
\ref{displacement}). Then, the Wigner function corresponding to the pure
state $\left\vert \psi \right\rangle $ can be readily written in the form:
\begin{equation}
\mathcal{W}(q,p)=2\left\langle \psi \right\vert D(q,p)\mathcal{P}%
D(q,p)^{\dagger }\left\vert \psi \right\rangle ,
\end{equation}
and for a given density state $\rho $ we obtain:
\begin{equation}
\mathcal{W}_{\rho }(p,q)=2\mathrm{Tr}\left[ \rho D(p,q)\mathcal{P}D^{\dagger
}(p,q)\right] =2\mathrm{Tr}\left[ \rho D(2p,2q)\mathcal{P}\right] .
\label{Wigpoldef}
\end{equation}

Because the displacement operators provide a specific irreducible
representation of the Weyl--Heisenberg group, the previous formula makes
apparent the possibility of generalizing Wigner's function as a function on
the space of irreducible representations of the Weyl--Heisenberg group. In
fact the Weyl--Heisenberg group $\mathrm{WH}(n),$ for $n=1,$ may be
presented as the group of triples of real numbers $(p,q,t)$ with the
composition law:
\begin{equation}
(q,p,t)\circ (q^{\prime },p^{\prime },t^{\prime })=(q+q^{\prime
},p+p^{\prime },t+t^{\prime }+\frac{1}{2}(pq^{\prime }-qp^{\prime })).
\end{equation}
The associated canonical operators with their commutation relations $%
[Q,P]=iI $ are a realization of the Lie algebra of the Weyl--Heisenberg
group, and a irreducible unitary representation is provided by:
\begin{equation}
U(q,p,t)=D(q,p)e^{itI}.  \label{repres}
\end{equation}

In the general case $n\geq 1$, the irreducible representations of $\mathrm{WH%
}(n)$ are parametrized up to a unitary equivalence by a real parameter $%
\gamma $ \cite{Folland}. Kirillov's theory of coadjoint orbits \cite%
{Kirillov} provides a natural way to construct them. In fact, Kirillov's
theorem establishes that for nilpotent groups there is a one-to-one
correspondence between coadjoint orbits of the group and equivalence classes
of unitary irreducible representations of it. It is easy to check that for
the Weyl--Heisenberg group the space of coadjoint orbits has two strata, the
regular one whose coadjoint orbits are copies of the symplectic linear space
$(V,\omega )$ and are labelled by $\gamma \neq 0$, and the singular stratum,
corresponding to the label $\gamma =0$ whose coadjoint orbits are points,
hence giving rise to trivial representations. The parameter $\gamma $
weights the central element of the group and it can be easily read out from
a given irreducible representation looking at $U_{\gamma }(0,t)=e^{i\gamma
t} $ and therefore the action of \textrm{Aut}$(\mathrm{WH}(n))$ on the set
of irreducible representations can be analyzed.

In order to introduce a generalized notion of Wigner functions \cite{Ib09}
for representations with $\gamma \neq 1$, we have to choose previously a
representative $U_{\gamma }$ out of any equivalence class $\left[ U\right]
_{\gamma }.$ We choose, for $n=1,$ the representatives for $\gamma >0$ as:
\begin{equation}
U_{\gamma }(q,p,t):=U_{\gamma =1}(\sqrt{\gamma }q,\sqrt{\gamma }p,\gamma
t)=D(\sqrt{\gamma }q,\sqrt{\gamma }q)e^{i\gamma t}.
\end{equation}

Once a representation $U_{\gamma }$ has been chosen, the parity operator $%
\mathcal{P}$ given by Eq. (\ref{parity}) may be expressed as:
\begin{equation}
\mathcal{P}=\frac{\gamma }{2}\int \frac{dqdp}{2\pi }D(\sqrt{\gamma }q,\sqrt{%
\gamma }p)=\frac{1}{2}\int \frac{dqdp}{2\pi }D(q,p).  \label{parity2}
\end{equation}
>From this expression the properties:
\begin{equation}
\mathcal{P}=\mathcal{P}^{\dagger };\qquad \mathcal{P}U_{\gamma }(q,p,t)%
\mathcal{P}=U_{\gamma }(-q,-p,t),
\end{equation}
readily follow.

Now, given $\gamma ,$ we define the associated (generalized) Wigner function
of a density state $\rho $ as
\begin{eqnarray}
\mathcal{W}_{\rho }(q,p;\gamma ) &:&=2\,\mathrm{Tr}\left[ \rho U_{\gamma
}(q,p,t)\mathcal{P}U_{\gamma }^{\dagger }(q,p,t)\right]  \notag
\label{WigGam} \\
&=&2\,\mathrm{Tr}\left[ \rho D(2\sqrt{\gamma }q,2\sqrt{\gamma }p)\mathcal{\ P%
}\right]  \notag \\
&=&\mathcal{W}_{\rho }(\sqrt{\gamma }q,\sqrt{\gamma }p;1).
\end{eqnarray}
We remark that, while the dependence on the parameter $t$ disappears and the
function is invariant on the subgroup $(0,t)$, a new dependence on the
representation label $\gamma $ appears.

We now consider the action of a dilation $\phi _{\lambda }:\phi _{\lambda }
(q,p,t) = (\lambda q,\lambda p,\lambda^2 t)$. Then, as a result of our
choice of the representatives $U_{\gamma },$ we get
\begin{equation}
U_{\gamma }\left( \phi _\lambda (q,p,t) \right) = U_{\gamma }(\lambda q,
\lambda p,\lambda^2 t) = U_{\lambda^2 \gamma } (q,p,t) .
\end{equation}
So, the Wigner function transforms as:
\begin{eqnarray}
\mathcal{W}_\rho (\phi _\lambda \left( q,p \right) ; \gamma ) & = & \mathcal{%
W}_\rho (\lambda q, \lambda p ; \gamma )  \notag \\
& = & 2\, \mathrm{Tr} \left[ \rho U_{\lambda ^2\gamma }(q,p,t)\mathcal{P}
U_{\lambda^2\gamma }(q,p,t) \right]  \notag \\
& = & \mathcal{W}_\rho (q,p; \lambda^2 \gamma ),
\end{eqnarray}
while
\begin{equation}
\int \frac{\lambda^2 \gamma dqdp}{2\pi} \mathcal{W}_\rho (\lambda q, \lambda
p ; \gamma ) = \int \frac{\lambda^2\gamma dqdp} {2\pi} \mathcal{W}_\rho
(q,p;\lambda^2 \gamma ) = \mathrm{Tr}\, \rho .
\end{equation}
The dilation transformation may be more interestingly written as:
\begin{equation}
\mathcal{W}_\rho \left( \lambda q, \lambda p ; \frac{\gamma }{\lambda^2}
\right) = \mathcal{W}_\rho (q,p;\gamma ).
\end{equation}
We observe that the dilation $(\lambda q, \lambda p,\lambda^2 t)$ yields the
expected dilation $\gamma /\lambda^2$ on the label $\gamma$, which is `dual'
of the parameter $t$. For an infinitesimal dilation $\lambda =1+\epsilon $
we may expand:
\begin{eqnarray}
\mathcal{W}_\rho (q,p;\gamma ) &=& \mathcal{W}_\rho \left( \left( 1+\epsilon
\right) (q,p) ; \frac{\gamma }{\left( 1+\epsilon \right) ^{2}}\right) \\
& = & \mathcal{W}_\rho (q,p; \gamma ) + \epsilon \left[ \mathbf{v}\frac{%
\partial \mathcal{W}_{\rho }}{\partial \mathbf{v}}(q,p;\gamma ) - 2 \gamma
\frac{\partial \mathcal{W}_\rho}{\partial \gamma }(q,p ; \gamma )\right] + O
(\epsilon ^{2}) ,  \notag
\end{eqnarray}
where we have used the notation $\mathbf{v} = (q,p)$ and
\begin{equation}
\mathbf{v}\frac{\partial \mathcal{W}_{\rho }}{\partial \mathbf{v}} = q \frac{%
\partial \mathcal{W}_{\rho }}{\partial q}+ p\frac{\partial \mathcal{W}_{\rho
}}{\partial p} .
\end{equation}
Then we obtain the following differential equation for the Wigner function:
\begin{equation}  \label{dilatdifferential}
\mathbf{v}\frac{\partial \mathcal{W}_{\rho }}{\partial \mathbf{v}}(\mathbf{v}%
;\gamma )-2\gamma \frac{\partial \mathcal{W}_{\rho }}{\partial \gamma }(%
\mathbf{v};\gamma )=0.
\end{equation}

So far, we have put $\hbar =1$. It is possible however to study the
dependence on $\hbar $ by using the displacement operators given, instead of
Eq. (\ref{displacement}), by the expressions:
\begin{equation}
D_{\hbar }(q,p)=\exp \left[ i\left( \frac{p}{\sqrt{\hbar }}\frac{Q}{\sqrt{%
\hbar }}-\frac{q}{\sqrt{\hbar }}\frac{P}{\sqrt{\hbar }}\right) \right] ,
\end{equation}
and the canonical commutation relations:
\begin{equation}
\frac{1}{\hbar }\left[ Q,P\right] =iI,
\end{equation}
while $t$ gives place to $t/\hbar $ and the unitary representation given by
Eq. (\ref{repres}) becomes:
\begin{equation}
U\left( \frac{q}{\sqrt{\hbar }},\frac{p}{\sqrt{\hbar }},\frac{t}{\hbar }%
\right) =D_{\hbar }(q,p)e^{i\frac{t}{\hbar }I}=\exp \left( i\frac{R(q,p)}{%
\hbar }\right) e^{i\frac{t}{\hbar }I},
\end{equation}
so that eventually we get the above formulae with $\gamma $ replaced by $%
\gamma /\hbar $ everywhere. In particular, for the Wigner function we have:
\begin{equation}
\mathcal{W}_{\rho }\left( \sqrt{\gamma }\mathbf{v};1\right) =\mathcal{W}%
_{\rho }(\mathbf{v};\gamma )\longrightarrow \mathcal{W}_{\rho }\left( \sqrt{%
\frac{\gamma }{\hbar }}\mathbf{v};1\right) =\mathcal{W}_{\rho }\left(
\mathbf{v};\frac{\gamma }{\hbar }\right) .  \label{scaling_W}
\end{equation}

Under the action of a dilation,
\begin{equation}
(\lambda \mathbf{v},\lambda ^{2}t)\rightarrow (\lambda \mathbf{v};\gamma
/\lambda ^{2})\rightarrow (\lambda \mathbf{v};\gamma /\lambda ^{2}\hbar )
\label{scaling}
\end{equation}
and we may choose $\gamma =1,$ to get a differential equation for the Wigner
function $\mathcal{W}_{\rho }\left( \mathbf{v};\frac{1}{\hbar }\right) $
corresponding to the infinitesimal `dilation' $(\lambda \mathbf{v};1/\lambda
^{2}\hbar )$.

Notice that the scaling properties Eq.s (\ref{scaling_W}, \ref{scaling}) is
consistent with the dependence on $\hbar $ of the Wigner function. We recall
that the density state $\rho $ has the dimension of an inverse length $\ell
^{-1}$ , where $\ell \sim \sqrt{\hbar }/\sqrt{m\omega }:$ so \ $\rho
(x,x^{\prime })=\ell ^{-1}\rho ^{\prime }(x/\ell ,x^{\prime }/\ell )$. Then
it is easy to check that the following property holds:
\begin{equation}
\mathcal{W}(\frac{q}{\lambda },\frac{p}{\lambda },\frac{1 }{\lambda ^{2}\hbar%
}) = \mathcal{W}(q,p,\frac{1}{\hbar} )\ .  \label{scal wign}
\end{equation}
We refer to \cite{Ib09} for more details and further results on this
direction.

\subsection{Tomograms, Wigner functions and Radon Transform}

By means of the reconstruction formula (\ref{symprec}) the Wigner function $%
\mathcal{W}(p,q)$ of a density state $\rho (q,q^{\prime })$ may be recast in
the form:
\begin{equation}
\mathcal{W}(p,q)=\frac{\hbar }{2\pi }\int \mathcal{T}(X,\mu ,\nu )\exp \left[
i(X-\mu q-\nu p)\right] dXd\mu d\nu .  \label{Wigrec}
\end{equation}
The above equation explicitely contains the Planck constant $\hbar $, to be
coherent with Eq. (\ref{symprec}). Hereafter $\hbar =1,$ however. We recall
that in general the Wigner function is not a fair probability distribution
as it is not non-negative; nevertheless, it is a function on the phase space
of the system, and its reconstruction formula (\ref{Wigrec}) is just the
Radon anti-transform of the tomogram.

The Radon transform \cite{Radon1917} originally was formulated to solve the
problem of reconstructing a function $f(p,q)$ from its integrals on
arbitrary straight lines $\mu q+\nu p=X$\ in the $(q,p)$-plane
\begin{equation}
\int f(p,q)\delta (X-\mu q-\nu p)dpdq=:(\mathcal{R}f)(X,\mu ,\nu ).
\label{RadTra}
\end{equation}
Here $\delta $ is the Dirac delta function and the parameters $X,\mu ,\nu $
are real. The homogeneity property follows from the properties of the delta
function. The inverse transform reads:
\begin{equation}
f(p,q)=\frac{1}{\left( 2\pi \right) ^{2}}\int (\mathcal{R}f)(X,\mu ,\nu
)\exp \left[ i(X-\mu q-\nu p)\right] dXd\mu d\nu .
\end{equation}
\textbf{Remark}: Additional hypotheses, such as global integrability
conditions, are required to guarantee the uniqueness of the inverse
transform \cite{armitage}. We point out that here a subclass of functions is
selected by requiring that our `manipulations' provide us with an injective
map.

In a general sense, we may call the Radon transform (\ref{RadTra}), the
tomogram of the function $f(p,q),$ therefore the ambient space for
tomographic states is provided by the range of the Radon transform when it
is applied to properly chosen functions on phase space.

The problem we address now is the following. Let us select two classes of
functions on the phase space satisfying special conditions. The first class
of functions consists of all probability distribution densities on phase
space (the $q-p$ plane) describing states of classical particles. The second
class consists of all the Wigner functions describing quantum states thought
of as rank-one projectors. We study tomogram properties of these two
classes. There exists also a class of tomograms which are Radon transforms
of the Weyl symbol of observables. These tomograms are not probability
densities.

So, the symplectic tomogram we dealt with in section \ref{Pauli_problem} may
be eventually interpreted as the Radon transform of the Wigner function
\begin{eqnarray}
\mathcal{T}(X,\mu ,\nu ) &=&\int \frac{1}{2\pi }\mathcal{W}(p,q)\delta
(X-\mu q-\nu p)dpdq  \label{tomwig} \\
&=&\int \rho (y,y^{\prime })\varphi _{X\mu \nu }^{\ast }(y)\varphi _{X\mu
\nu }(y^{\prime })dydy^{\prime }\ .  \notag
\end{eqnarray}

The standard description of classical states with fluctuations is given by a
non-negative joint probability distribution function $f(p,q)$ on the phase
space (a plane, for a particle with one degree of freedom). The function is
normalized, i.e.
\begin{equation}
\int f(p,q)dpdq=1\ .
\end{equation}
The classical state tomogram $(\mathcal{R}f)(X,\mu ,\nu )$ can be written in
the form
\begin{equation}
(\mathcal{R}f)(X,\mu ,\nu )=\left\langle \delta (X-\mu q-\nu p)\right\rangle
_{f}  \label{tomclas}
\end{equation}
where the average is done using the probability distribution $f(p,q)$ in the
phase space \cite{manman, manmen2}. The tomogram is the probability
distribution function in a rotated and scaled reference frame on the phase
space. It can be expressed in terms of a scaling parameter $s$ and a
rotation parameter $\theta :$%
\begin{equation}
\mu =s\cos \theta \ ,\ \nu =s^{-1}\sin \theta .
\end{equation}
For fixed $\mu $ and $\nu $ one then gets a line $X=\mu q+\nu p$ in the
plane $(q,p)$ with an orientation $\theta $ from the position axis. Thus the
physical meaning of the variable $X$ is that it is the `position'\ of the
particle measured in the reference frame of the phase-space whose axes are
rotated by an angle $\theta $ with respect to the old reference frame, after
preliminary canonical scaling of the initial position $q\rightarrow sq$ and
momentum $p\rightarrow s^{-1}p$. The coordinates $X$ and $Y=-s^{2}\nu
q+s^{-2}\mu p$ provide a canonical transformation preserving the symplectic
form in the phase space. For that reason the classical tomogram is called
`symplectic'.

In the quantum case, Eq. (\ref{tomwig}) can be written in a form similar to
Eq. (\ref{tomclas}):
\begin{equation}
\mathcal{T}_{\rho }(X,\mu ,\nu )=\left\langle \delta (X-\mu Q-\nu
P)\right\rangle _{\rho }.  \label{tomwigdel}
\end{equation}
The difference with Eq. (\ref{tomclas}) is that here the position and
momentum are quantum operators $Q$ and $P$, and therefore we have to take
into account uncertainty relations. The averaging in Eq. (\ref{tomwigdel})
is done using a density state $\rho ,$ i.e.
\begin{equation}
\left\langle A\right\rangle _{\rho }:=\text{Tr}(\rho A).
\end{equation}

For fixed $\mu $ and $\nu $, the operator
\begin{equation}
X(\mu ,\nu )=\mu Q+\nu P
\end{equation}
together with its conjugate
\begin{equation}
Y(\mu ,\nu )=-s^{2}\nu Q+s^{-2}\mu P
\end{equation}
satisfies the canonical commutation relations of the Weyl--Heisenberg
algebra: $[X(\mu ,\nu ),Y(\mu ,\nu )]=[Q,P].$ The observable $X(\mu ,\nu )$
is a new position operator, i.e. the position after a symplectic (linear
canonical) transformation in the quantum non-commutative phase space $(Q,P)$
of the particle. The real variable $X$ gives the possible results of a
measure of $X\left( {\mu ,\nu }\right) $ and runs over the spectrum of $%
X\left( {\mu ,\nu }\right) $. In this way a description of quantum tomograms
is recovered in complete analogy with the classical case. So the tomogram is
also `symplectic' in the quantum case, it is associated with an automorphism
of the Weyl--Heisenberg algebra.

In classical mechanics, the transition from the distribution function of two
canonically conjugate variables (position $q$ and momentum $p$) to the
distribution function of a $(\mu ,\nu )-$family of position variables
(position $X$) does not play a crucial role, due to absence of quantum
mechanical constraints like the uncertainty relations of Heisenberg \cite%
{Heis27} and Schr\"{o}dinger-Robertson \cite{Rob29,Shr30,SudBh,Dod}. On the
contrary, the use of tomograms in quantum mechanics should encode the
properties required to allow for the uncertainty relations.

So, the tomograms of all admissible functions $f(p,q)$ form an ambient space
(here admissibility means only that the Radon transform exists and is
one-to-one). This space contains the subset of probability densities. In
turn, the subset of the probability densities contains two subsets. One
subset contains the Radon transforms of Wigner functions which are
probability densities (quantum domain). The other one contains the Radon
transforms of\ classical probability distributions on phase space. These two
subsets have a not empty intersection. Both these subsets are embedded into
the total set of tomograms, which therefore contains tomographic functions
corresponding neither to classical nor to quantum states.

\subsection{Distributions and quasi-distributions: classical and quantum}

\label{Distrib_quasi}

As we have argued, on the same space an object like a symplectic tomogram $%
\mathcal{T}(X,\mu ,\nu )$ may determine a state both in classical and in
quantum domain. Let us discuss some difference which exist for these two
domains in the context of the tomographic description. State tomograms in
both domains must satisfy the following common requirements:

\begin{enumerate}
\item Nonnegativity: $\mathcal{T}(X,\mu ,\nu )\geq 0$.

\item Integrability: $\int \mathcal{T}(X,\mu ,\nu )dX<\infty$.

\item Homogeneity: $\mathcal{T}(\lambda X,\lambda \mu ,\lambda \nu )=\frac{1%
}{\left\vert \lambda \right\vert }\mathcal{T}(X,\mu ,\nu ) $.
\end{enumerate}

Other properties of the symplectic state tomograms are required to
distinguish the states in quantum and classical domains. For example, the
necessary condition for the tomogram of a classical state is the
nonnegativity of its Radon anti-transform, i. e.
\begin{equation}
\int \mathcal{T}(X,\mu ,\nu )\exp \left[ i(X-\mu q-\nu p)\right] dXd\mu d\nu
\geq 0\ .  \label{Class}
\end{equation}
The violation of this inequality means that the tomogram does not describe a
classical state. On the other side the condition for a symplectic tomogram
to describe a quantum state can be formally written in an analogous way, as
\begin{equation}
\int \mathcal{T}(X,\mu ,\nu )\exp \left[ i(X-\mu Q-\nu P)\right] dXd\mu d\nu
\geq 0\ ,  \label{Quantum}
\end{equation}
which means that the operator obtained from the above Radon anti-transform,
being a density state, must be a non-negative normalized functional on
observables. If the inequality is violated, the tomogram does not describe a
quantum state. Tomograms satisfying both conditions belong to the
intersection of quantum and classical domain, so they may be chosen as
starting Cauchy datum of either quantum or classical time evolutions. For
example, the gaussian tomogram of coherent or squeezed and correlated states
\cite{Dod, Annast, Hall} satisfies both inequalities.

Now, the question arises if the previous requirements of nonnegativity,
integrability and homogeneity are sufficient to select only classical or
quantum tomograms. In other words, there exist tomograms satisfying the
three requirements, but violating both inequalities? It would mean that
these tomograms need not describe a state, neither a classical one nor a
quantum one if additional requirements are not met. An example of such a
kind of tomograms may be manufactured to answer in the affirmative this
question. The example is provided by scaling the parameters $\mu ,\nu $ of
the tomogram of the first excited state of an harmonic oscillator
\begin{equation}
\mathcal{T}_{1}(X,\mu ,\nu )=\frac{2 e^{-\frac{1}{\mu ^{2}+\nu ^{2}}X^{2}}}{%
\sqrt{\pi (\mu ^{2}+\nu ^{2})}}\frac{X^{2}}{\mu ^{2}+\nu ^{2}}
\end{equation}
by means of a real parameter $\lambda $ obtaining
\begin{equation}
\mathcal{T}_{\lambda }(X,\mu ,\nu )=\frac{2 e^{-\frac{1}{\left( \lambda \mu
\right) ^{2}+\left( \lambda \nu \right) ^{2}}X^{2}}}{\sqrt{\pi (\left(
\lambda \mu \right) ^{2}+\left( \lambda \nu \right) ^{2})}}\frac{X^{2}}{%
\left( \lambda \mu \right) ^{2}+\left( \lambda \nu \right) ^{2}} .
\end{equation}
This new tomogram is still positive, integrable and homogeneous. But the
quantum inequality is not fulfilled by $\mathcal{T}_{\lambda }$\ for $%
\lambda \neq 1,$ as discussed in \cite{Beppe george olga PL} where, in a
different contest, it is shown that the fidelity of the scaled tomogram and
the genuine tomogram of the harmonic oscillator ground state is negative: $%
f=-2\left| \lambda \right| ^{2}$ for small $\lambda .$ Also, the classical
inequality is violated by the scaled tomogram, because its Radon
anti-transform yields a distribution (generalized function) which is
negative for small $q,p.$

In conclusion, the set of non-negative, integrable and homogeneous
tomographic functions is divided into three parts: one containing tomograms
of quantum states, another one containing those of classical states and a
third part containing neither. The first two parts intersect each other in
the domain of tomograms satisfying both quantum and classical inequalities.
The third part does not intersect the others, and contains tomographic
functions describing neither quantum nor classical states therefore they
violate both inequalities. As a consequence, more constraints than the
previous ones are needed to unambiguously select quantum, or classical,
tomograms in order to give a tomographic version of quantum mechanics fully
equivalent to the usual formulations and yielding the classical mechanics in
an appropriate limit. At least in principle, for the symplectic case they
are sufficiently described by the quantum, or classical, inequality.

As intrinsic characterization of classical tomogram is the property that the
Fourier components of the classical tomograms is non-negative (see Eq. (\ref%
{Class})). The property of quantum tomograms given by inequality (\ref%
{Quantum}) is also intrinsic but needs the additional construction of
operators $Q$ and $P$ on a Hilbert space. One can formulate the inequality
in a form equivalent to Eq. (\ref{Quantum}) without introducing these
operators. This is accomplished replacing the nonnegativity condition of the
operator with Sylvester criterion for the matrix of the operator
corresponding to the integral kernel of Eq. (\ref{Quantum}). It means that
in any basis the principal minors of such a matrix are non-negative. This
condition is given in form of algebraic integral inequalities, not
containing formally mention of a Hilbert space.

Another intrinsic description of the set of quantum tomograms can be
formulated in the following way. The tomograms of coherent states are
gaussian and belong to the intersection of quantum and classical sets. The
intersection is invariant under the operation of taking convex sums of the
tomograms. Due to the completeness property of coherent states, the
composition of all coherent state tomograms given by Eq.s (\ref{L}), (\ref%
{L1}) provides the set of all quantum tomograms corresponding to pure
quantum states. Then, combining with convex sums these pure quantum states,
we get any mixed state tomogram. In this way the whole set of quantum
tomograms is recovered.

\subsection{Tomographic sets and state reconstruction}

>From the conceptual point of view, Pauli's problem raises a central
question in the formulation of quantum mechanics. In general, any system may
be described by a set $\mathcal{O}$ of observables and a `dual'\ set of
density states $\mathcal{S},$ which together give rise to probability
measures on the real axis, they are just the probability distributions of
the values of the observables in the states.

In the usual formulation of quantum mechanics, any observable $A$, i.e., an
hermitian operator, uniquely determines a projector valued measure (PVM) $%
P_{A}(E)$ on the sets $E$ of the Borel $\sigma $-algebra of the real line
\cite{reed-simon, Stulpe book}, so that from a density state $\rho $ a
probability measure $m_{\rho ,A}$ can be defined as
\begin{equation}
m_{\rho ,A}(E):=\mathrm{Tr}(\rho P_{A}(E))\ .
\end{equation}
It is just the probability that the mean value of the observable $A$ in the
state $\rho $ belongs to $E$. As a consequence:
\begin{equation}
m_{\rho ,A}(\mathbb{R})=1\ .
\end{equation}
The mean value of $A$\ on the state $\rho ,$\ when it exists, can be written
as an integral over a real variable $\lambda $ with respect to that
probability measure:
\begin{equation}
\mathrm{Tr}(\rho A)=\int \lambda m_{\rho ,A}\ .
\end{equation}
This for pure states $\rho =\left| \psi \right\rangle \left\langle \psi
\right| $\ reads
\begin{equation}
\left\langle \psi \left| A\right| \psi \right\rangle =\int \lambda m_{\rho
,A}\ .
\end{equation}
and a functional calculus for hermitian operators can be constructed by
defining the operator $f(A)$ as:
\begin{equation}  \label{fA}
\left\langle \psi \left| f(A) \right| \psi \right\rangle =\int f(\lambda)
m_{\rho ,A}\ .
\end{equation}
for any integrable function $f$. So, the knowledge of $m_{\rho ,A}$ for any
state $\rho $ and fixed $A$ allows for the reconstruction of $\left\langle
\psi \left| A\right| \psi \right\rangle $ for all $\psi $ in the domain of $%
A $ and therefore of $A.$ In fact, the matrix elements of $A$ in any chosen
basis of its domain are given by the polarization identity:
\begin{equation}
\left\langle \psi |\varphi \right\rangle =\frac{1}{4}\left[ \left| \left|
\psi +\varphi \right| \right| ^{2}+\left| \left| \psi -\varphi \right|
\right| ^{2}-i\left| \left| \psi +i\varphi \right| \right| ^{2}+i\left|
\left| \psi -i\varphi \right| \right| ^{2}\right] \ .
\end{equation}

Viceversa, when $\rho $ is fixed, the knowledge of $m_{\rho ,A}$ for any
observable $A,$ in particular for all projectors $\left| \psi \right\rangle
\left\langle \psi \right| ,$ allows for the reconstruction of $\ \mathrm{Tr}%
(\rho \left| \psi \right\rangle \left\langle \psi \right| )=\left\langle
\psi \left| \rho \right| \psi \right\rangle $ and therefore, by
polarization, of $\rho .$

For instance, when $A$ is the position $Q,$ the associated projectors $%
P_{Q}(E)$ act on wave functions as a multiplication by the characteristic
function of the Borel set $E$:
\begin{equation}
(P_{Q}(E)\psi )(x)=\chi _{E}(x)\psi (x)
\end{equation}
Now, a density state $\rho $ can be spectrally decomposed in terms of
rank-one projectors as (because selfadjoint compact operator):
\begin{equation}
\rho =\sum\nolimits_{k}\alpha _{k}P_{k}\ ;\ \ \alpha _{k}\geq 0,\
\sum\nolimits_{k}\alpha _{k}=1,
\end{equation}
therefore the previous formulae became:
\begin{eqnarray}
m_{\rho ,Q}(E)&:=&\mathrm{Tr}(\rho P_{Q}(E))=\sum\nolimits_{k}\alpha _{k}%
\mathrm{Tr}(P_{k}P_{Q}(E))  \notag \\
&=&\sum\nolimits_{k}\alpha _{k}\int \chi _{E}(x)\left| \psi _{k}(x)\right|
^{2}dx=\sum\nolimits_{k}\alpha _{k}\int_{E}\left| \psi _{k}(x)\right| ^{2}dx
\notag \\
&=&\int_{E}\sum\nolimits_{k}\alpha _{k}\left| \psi _{k}(x)\right|
^{2}dx=\int_{E}\rho (x,x)dx.
\end{eqnarray}
This shows that the probability measure $m_{\rho ,Q}(E)$ is absolutely
continuous with respect to the Lebesgue measure on $\mathbb{R},$ with
Radon-Nykodim derivative
\begin{equation}
\rho (x,x)=\sum\nolimits_{k}\alpha _{k}\left| \psi _{k}(x)\right| ^{2}\
\end{equation}
which is just the diagonal part of the density matrix in the position
representation. Then the mean value of the position operator is:
\begin{equation}
\mathrm{Tr}(\rho Q)=\int x\rho (x,x)dx
\end{equation}
or, for pure states $|\psi \rangle $:
\begin{equation}
\left\langle \psi \left| Q\right| \psi \right\rangle =\int x\left| \psi
(x)\right| ^{2}dx\ .
\end{equation}

Thus, Pauli's problem may be reformulated as: To determine the state from
the knowledge of a pair of probability measures $m_{\rho ,A}$, i.e. when $A$
is the position $Q$ or the momentum $P$. This set is not sufficient, while
the symplectic tomography provides a set of observables $X(\mu ,\nu )=\mu
Q+\nu P$ which is sufficient for the reconstruction. Notably, the symplectic
set is generated by the position operator $Q$ under the action of a family
of unitary transformations $S(\mu ,\nu ).$ Introducing the auxiliary
parameters $\lambda ,\theta $ as
\begin{equation}
\mu =e^{\lambda }\cos \theta \ ,\nu =e^{-\lambda }\sin \theta \ ,
\label{Symp1}
\end{equation}
we can write
\begin{equation}
S(\mu ,\nu )=\exp \left[ \frac{i\lambda }{2}\left( QP+PQ\right) \right] \exp %
\left[ \frac{i\theta }{2}\left( Q^{2}+P^{2}\right) \right] \ ,  \label{Symp2}
\end{equation}
so that
\begin{equation}
S(\mu ,\nu )QS^{\dagger }(\mu ,\nu )=\mu Q+\nu P\ .  \label{Symp3}
\end{equation}

Thus, the transformation $S(\mu ,\nu )$ yields the appropriate probability
measure associated to the observable $X(\mu ,\nu )$
\begin{equation}
m_{\rho ,X(\mu ,\nu )}(E):=\mathrm{Tr}(\rho S(\mu ,\nu )P_{Q}(E)S^{\dagger
}(\mu ,\nu ))=\int_{E}\left\langle X\left| S^{\dagger }(\mu ,\nu )\rho S(\mu
,\nu )\right| X\right\rangle dX
\end{equation}
whose density is just the tomographic probability distribution
\begin{equation}
\mathcal{T}_{\rho }(X,\mu ,\nu ):=\left\langle X\left| S^{\dagger }(\mu ,\nu
)\rho S(\mu ,\nu )\right| X\right\rangle =\mathrm{Tr}(\rho S(\mu ,\nu
)\left| X\right\rangle \left\langle X\right| S^{\dagger }(\mu ,\nu ))
\label{project}
\end{equation}
where the kets $\left| X\right\rangle $ are the eigenkets of $Q:Q\left|
X\right\rangle =X\left| X\right\rangle .$

\subsection{Rank-one projectors as tomographic sets}

In view of formula (\ref{project}) we may consider in general tomographic,
i.e. (possibly over-) complete, sets of rank-one projectors. In a sense,
they are the elementary `building blocks'\ of any tomography. Moreover, as
we will show in the following, tomographic family of rank-one projectors
allow to clarify readily the ingredients of a tomographic reconstruction
formula.

We start with an abstract finite dimensional case. Assume the Hilbert space
of the vector states $\mathcal{H}$ to be $n-$dimensional, so that rank-one
projectors span a $n^{2}-$dimensional Hilbert space $\mathbb{H=}\mathcal{%
H\otimes H^{\ast }},$ containing all the density states as well as the
(bounded) operators on $\mathcal{H},$ i.e. $\mathbb{H}=B(\mathcal{H}).$ The
scalar product is given by the trace: $\left\langle A|B\right\rangle =%
\mathrm{Tr}(A^{\dagger }B).$ A \textit{minimal} tomographic set is a basis $%
\left\{ P_{k}\right\} ,{k\in \left\{ 1,...,n^{2}\right\} },$ of rank-one
projectors, which may be orthonormalized by a Gram-Schmidt procedure:
\begin{equation*}
\left\vert V_{j}\right\rangle =\sum\limits_{k=1}^{n^{2}}\gamma
_{jk}\left\vert P_{k}\right\rangle \quad ,\quad \left\langle
V_{i}|V_{j}\right\rangle =\delta _{ij}\ .
\end{equation*}
In general, every $\left\vert V_{j}\right\rangle $ is a linear combination
of projectors, rather than a single projector like $\left\vert
P_{k}\right\rangle .$ Then a resolution of the super-unity on $\mathbb{H}$
in terms of the $P$'s reads as
\begin{eqnarray}
\mathbb{\hat{I}}_{n^{2}} &=&\sum\limits_{i=1}^{n^{2}}\left\vert
V_{i}\right\rangle \left\langle V_{i}\right\vert
=\sum\limits_{i,j,l=1}^{n^{2}}\gamma _{il}^{\ast }\gamma _{ij}P_{j}\mathrm{Tr%
}(P_{l}\bullet )  \notag \\
&=&\sum\limits_{l=1}^{n^{2}}\left\vert G_{l}\right\rangle \left\langle
P_{l}\right\vert =\sum\limits_{j=1}^{n^{2}}\left\vert P_{j}\right\rangle
\left\langle G_{j}\right\vert  \label{dualid}
\end{eqnarray}
where the dual set of Gram-Schmidt operators $\left\{ G_{k}\right\} ,{k\in
\left\{ 1,...,n^{2}\right\} ,}$ has been introduced:
\begin{equation}
\left\vert G_{l}\right\rangle =\sum\limits_{i=1}^{n^{2}}\gamma _{il}^{\ast
}\left\vert V_{i}\right\rangle =\sum\limits_{i,j=1}^{n^{2}}\gamma
_{il}^{\ast }\gamma _{ij}\left\vert P_{j}\right\rangle .
\end{equation}
We observe that $G_{l}$ is a nonlinear function of the projectors $\left\{
P_{k}\right\} $, because also the coefficients $\gamma $'s depend on the
projectors. Moreover:
\begin{eqnarray}
\left\langle P_{i}|G_{l}\right\rangle &=&\sum\limits_{j=1}^{n^{2}}\gamma
_{jl}^{\ast }\left\langle P_{i}|V_{j}\right\rangle
=\sum\limits_{j,k=1}^{n^{2}}\gamma _{jl}^{\ast }(\gamma ^{\ast
})_{ik}^{-1}\left\langle V_{k}|V_{j}\right\rangle  \notag \\
&=&\sum\limits_{j=1}^{n^{2}}\gamma _{jl}^{\ast }(\gamma ^{\ast
})_{ij}^{-1}=\delta _{il}.
\end{eqnarray}

Thus, Eq. (\ref{dualid}) shows that a resolution of the (super-) identity,
associated with a tomographic reconstruction formula, is determined by a
pair of dual sets, the $P^{\prime }$s and the $G^{\prime }$s. Besides the
role of the dual operators may be interchanged in such a formula, so that a
tomography would be better defined in terms of a pair of dual sets. In the
context of the harmonic analysis and of the wavelet signal analysis, these
dual sets are known as dual frames \cite{daubechies}.

Formulae similar to the previous ones hold for any tomographic, i.e. (over-)
complete, set of rank-one projectors. In fact, along with the minimal
tomographic set discussed above, that is a quorum of rank-one projectors, it
is very useful to deal with over-complete or even maximal sets of rank-one
projectors, obtained for instance by acting on a fiducial projector $P_{0}$
with a unitary representation of a Lie group. The previous summations became
then integrations over the orbit $\Omega $ through $P_{0}$.

A suitable illustration of such tomographic sets is given by the qubit (spin
$1/2$) tomography over a $2-$dimensional Hilbert space of the vector states $%
\mathcal{H}.$ Out of the standard basis vectors $\{|m\rangle \},$ where $%
m=\pm 1/2$ is the spin projection on the $z$-xis, a fiducial projector $%
P_{0}=|m\rangle \langle m|$ is rotated by means of the operators $U$ of an
irreducible representation of the group $SU(2):$ the tomogram of any
operator $\rho $ is defined as
\begin{equation*}
{\mathcal{T}}_{\rho }(m,U)=\text{Tr}(U|m\rangle \langle m|U^{\dag }\rho
)=\langle m|U^{\dag }\rho U|m\rangle \ .
\end{equation*}
Then $\Omega $ is the orbit of the co-adjoint group action in the dual of
the algebra, that is the Bloch sphere $S^{2}$ of all rank-one projectors.
The projector $U|m\rangle \langle m|U^{\dag }$, with $U$ parametrized by the
usual Euler angles $(\theta ,\phi ,\psi )$, corresponds to a point on $S^{2}$%
determined by a unit vector $\vec{n}=(\sin \theta \cos \phi ,\cos \theta
\cos \phi ,\cos \theta ).$ In other words, we can use the $(\theta ,\phi )$
parametrization to write a generic projector in matrix form as $(m=1/2)$
\begin{equation}
P(\theta ,\phi )=\frac{1}{2}\left[
\begin{array}{cc}
1+\cos \theta & e^{-i\phi }\sin \theta \\
e^{i\phi }\sin \theta & 1-\cos \theta%
\end{array}
\right] \ .
\end{equation}
Then, the corresponding resolution of the identity reads \cite{Ventriglia2}:
\begin{equation}
\mathbb{\hat{I}}=\int_{0}^{2\pi }\int_{0}^{\pi }\left| G(\theta ,\phi
)\right\rangle \left\langle P(\theta ,\phi )\right| \sin \theta d\theta
d\phi \ ,
\end{equation}
where in matrix form:
\begin{equation}
G(\theta ,\phi )=\frac{1}{4\pi }\left[
\begin{array}{cc}
1+3\cos \theta & 3e^{-i\phi }\sin \theta \\
3e^{i\phi }\sin \theta & 1-3\cos \theta%
\end{array}
\right] \ ,
\end{equation}
so that, for any density state $\rho $ we get the dual reconstruction
formulae:
\begin{eqnarray}
\rho &=&\int_{0}^{2\pi }\int_{0}^{\pi }G(\theta ,\phi )\mathrm{Tr}(P(\theta
,\phi )\rho )\sin \theta d\theta d\phi \\
&=&\int_{0}^{2\pi }\int_{0}^{\pi }P(\theta ,\phi )\mathrm{Tr}(\hat{G}(\theta
,\phi )\rho )\sin \theta d\theta d\phi \ ,
\end{eqnarray}

We remark that, however, in the infinite dimensional case the relation $%
\mathbb{H}=B(\mathcal{H})$ is no more valid and there are several relevant
spaces. In particular the Hilbert space at our disposal is the space of the
Hilbert-Schmidt operators $\mathbb{H}\subset B(\mathcal{H})$, which is the
typical setting of the frame theory \cite{Mankaniello}, while the finest
tomographic sets of rank-one projectors have to be complete both in such a
Hilbert space and in the Banach space of the trace class operators. We will
not insist here on the topological subtleties of the infinite dimensional
case, they are discussed, e.g., in \cite{Ventriglia3, CSTom}.

\subsection{General aspects of tomography}

\label{gen_aspect}

The Pauli's problem and a possible solution have been previously analyzed
within the machinery of spectral analysis of selfadjoint operators. In this
section, inspired by that analysis, we provide a general setting for
tomography together with some\ general considerations.

When $\mathcal{S}$ is the set of states of a physical system and $\mathcal{A}
$ a suitable subset of the observables $\mathcal{O},$ a tomography $\mathcal{%
T}$\ is a map\ from $\mathcal{S\times }\left( \mathcal{A\subset O}\right) $\
into the set of probability measures on the real line $\mathbb{R}.$ It is
required that $\mathcal{T}$\ is such that if $\ $the probability measures $%
\mathcal{T}(\rho ,A)$ are known for all $A\in $ $\mathcal{A}$ it is
possible, at least in principle, to reconstruct $\rho .$

If $\mathcal{A}=\mathcal{O}$ a tomography is available, via spectral
analysis, as shown previously. In this case $\mathcal{A}$ is a huge linear
space of selfadjoint operators, bounded or not.

However, as we have seen, a tomography is available even if $\mathcal{A}$ is
restricted to a subset $\mathcal{O}^{\prime }\subset \mathcal{O}$ of all the
rank-one projectors. Now $\mathcal{O}^{\prime }$\ is a unit spherical
surface in the Hilbert space, in general infinite dimensional. It may be
more useful to restrict $\mathcal{A}$ to a subset of $\mathcal{O}$ specified
by some (multi-) parameter $\mu $ which varies in some index set $\mathcal{M}
$. $\mathcal{A}$ must still be such that the reconstruction of any state $%
\rho $ is possible. The tomogram $\mathcal{T}(\rho ,A)$ appears now as $%
\mathcal{T}(\rho ,\mu )(E)=\mathcal{T}(\rho ,A_{\mu })(E)=\mathrm{Tr}(\rho
P_{A_{\mu }}(E))$ and is a probability measure: $0\leq \mathcal{T}(\rho ,\mu
)(E)\leq 1,$ for any Borel set $E$ of reals.

In principle the elements in $\mathcal{A}$ can be selfadjoint operators with
different spectra but it is more convenient to deal with iso-spectral
operators, so that a spectrum $\sigma \subset \mathbb{R}$ is associated with
$\mathcal{A}$. In this case, for any $\rho ,\mu $ we have that $\mathcal{T}%
(\rho ,\mu )(E)=0$ for all sets $E$ which do not intersect the spectrum $%
\sigma $ and $\mathcal{T}(\rho ,\mu )(E)=1$ for sets $E$ containing the
spectrum $\sigma .$

If $\sigma $ is a purely continuous spectrum, as in the symplectic
tomography, and $\mathcal{T}(\rho ,\mu )$ is absolutely continuous with
respect to the Lebesgue measure, $\mathcal{T}(\rho ,\mu )$ will appear as $%
\mathcal{T}(X,\rho ,\mu )dX,$ where $X\in \sigma $ is the spectral variable,
with the property $\int_{\sigma }\mathcal{T}(X,\rho ,\mu )dX=1.$ In other
terms, $\mathcal{T}(X,\rho ,\mu )$ is a probability density function on the
real line concentrated on the spectrum $\sigma $.

If $\sigma $ is a pure point spectrum, $\mathcal{T}(\rho ,\mu )$ will be
concentrated on the points $X_{k}$ of $\sigma ,$ with $\mathcal{T}(\rho ,\mu
)(X_{k})=\mathrm{Tr}(\rho P_{X_{k},\mu })=\mathcal{T}(k,\rho ,\mu )$ a
positive function of $k$ such that $\sum_{k}\mathcal{T}(k,\rho ,\mu )=1.$
Now $\left\{ \mathcal{T}(k,\rho ,\mu )\right\} _{k}$ can be regarded as a
probability vector $\vec{\mathcal{T}}$ with a finite or countable number of
components, i.e. a vector with non-negative components whose sum is one.
When the spectrum is non-degenerate finite, the number $n$ of components of $%
\vec{\mathcal{T}}$ is just the dimension of the Hilbert space.
Geometrically, these $n$ components may be considered as the coordinates $%
\{\tau _{k}\}$ of a point belonging to the simplex $\tau _{1}+\tau
_{2}+\dots +\tau _{n}=1$ of $\mathbb{R}^{n}$. If the spectrum is degenerate,
the number of components of $\vec{\mathcal{T}}$ is less than the dimension
of the Hilbert space. For instance, if $\mathcal{A}$ is composed by rank-one
projectors, $\sigma =\{0,1\}$ and $\vec{\mathcal{T}}$ has two components.
However, we notice that the qubit tomographic set presented in the previous
subsection is a non-degenerate case: given the component $\tau _{1}={%
\mathcal{T}}_{\rho }(m,u),$ the other one is $\tau _{2}=1-{\mathcal{T}}%
_{\rho }(m,u)$ and the\ probability vector belongs to the simplex $\tau
_{1}+\tau _{2}=1.$

\subsection{Weyl systems and tomography}

\label{weyl_tom}

As we have seen, tomographic families of iso-spectral observables can be
obtained by conjugation of a fiducial observable $A_{0}$ by a parameterized
set of unitary operators $U_{\mu }$, which may be the elements of a unitary
(possibly square integrable \cite{Cassinelli}) representation of some Lie
group $\mathcal{G}$ and $\mu $ will then be a point on a finite dimensional
manifold. In this section the most common tomographic schemes are obtained
in an unified approach, by using a particular representation of a Lie group $%
\mathcal{G}$.

We discussed in section \ref{weyl_aut} how the Weyl map allows to associate
inner automorphisms on the space of unitary operators with symplectic
isomorphisms of the symplectic linear space $(V,\omega )$. Moreover if $%
\mathcal{G}$ is a Lie group and $T\colon \mathcal{G} \to Sp(V,\omega )$ is a
linear representation of the elements of the group by symplectic maps, then
we can associate an inner automorphism $\nu_{T_g}$ of the group of unitary
operators to any element $g$ of the Lie group by using the analog of Eq. (%
\ref{inner_aut}):
\begin{equation}
\nu _{T_{g}}(W(v))=W(T_{g}v)=U_{g}^{\dagger }W(v)U_{g}\ .
\end{equation}
where now $U_g$ denotes the unitary operator associated to $T_g$ realizing
the inner autormorphism $\nu_{T_g}$. In other words the automorphism $\nu
_{T_{g}},$ corresponding to the symplectic linear transformation $T_{g}$ of $%
V,$ is a inner automorphism of the unitary operators, as it belongs to the
component of the automorphism group which is connected with the identity
\cite{Kadison}. At the level of the infinitesimal generators of the unitary
group, we have
\begin{equation}
U_{g}^{\dagger }R(v)U_{g}=R(T_{g}v)\,.
\end{equation}

The equation above allows to make contact with most used tomographic
schemes. For instance, by acting with a rotation of an angle $\theta $
followed by a squeezing $e^{\lambda }$ we have
\begin{equation}
T_{g}=\left[
\begin{array}{cc}
e^{\lambda } & 0 \\
0 & e^{-\lambda }%
\end{array}
\right] \left[
\begin{array}{cc}
\cos \theta & \sin \theta \\
-\sin \theta & \cos \theta%
\end{array}
\right] =\left[
\begin{array}{cc}
e^{\lambda }\cos \theta & e^{\lambda }\sin \theta \\
-e^{-\lambda }\sin \theta & e^{-\lambda }\cos \theta%
\end{array}
\right]
\end{equation}
Then, by
\begin{equation}
U_{g}R(v)U_{g}^{\dagger }=R(T_{g}^{\dagger }v)\,.
\end{equation}
and
\begin{equation}
T_{g}^{\dagger }=\left[
\begin{array}{cc}
e^{\lambda }\cos \theta & -e^{-\lambda }\sin \theta \\
e^{\lambda }\sin \theta & e^{-\lambda }\cos \theta%
\end{array}
\right]
\end{equation}
we obtain \cite{Barg, Brunet}:
\begin{equation}
\left[
\begin{array}{c}
U_{g}QU_{g}^{\dagger } \\
U_{g}PU_{g}^{\dagger }%
\end{array}
\right] =\left[
\begin{array}{cc}
e^{\lambda }\cos \theta & e^{-\lambda }\sin \theta \\
-e^{\lambda }\sin \theta & e^{-\lambda }\cos \theta%
\end{array}
\right] \left[
\begin{array}{c}
Q \\
P%
\end{array}
\right] =\left[
\begin{array}{c}
e^{\lambda }\cos \theta Q+e^{-\lambda }\sin \theta P \\
-e^{\lambda }\sin \theta Q+e^{-\lambda }\cos \theta P%
\end{array}
\right] .
\end{equation}
In this way, bearing in mind Eq.s (\ref{Symp1}$\div $\ref{Symp3}), we
recover the symplectic tomographic family of operators $X(\mu ,\nu ):$
\begin{equation}
U_{g}QU_{g}^{\dagger }=e^{\lambda }\cos \theta Q+e^{-\lambda }\sin \theta
P=\mu Q+\nu P=S(\mu ,\nu )QS^{\dagger }(\mu ,\nu )=X(\mu ,\nu )\ .
\end{equation}

We get a general, unified approach by using the inhomogeneous symplectic
linear group, which is a semi-direct product of the translation group and
the linear symplectic group. In that case we have only projective
representations, whose generators are $I,Q,P,\frac{1}{2}(P^{2}+Q^{2}),\frac{1%
}{2}(PQ+QP),\frac{1}{2}(P^{2}-Q^{2}).$ Then most common tomographic schemes
may be obtained by acting with the group on its generators as fiducial
operators $A_{0}.$

For instance, by acting on $Q,\ \sigma =\mathbb{R},$ we get
\begin{eqnarray}
&&D(q,p)\tilde{S}(\alpha ,\mu ,\nu )Q\tilde{S}^{\dagger }(\alpha ,\mu ,\nu
)D^{\dagger }(q,p) \\
&&=D(q,p)(\xi Q+\eta P)D^{\dagger }(q,p)=(\xi Q+\eta P)+(\xi q-\eta p)I
\notag
\end{eqnarray}
with
\begin{equation}
\xi =\mu \cosh \alpha +\nu \sinh \alpha ;\ \eta =\nu \cosh \alpha +\mu \sinh
\alpha \ ,
\end{equation}
and where, restoring the explicit dependence of $\mu ,\nu $ on the
parameters $\lambda ,\theta ,$ the transformation $\tilde{S}(\alpha ,\mu
,\nu )\leftrightarrow \tilde{S}(\alpha ,\lambda ,\theta )$ reads:
\begin{equation*}
\tilde{S}(\alpha ,\lambda ,\theta )=\exp \left[ i\frac{\alpha }{2}%
(P^{2}-Q^{2})\right] \exp \left[ i\frac{\lambda }{2}(PQ+QP)\right] \exp %
\left[ i\frac{\theta }{2}(P^{2}+Q^{2})\right] \ .
\end{equation*}
\noindent \textbf{Remark}: Equivalently we could have acted on $P,\ \sigma =%
\mathbb{R},$ obtaining again the same tomographic family:
\begin{eqnarray}
&&D(q,p)\tilde{S}(\alpha ,-\lambda ,\theta -\frac{\pi }{2})P\tilde{S}%
^{\dagger }(\alpha ,-\lambda ,\theta -\frac{\pi }{2})D^{\dagger }(q,p) \\
&&=D(q,p)(\xi Q+\eta P)D^{\dagger }(q,p)=(\xi Q+\eta P)+(\xi q-\eta p)I\ .
\notag
\end{eqnarray}

We may recover from the above tomographic family of observables the usual
symplectic tomographic scheme \cite{QSO1996, QSO1997}. We do that by
quotienting the whole inhomogeneous group with respect to its subgroup
generated by the displacement operators times the hyperbolic rotations, \
and choosing a section
\begin{equation}
q=0,p=0,\alpha =0.
\end{equation}

Besides, by acting on the number operator
\begin{equation}
\frac{1}{2}(P^{2}+Q^{2})-\frac{1}{2}=a^{\dagger }a=N,\ \sigma =\mathbb{N}%
_{0},
\end{equation}
we get the photon number tomography \cite{Vog, Banag, Manc} on a section
\begin{equation}
\alpha =0,\lambda =0,\theta =0,
\end{equation}
while on the other section
\begin{equation}
q=0,p=0,\alpha =0,
\end{equation}
the so called squeeze tomography \cite{squeeze} is recovered.

In the coherent state tomography the fiducial operator is the projector on
the vacuum state: $a\left\vert 0\right\rangle =0,$
\begin{equation}
A_{0}=\left\vert 0\right\rangle \left\langle 0\right\vert ,\ \sigma =\left\{
0,1\right\}
\end{equation}
and the\ tomographic family of operators is generated by the displacement
operators depending on a complex parameter $z$, $\mathcal{D}(z)=\exp \left[
za^{\dagger }-z^{\ast }a\right]$, eq (\ref{displacement_complex}), or
equivalently, two real parameters. Thus one obtains a tomogram which is
formally equivalent to the Husimi function, while one of the dual
reconstruction formulae is the Sudarshan's diagonal coherent state
representation of an operator \cite{CSTom}.

We can also obtain spin tomography. For this, we observe that unitary
irreducible representations of $SU(2)$ of dimension $m\geq 2$\ are a
subgroup of $SU(m),$ which in turn is isomorphic to $SO(2m)\cap Sp(2m,%
\mathbb{R}),$ the intersection of $2m-$dimensional rotations and linear
symplectic transformations. Now, for definiteness, assume $V=\mathbb{R}^{4}$
with coordinates $(x_{1},x_{2},p_{1},p_{2})$ in the Weyl diagram, with $T$ a
$2-$dimensional unitary irreducible representations of $SU(2)$, then for any
$g \in SU(2)$:
\begin{equation}
\begin{tabular}{ccc}
$V$ & $\overset{W}{\longrightarrow }$ & $\mathcal{U}(\mathcal{H}\mathbb{)}$
\\
$T_{g}\downarrow $ &  & $\downarrow \nu _{T_{g}}$ \\
$V$ & $\overset{W}{\longrightarrow }$ & $\mathcal{U}(\mathcal{H}\mathbb{)}$%
\end{tabular}%
\end{equation}
The matrices $T_{g}$ are real, orthogonal symplectic $4\times 4$ matrices
while the unitary operators $U_{g}$ corresponding to the inner automorphism $%
\nu _{T_{g}}$ are generated by the hermitian operators
\begin{eqnarray}
J_{1} &=&\frac{1}{2}\left( P_{1}Q_{2}-P_{2}Q_{1}\right) \\
J_{2} &=&\frac{1}{2}\left( P_{1}P_{2}+Q_{1}Q_{2}\right)  \notag \\
J_{3} &=&\frac{1}{2}\left[ \frac{1}{2}\left( P_{1}^{2}+Q_{1}^{2}\right)-%
\frac{1}{2}\left(P_{2}^{2}+Q_{2}^{2}\right) \right]  \notag \\
&=&\frac{1}{2}\left[ H(P_{1},Q_{1})-H(P_{2},Q_{2})\right]  \notag
\end{eqnarray}
which satisfy the commutation relations $[J_{h},J_{k}]=i\epsilon
_{hkl}J_{l}. $ The operators $Q_{a},P_{a},a=1,2,$ generate the displacement
operator of the Weyl system.

In fact, choosing the lagrangian manifold to be $\left\{ \left(
x_{1},x_{2}\right) \right\} ,$ or $\left\{ \left( \rho ,\varphi \right)
\right\} $ in polar coordinates, then $J_{1}$ acts as $\frac{1}{2}(i\partial
/\partial \varphi) $ on $\mathcal{H}={\mathcal{L}}^{2}(\mathbb{R}^{2})$ and
its eigenfunctions are $\psi (\rho )e^{in\varphi }$ so that its spectrum is
the set of the integer and semi-integer numbers. $J_{3}$ is a
semi-difference of two isotropic harmonic oscillators, so it has the same
spectrum $\frac{1}{2}\left\{ n_{1}-n_{2}\right\}.$ Finally, by a rotation of
coordinates:
\begin{eqnarray}
\xi &=&\frac{1}{\sqrt{2}}\left( x_{1}+x_{2}\right) \quad;\quad \eta =\frac{1%
}{\sqrt{2}}\left( x_{1}-x_{2}\right) , \\
x_{1} &=&\frac{1}{\sqrt{2}}\left( \xi +\eta \right) \quad;\quad x_{2}=\frac{1%
}{\sqrt{2}}\left( \xi -\eta \right) ,  \notag \\
\frac{\partial }{\partial x_{1,2}} &=&\frac{\partial \xi }{\partial x_{1,2}}%
\frac{\partial }{\partial \xi }+\frac{\partial \eta }{\partial x_{1,2}}\frac{%
\partial }{\partial \eta }=\frac{1}{\sqrt{2}}\left( \frac{\partial }{%
\partial \xi }\pm \frac{\partial }{\partial \eta }\right) ,  \notag
\end{eqnarray}
also $J_{2}$ becomes a semi-difference of two isotropic harmonic
oscillators:
\begin{equation}
J_{2}=\frac{1}{2}\left[\frac{1}{2}\left( \frac{\partial ^{2}}{\partial \xi
^{2}}-\frac{\partial ^{2}}{\partial \eta ^{2}}\right) +\frac{1}{2}\left( \xi
^{2}-\eta ^{2}\right)\right] =\frac{1}{2}[H(P_{\xi },\xi )-H(P_{\eta },\eta
)]
\end{equation}
and obviously the spectrum is the same.

To analyze the decomposition of the infinite dimensional representation of $%
SU(2)$ in irreducible ones we evaluate the Casimir $J^{2}.$ We start with
\begin{equation}
J_{1}^{2}=\frac{1}{4}\left\{ P_{1}^{2}Q_{2}^{2}+P_{2}^{2}Q_{1}^{2}-\left[
P_{1}Q_{2}P_{2}Q_{1}+P_{2}Q_{1}P_{1}Q_{2}\right] \right\}
\end{equation}
and from $\left[ Q_{a},P_{b}\right] =i\delta _{ab},a,b=1,2,$ we have:
\begin{eqnarray}
J_{1}^{2} &=&\frac{1}{4}\left\{ P_{1}^{2}Q_{2}^{2}+P_{2}^{2}Q_{1}^{2}-\left[
P_{1}\left( P_{2}Q_{2}+i\right) Q_{1}+Q_{1}P_{1}\left( Q_{2}P_{2}-i\right) %
\right] \right\}  \notag \\
&=&\frac{1}{4}\left\{ P_{1}^{2}Q_{2}^{2}+P_{2}^{2}Q_{1}^{2}-\left(
P_{1}P_{2}Q_{2}Q_{1}+Q_{1}P_{1}Q_{2}P_{2}\right) +i[Q_{1},P_{1}]\right\}
\end{eqnarray}
while
\begin{equation}
J_{2}^{2}=\frac{1}{4}\left\{ P_{1}^{2}P_{2}^{2}+Q_{1}^{2}Q_{2}^{2}+\left(
P_{1}P_{2}Q_{1}Q_{2}+Q_{1}Q_{2}P_{1}P_{2}\right) \right\}
\end{equation}
so that
\begin{equation}
J_{1}^{2}+J_{2}^{2}=H(P_{1},Q_{1})H(P_{2},Q_{2})-\frac{1}{4}
\end{equation}
Then
\begin{eqnarray}
J^{2} &=&\frac{1}{4}\left[ H(P_{1},Q_{1})-H(P_{2},Q_{2})\right]
^{2}+H(P_{1},Q_{1})H(P_{2},Q_{2})-\frac{1}{4}  \notag \\
&=&\frac{1}{4}\left[ H(P_{1},Q_{1})+H(P_{2},Q_{2})\right] ^{2}-\frac{1}{4}.
\end{eqnarray}
In the number representation $H(P_{a},Q_{a})\rightarrow N_{a}+\frac{1}{2},$
with $N_{1}+N_{2}=N,$ it is
\begin{equation}
J^{2}=\frac{1}{4}\left( N_{1}+N_{2}+1\right) ^{2}-\frac{1}{4}=\frac{1}{2}%
N\left( \frac{1}{2}N+1\right)
\end{equation}
This shows that the representation generated by $J_{1},J_{2},J_{3}$ contains
all integer and semi-integer spins without degeneration, so it is a
Schwinger representation \cite{Schwinger}. Finally, the basis $\{\left\vert
j,m\right\rangle \},-j\leq m\leq j,$\ which diagonalizes $\{J^{2},J_{3}\}$
is $\{\left\vert n_{1}+n_{2},n_{1}-n_{2}\right\rangle =\left\vert
n_{+},n_{-}\right\rangle \},-n_{+}\leq n_{-}\leq n_{+}$ and is obtained with
operators $N_{\pm }=N_{1}\pm N_{2}$ of the isotropic bidimensional harmonic
oscillator, by rotating the old basis $\{\left\vert n_{1},n_{2}\right\rangle
=\left\vert n_{1}\right\rangle \left\vert n_{2}\right\rangle \}$ (see, e.g.,
\cite{Peppe's book}).

The whole Hilbert space $\mathcal{H}$ decomposes in a direct sum of finite
dimensional subspaces $\mathcal{H}^{j}$ of dimension $2j+1,$ which are the
carrier spaces of the spin tomography. In each $\mathcal{H}^{j},$ the $%
(2j+1)-$dimensional spin tomography uses a set of fiducial operators, the
eigenstates of the spin projection on $z-$axis $J_{3}$, a generator of an
irreducible unitary ($2j+1)-$dimensional representation $\tau $ of $SU(2):$
\begin{equation}
\{A_{0}(j,m)\}_{m\in\sigma}=\{|jm \rangle\langle jm|\}_{m\in\sigma},\ \sigma
=\left\{ -j,.......,j\right\}
\end{equation}

The unitary transformations may be either the operators of the unitary group
$U(2j+1)$ or simply those of the representation $\tau $ of $SU(2),$
depending only on the angle parameters which determine a point on the sphere
$S^{2}:$
\begin{equation}
U(g),\quad g\in U(2j+1); \quad \mathrm{or} \quad U(g)=\tau (g),\quad g\in
SU(2).
\end{equation}
In the last case the tomogram describing a spin state is a probability
distribution depending on random discrete spin projection along the
direction determined by the angle parameters. General reconstruction
formulae for the spin tomography were obtained in \cite{OlgaJetph}.

\newpage

\section{Quantum Mechanics in the tomographic picture}

\subsection{Superposition rule}

Having put the tomographic approach within a general setting of states $%
\mathcal{S}$ and observables $\mathcal{O}$, we may now ask how one can add
probabilities to describe the superposition of pure quantum states which
allows for the description of interference phaenomena. For the case of two
(orthonormalized) state vectors $|\psi _{1}\rangle $ and $|\psi _{2}\rangle
, $ given the probabilities (non-negative numbers) $p_{1}$ and $p_{2}$, with
$p_{1}+p_{2}=1$, the superposition rule states that the linear combination
\begin{equation}
|\psi \rangle =\sqrt{p_{1}}|\psi _{1}\rangle +\mathrm{e}^{i\varphi }\sqrt{%
p_{2}}|\psi _{2}\rangle  \label{A}
\end{equation}
is also a (pure) state vector. Here the factor $\mathrm{e}^{i\varphi }$
corresponds to the relative phase of the two state vectors. In the
tomographic picture the formulation of the superposition principle should be
discussed in terms of density states. In fact, since density states depend
nonlinearly on state vectors:
\begin{equation}
\rho _{1}=\left\vert \psi _{1}\right\rangle \left\langle \psi
_{1}\right\vert \ ,\ \rho _{2}=\left\vert \psi _{2}\right\rangle
\left\langle \psi _{2}\right\vert ,  \label{B}
\end{equation}
a naive way to add them, that is $p_{1}\rho _{1}+p_{2}\rho _{2},$ yields an
operator corresponding to a mixed quantum state, different from a rank-one
projector $\rho =\left\vert \psi \right\rangle \left\langle \psi \right\vert
$ corresponding to the pure state of Eq. (\ref{A}). However, a way to add
rank-one projectors to obtain a rank one projector has been proposed in \cite%
{SudPL} and is given by
\begin{equation}
\rho =p_{1}\rho _{1}+p_{2}\rho _{2}+\sqrt{p_{1}p_{2}}\frac{\rho
_{1}P_{0}\rho _{2}+\rho _{2}P_{0}\rho _{1}}{\sqrt{\mathrm{Tr}(\rho
_{1}P_{0}\rho _{2}P_{0})}}\quad ,  \label{D}
\end{equation}
where $P_{0}$ is a fiducial rank-one projector corresponding to the
arbitrary phase factor $\mathrm{e}^{i\varphi }$ of Eq. (\ref{A}). The
previous formula can be viewed\ as a purification of the mixed quantum state
$p_{1}\rho _{1}+p_{2}\rho _{2}$ . If $\rho _{1}$ and $\rho _{2}$ are not
orthogonal, the formula yields a non-normalized $\rho ,$ so it requires a
normalization factor $\left( \mathrm{Tr}\rho \right) ^{-1}.$ For the
denominator to be different from zero we need to superpose only states which
are not orthogonal to $P_{0}.$

The composition rule for tomographic probabilities corresponding to the
superposition of density states (Eq. (\ref{D})) reads, using $\vec{x}=(X,\mu
,\nu )$%
\begin{equation}
\mathcal{T}_{\rho }(\vec{x})=p_{1}\mathcal{T}_{\rho _{1}}(\vec{x})+p_{2}%
\mathcal{T}_{\rho _{2}}(\vec{x})+\frac{\sqrt{p_{1}p_{2}}}{\sqrt{\mathrm{Tr}%
(\rho _{1}P_{0}\rho _{2}P_{0})}}(\mathcal{T}_{\rho _{1}P_{0}\rho _{2}}(\vec{x%
})+\mathcal{T}_{\rho _{2}P_{0}\rho _{1}}(\vec{x})).
\end{equation}
This equations contains tomograms such as $\mathcal{T}_{\rho _{1}P_{0}\rho
_{2}}(\vec{x})$ which is possible to express in terms of tomograms of $\rho
_{1}$ and $\rho _{2},$ by means of the star-product formalism.

In a star-product procedure, operators are replaced by their symbols, which
are functions along the lines of section \ref{Wigner_functions}:
\begin{equation}
\hat{A}\longleftrightarrow f_{A}(\vec{x})\ .
\end{equation}
The associative product of operators is mapped onto an associative product
of symbols
\begin{equation}
\hat{A}\hat{B}\longleftrightarrow f_{AB}(\vec{x}):=f_{A}(\vec{x})\ast f_{B}(%
\vec{x})\ .
\end{equation}
This star-product is described through a nonlocal integral kernel as
\begin{equation}
f_{AB}(\vec{x})=\int f_{A}(\vec{x}_{1})K(\vec{x}_{1},\vec{x}_{2},\vec{x}%
)f_{B}(\vec{x}_{2})d\vec{x}_{1}d\vec{x}_{2}\ .
\end{equation}
Tomographic probabilities (tomograms) can be considered as symbols of
operators in a specific kind of star-product scheme.

In \cite{BeppeOlga, OlgaBeppe} it is shown that the symplectic tomogram $%
\mathcal{T}_{A}(X,\mu ,\nu )$ is the symbol of an operator $A$ in a specific
instance of star-product, with star-product kernel depending on the
continuous variables $(X,\mu ,\nu )=\vec{x}.$ It is given by \cite{OlgaKap}:
\begin{eqnarray}
&&K(X_{1},\mu _{1},\nu _{1},X_{2},\mu _{2},\nu _{2},X,\mu ,\nu )=\frac{1}{%
4\pi ^{2}}\delta (\mu (\nu _{1}+\nu _{2})-\nu (\mu _{1}+\mu _{2}))  \label{H}
\\
&&\times \exp \left\{ \frac{i}{2}\left[ \nu _{1}\mu _{2}-\nu _{2}\mu
_{1}+2X_{1}+2X_{2}-\left( \frac{\nu _{1}+\nu _{2}}{\nu }+\frac{\mu _{1}+\mu
_{2}}{\mu }\right) X\right] \right\} \ .  \notag
\end{eqnarray}

Thus, the addition rule for tomographic probabilities, corresponding to the
superposition of density states Eq.(\ref{D}), reads, using again $\vec{x}%
=(X,\mu ,\nu )$%
\begin{eqnarray}
\mathcal{T}_{\rho }(\vec{x})&=&p_{1}\mathcal{T}_{\rho _{1}}(\vec{x})+p_{2}%
\mathcal{T}_{\rho _{2}}(\vec{x})  \label{L} \\
&+&\sqrt{p_{1}p_{2}}\frac{\mathcal{T}_{\rho _{1}}(\vec{x})\ast \mathcal{T}%
_{0}(\vec{x})\ast \mathcal{T}_{\rho _{2}}(\vec{x})+\mathcal{T}_{\rho _{2}}(%
\vec{x})\ast \mathcal{T}_{0}(\vec{x})\ast \mathcal{T}_{\rho _{1}}(\vec{x})}{%
\sqrt{C(\mathcal{T}_{\rho _{1}}(\vec{x})\ast \mathcal{T}_{0}(\vec{x})\ast
\mathcal{T}_{\rho _{2}}(\vec{x})\ast \mathcal{T}_{0}(\vec{x}))}}  \notag
\end{eqnarray}
where, restoring $(X,\mu ,\nu ):$%
\begin{equation}
\mathrm{Tr}(A)=C(\mathcal{T}_{A}(X,\mu ,\nu ))=\int \mathcal{T}_{A}(X,\mu
,\nu )\mathrm{e}^{iX}\delta (\mu )\delta (\nu )dXd\mu d\nu \ .  \label{L1}
\end{equation}
We can describe the addition rule of Eq.(\ref{L}) in the following terms.
Given the tomograms $\mathcal{T}_{\rho _{1}}$ and $\mathcal{T}_{\rho _{2}}$
of the pure states $\rho _{1},\rho _{2}$, and the tomogram $\mathcal{T}_{0}$
of a fiducial chosen pure state $P_{0}$, then the tomogram $\mathcal{T}%
_{\rho }$ is the tomogram of the pure state $\rho =\left\vert \psi
\right\rangle \left\langle \psi \right\vert $ of Eq.(\ref{D}) corresponding
to the linear superposition of Eq.(\ref{A}). The formula Eq.(\ref{L})
guarantees that if $\mathcal{T}_{\rho _{1}},\mathcal{T}_{\rho _{1}}$ and $%
\mathcal{T}_{0}$ are tomographic probabilities of rank-one projectors, the
result of the nonlinear summation $\mathcal{T}_{\rho }$ is again a
normalized tomographic probability corresponding to a superposition state
that can be realized in nature.

The composition rule Eq.(\ref{L}) has its partner in terms of Wigner
functions of pure states. Let the Wigner function $\mathcal{W}_{1}(p,q)$
corresponding to a pure state $\left\vert \psi _{1}\right\rangle
\left\langle \psi _{1}\right\vert $ be composed with theWigner function $%
\mathcal{W}_{2}(p,q)$ corresponding to another pure state $\left\vert \psi
_{2}\right\rangle \left\langle \psi _{2}\right\vert $ and let the
composition $\mathcal{W}(p,q)$ correspond to the superposition vector $|\psi
\rangle =\sqrt{p_{1}}|\psi _{1}\rangle +\mathrm{e}^{i\varphi }\sqrt{p_{2}}%
|\psi _{2}\rangle $\ with the parameters used in Eq.(\ref{L}). Then one has
\cite{SudarJRLR}, dropping the arguments $(p,q)$, the result:
\begin{equation}
\mathcal{W}=p_{1}\mathcal{W}_{1}+p_{2}\mathcal{W}_{2}+\sqrt{p_{1}p_{2}}\frac{%
\mathcal{W}_{1}\ast \mathcal{W}_{0}\ast \mathcal{W}_{2}+\mathcal{W}_{2}\ast
\mathcal{W}_{0}\ast \mathcal{W}_{1}}{\sqrt{C(\mathcal{W}_{1}\ast \mathcal{W}%
_{0}\ast \mathcal{W}_{2}\ast \mathcal{W}_{0})}}
\end{equation}
where, restoring the arguments $(p,q)$, for any operator $A$ and its Weyl
symbol $\mathcal{W}_{A}(p,q)$ one has
\begin{equation}
\mathrm{Tr}(A)=C(\mathcal{W}_{A}(p,q))=\frac{1}{2\pi }\int_{A}\mathcal{W}%
(p,q)dpdq\ .
\end{equation}
The star product of two Weyl symbols is given by the formula
\begin{eqnarray}
&&\mathcal{W}_{1}(p,q)\ast \mathcal{W}_{0}(p,q)  \notag \\
&=&\int \mathcal{W}_{1}(p_{1},q_{1})\mathcal{W}%
_{0}(p_{2},q_{2})K(p_{1},q_{1},p_{2},q_{2},p,q)dp_{1}dq_{1}dp_{2}dq_{2}
\end{eqnarray}
with the Gr\"{o}newold kernel
\begin{equation*}
K(p_{1},q_{1},p_{2},q_{2},p,q)=\frac{1}{\pi ^{2}}\exp \left[
2i(qp_{1}-q_{1}p+q_{1}p_{2}-q_{2}p_{1}+q_{2}p-p_{2}q)\right] .
\end{equation*}
One can easily see that the exponent in the r.h.s. of the above formula may
be written as $4iS$, where $S$ is the area of a triangle in the phase space
with vertices in the points $(q_{1},p_{1}),(q_{2},p_{2}),(q,p).$

\subsection{ Uncertainty relations}

Having considered the superposition rule within the tomographic scheme we
consider now the formulation of uncertainty relations. We present below the
derivation of uncertainty relations in a general framework. Given an
operator $A,$ then it is obvious that
\begin{equation}
\left\langle A^{\dagger }A\right\rangle _{\rho }=\mathrm{Tr}\left( \rho
A^{\dagger }A\right) \geq 0
\end{equation}
for any density state $\rho $ which is a non-negative hermitian operator
with unity trace. Let $A$ be, \textsl{for systems with one degree of freedom,%
} a linear combination of operators $\{B_{k}\}$ to be suitably chosen.
\begin{equation}
A=\sum\nolimits_{k}c_{k}B_{k}\ .
\end{equation}
Then
\begin{equation}
\left\langle A^{\dagger }A\right\rangle _{\rho
}=\sum\nolimits_{h,k}c_{h}^{\ast }c_{k}\left\langle B_{h}^{\dagger
}B_{k}\right\rangle _{\rho }\geq 0
\end{equation}
that can be rewritten, dropping the label $\rho $ and using commutator and
anti-commutator, as
\begin{equation}
\sum\nolimits_{h,k}c_{h}^{\ast }c_{k}\left\langle \frac{1}{2}\left[
B_{h}^{\dagger },B_{k}\right] _{+}+\frac{1}{2}\left[ B_{h}^{\dagger },B_{k}%
\right] \right\rangle \geq 0.
\end{equation}
Since $c_{h}^{\ast },c_{k}$\ are arbitrary complex numbers, the positivity
of the quadratic form means the positivity of the matrix
\begin{equation}
\left( S\right) _{hk}=\left\langle \frac{1}{2}\left[ B_{h}^{\dagger },B_{k}%
\right] _{+}+\frac{1}{2}\left[ B_{h}^{\dagger },B_{k}\right] \right\rangle
\end{equation}
The Sylvester criterion provides the necessary and sufficient condition for
the positivity of the matrix in terms of its principal minors
\begin{equation}
M_{1}(S)\geq 0\ ,\ M_{2}(S)\geq 0\ ,\ \ldots \ ,\ \det S\geq 0\ .
\end{equation}
This general scheme allows to obtain uncertainty relations.

Thus, for position $Q$ and momentum $P$ we choose the error operators as $B$%
's
\begin{equation}
B_{1}=Q-\left\langle Q\right\rangle =\Delta Q\ ,\ B_{2}=P-\left\langle
P\right\rangle =\Delta P\ .
\end{equation}
So, we get
\begin{equation}
S=\left(
\begin{array}{cc}
\sigma _{QQ} & \sigma _{QP} \\
\sigma _{QP} & \sigma _{PP}%
\end{array}
\right) +\frac{1}{2}\left(
\begin{array}{cc}
0 & \left\langle \left[ Q,P\right] \right\rangle \\
-\left\langle \left[ Q,P\right] \right\rangle & 0%
\end{array}
\right)
\end{equation}
where we have introduced the variances
\begin{equation}
\sigma _{QQ}=\left\langle Q^{2}\right\rangle -\left\langle Q\right\rangle
^{2}\ ,\ \sigma _{PP}=\left\langle P^{2}\right\rangle -\left\langle
P\right\rangle ^{2}
\end{equation}
and the covariance
\begin{equation}
\sigma _{QP}=\frac{1}{2}\left\langle QP+PQ\right\rangle -\left\langle
Q\right\rangle \left\langle P\right\rangle \ .
\end{equation}
We then obtain from the positivity of the $M_{1}$ principal minors the
obvious conditions
\begin{equation}
\sigma _{QQ}\geq 0\ ,\ \sigma _{PP}\geq 0\ ,
\end{equation}
and from the positivity of the determinant
\begin{equation}
\det S=\sigma _{QQ}\sigma _{PP}-\left( \sigma _{QP}^{2}+\frac{1}{4}%
\left\langle -\left[ Q,P\right] \right\rangle ^{2}\right) \geq 0
\end{equation}
the Schr\"{o}dinger-Robertson uncertainty relation follows
\begin{equation}
\left\langle \Delta Q^{2}\right\rangle \left\langle \Delta
P^{2}\right\rangle \geq \frac{1}{4}\left\langle \left[ \Delta Q,\Delta P%
\right] _{+}^{2}\right\rangle +\frac{1}{4},
\end{equation}
this reduces to the weaker Heisenberg uncertainty relation for uncorrelated
states:
\begin{equation}  \label{varvar}
\left( \left\langle P^{2}\right\rangle -\left\langle P\right\rangle
^{2}\right) \left( \left\langle Q^{2}\right\rangle -\left\langle
Q\right\rangle ^{2}\right) \geqslant \frac{1}{4}\quad ,\quad (\hbar =1) .
\end{equation}
In view of Eq. (\ref{fA}), we can write Eq. (\ref{varvar}) as:
\begin{eqnarray}
&&\left( \int X^{2}\mathcal{T}(X,0,1)dX-\left[ \int X\mathcal{T}(X,0,1)dX%
\right] ^{2}\right) \times  \label{beta} \\
&&\left( \int X^{2}\mathcal{T}(X,1,0)dX-\left[ \int X\mathcal{T}(X,1,0)dX%
\right] ^{2}\right) \geq \frac{1}{4}  \notag
\end{eqnarray}
because $\mathcal{T}(X,0,1)$ is the momentum probability density and $%
\mathcal{T}(X,1,0)$ is the position probability density. The uncertainty
relation in \textsl{covariant form} reads
\begin{eqnarray}
&&\left( \int X^{2}\mathcal{T}(X,\cos \theta ,\sin \theta )dX-\left[ \int X%
\mathcal{T}(X,\cos \theta ,\sin \theta )dX\right] ^{2}\right) \times
\label{gamma} \\
&&\left( \int X^{2}\mathcal{T}(X,-\sin \theta ,\cos \theta )dX-\left[ \int X%
\mathcal{T}(X,-\sin \theta ,\cos \theta )dX\right] ^{2}\right) \geq \frac{1}{%
4}\ .  \notag
\end{eqnarray}
The obtained relation takes into account all the values of optical tomogram
for all angles $\theta .$ It allows to use the values of experimentally
measured optical tomogram to check the uncertainty relation of position and
momentum by using a homodyne detector. \ This formula, that can be written
in the short form $F(\theta )\geq 1/4,$ provides a constraint on admissible
quantum mechanical tomographic probabilities due to Heisenberg uncertainty
relations. The experimental check of the Heisenberg uncertainty relations
can be done using the positivity condition of the function (uncertainty
function)
\begin{equation}
\Phi (\theta ):=F(\theta )-\frac{1}{4}\geq 0\ .
\end{equation}
Such a checking can be done \cite{Ad.Sci.Lett.} (see also the discussion in
\cite{Lahti}) using experimental data of \cite{Raymer, Mlynek, Lvovski,
Solimeno}.

The tomographic expression for Schr\"{o}dinger-Robertson relation requires
the star product to represent the mean value of the anti-commutator.

For the case of angular momentum we choose selfadjoint $B$'s operators as
\begin{equation}
B_{k}=\Delta J_{k}\quad ,\quad k=1,2,3
\end{equation}
and get, recalling that $\left[ J_{h},J_{k}\right] =i\varepsilon
_{hkl}J_{l}: $%
\begin{equation}
S=\left(
\begin{array}{ccc}
\sigma _{J_{1}J_{1}} & \sigma _{J_{1}J_{2}} & \sigma _{J_{1}J_{3}} \\
\sigma _{J_{1}J_{2}} & \sigma _{J_{2}J_{2}} & \sigma _{J_{2}J_{3}} \\
\sigma _{J_{1}J_{3}} & \sigma _{J_{2}J_{3}} & \sigma _{J_{3}J_{3}}%
\end{array}
\right) +\frac{i}{2}\left(
\begin{array}{ccc}
0 & \left\langle J_{3}\right\rangle & -\left\langle J_{2}\right\rangle \\
-\left\langle J_{3}\right\rangle & 0 & \left\langle J_{1}\right\rangle \\
\left\langle J_{2}\right\rangle & -\left\langle J_{1}\right\rangle & 0%
\end{array}
\right)  \label{spinunc3}
\end{equation}
where again variances and covariances appear.

The three first principal minors give the obvious conditions of positivity
of all the variances
\begin{equation}
\sigma _{J_{1}J_{1}}\geq 0\ ,\ \sigma _{J_{2}J_{2}}\geq 0\ ,\ \sigma
_{J_{3}J_{3}}\geq 0.  \label{spinunc1}
\end{equation}

The first of the three second principal minors yields a quadratic relation
with indices 123
\begin{equation}
\sigma _{J_{1}J_{1}}\sigma _{J_{2}J_{2}}-\left( \sigma _{J_{1}J_{2}}^{2}+%
\frac{1}{4}\left\langle J_{3}\right\rangle ^{2}\right) \geq 0
\label{spinunc2}
\end{equation}
while the others give quadratic relations obtained from the first by
circular permutation of the indices 231 and 312.

Finally, the positivity of $\det S$ yields a cubic relation connecting the
variances and covariances of all the components of the angular momentum, or
spin projections. The tomographic expression for a star-product kernel
providing such a cubic relation exists, even though cumbersome \cite{spin
cubic relation}.

\subsection{Classical and quantum distributions: examples}

Some observations are now in order to discuss the conditions the uncertainty
relations give for considering a `tomographic'\ normalized function the
quantum tomogram of a density state (the necessary and sufficient conditions
for a function on phase space to be a Wigner function are discussed, e.g, in
\cite{Na86}). As we have observed previously, an exact condition stems out
from the positivity of density states. Thus, given a `tomographic'\
probability $\mathcal{T}(\vec{x})$ (either for spin variables or for
continuous ones as position and momentum), the appropriate reconstruction
formula can be used to get an operator $\varrho _{\mathcal{T}}$. Then, we
can check wether its eigenvalues are non-negative. If they are, $\varrho _{%
\mathcal{T}}$ is a density state, \textsl{possibly non-normalized,} and $%
\mathcal{T}$ is a quantum tomogram. However, this condition is rather formal
and other, possibly more operative, conditions may be given. The `tomogram'\
$\mathcal{T}$ is assumed to satisfy the specific properties which hold for
the tomographic probabilities in the given tomographic scheme, such as the
homogeneity condition $\mathcal{T}(\lambda X,\lambda \mu ,\lambda \nu )=%
\mathcal{T}(X,\mu ,\nu )/\left\vert \lambda \right\vert $ in the symplectic
case, for instance. Altogether these properties guarantee the hermiticity of
$\varrho _{\mathcal{T}}$ and the normalization $\mathrm{Tr}(\varrho _{%
\mathcal{T}})=1$, but they are not sufficient for the nonnegativity of $%
\varrho _{\mathcal{T}}$. For that, there are necessary, in general not
sufficient, conditions like uncertainty relations. In fact, the
nonnegativity of the operator $\varrho _{\mathcal{T}}$ guarantees the
fulfilling of all available uncertainty relations because they stem from the
equation $\left\langle A^{\dagger }A\right\rangle _{\varrho _{\mathcal{T}}}=%
\mathrm{Tr}(A^{\dagger }A\varrho _{\mathcal{T}})/\mathrm{Tr}(\varrho _{%
\mathcal{T}})\geq 0$ . Of course a complete (usually infinite) set of
uncertainty inequalities is sufficient for the nonnegativity of $\varrho _{%
\mathcal{T}}$. However, such a completeness is a tautology, as in a sense it
stands for 'positivity'.

So, if the corresponding tomographic expressions of such inequalities are
satisfied by $\mathcal{T}$, we have the necessary, or sufficient, conditions
for $\mathcal{T}$ to be a genuine quantum tomogram.

This method requires the use of the dual tomographic map \cite{Patr} for the
operator symbols $f_{A}^{d}(\vec{x}):=\langle A|G(\vec{x})\rangle $, as for
an observable $A$ one has \cite{PatrBeppeOlga}:
\begin{eqnarray*}
\langle A\rangle _{\varrho _{\mathcal{T}}}=\text{Tr}(A\varrho _{\mathcal{T}%
})=\langle A|\varrho _{\mathcal{T}}\rangle =\int \langle A|G(\vec{x})\rangle
\langle P(\vec{x})|\varrho _{\mathcal{T}}\rangle d\vec{x}=\int \mathcal{T}(%
\vec{x})f_{A}^{d}(\vec{x})d\vec{x}
\end{eqnarray*}

Thus, the Schr\"{o}dinger-Robertson uncertainty relation can be written for
the symplectic case as:
\begin{eqnarray}
&&\int f_{Q^{2}}^{d}(X,\mu ,\nu )\mathcal{T}(X,\mu ,\nu )dXd\mu d\nu \times
\int f_{P^{2}}^{d}(X,\mu ,\nu )\mathcal{T}(X,\mu ,\nu )dXd\mu d\nu  \notag \\
&&-\left( \int f_{\frac{1}{2}[Q,P]_{+}}^{d}(X,\mu ,\nu )\mathcal{T}(X,\mu
,\nu )dXd\mu d\nu \right) ^{2}\geq \frac{1}{4}
\end{eqnarray}
where $\left\langle Q\right\rangle ,\left\langle P\right\rangle $ are
assumed to be zero, while the dual symbols are given by:
\begin{eqnarray}
&&f_{Q^{2}}^{d}(X,\mu ,\nu )=\frac{1}{2\pi }\text{\textrm{Tr}}(Q^{2}\exp %
\left[ i(X-\mu Q-\nu P)\right] , \\
&&f_{P^{2}}^{d}(X,\mu ,\nu )=\frac{1}{2\pi }\text{\textrm{Tr}}(P^{2}\exp %
\left[ i(X-\mu Q-\nu P)\right] ,  \notag \\
&&f_{\frac{1}{2}[Q,P]_{+}}^{d}(X,\mu ,\nu )=\frac{1}{2\pi }\text{\textrm{Tr}}%
(\frac{1}{2}[Q,P]_{+}\exp \left[ i(X-\mu Q-\nu P)\right] ),  \notag
\end{eqnarray}
and are known generalized functions \cite{GelfandShilov2, PilavetzJRLR}. The
Schr\"{o}dinger-Robertson relation, written in terms of these generalized
functions integrated with the $\mathcal{T}$ function, provides a necessary
condition for $\mathcal{T}$ to be a quantum tomogram. We illustrate this
analysis with the help of some examples.

\noindent \textbf{Example 1} Let us consider the probability distribution
\begin{equation*}
\frac{e^{-\frac{X^{2}}{\mu ^{2}+\nu ^{2}}}}{\sqrt{\pi (\mu ^{2}+\nu ^{2})}}
\end{equation*}
and check wether it fulfills or not the uncertainty relations.

To this aim one has to evaluate dual symbols like
\begin{equation}
f_{Q^{2}}^{d}(X,\mu ,\nu )=\frac{1}{2\pi }\mathrm{Tr}(Q^{2}e^{i(X-\mu Q-\nu
P)}).
\end{equation}
By using the Weyl symbols
\begin{equation}
A\rightarrow W_{A}(p,q)
\end{equation}
the trace is readily evaluated as
\begin{equation}
\mathrm{Tr}(AB)=\int \frac{W_{A}W_{B}}{2\pi }dpdq.
\end{equation}
We have
\begin{eqnarray}
W_{Q^{2}} &=&q^{2}, \\
W_{P^{2}} &=&p^{2},  \notag \\
W_{\frac{1}{2}(PQ+QP)} &=&pq,  \notag \\
W_{e^{i(X-\mu Q-\nu P)}} &=&e^{i(X-\mu q-\nu p)},  \notag
\end{eqnarray}
so that
\begin{eqnarray}
f_{Q^{2}}^{d}(X,\mu ,\nu ) &=&\frac{1}{2\pi }\mathrm{Tr}(Q^{2}e^{i(X-\mu
Q-\nu P)})=\frac{1}{2\pi }\int \frac{q^{2}e^{i(X-\mu q-\nu p)})}{2\pi }dpdq
\notag \\
&=&e^{iX}\delta (\nu )\int \frac{q^{2}e^{-i\mu q}}{2\pi }dq=-e^{iX}\delta
(\nu )\delta ^{\prime \prime }(\mu )
\end{eqnarray}
and analogously
\begin{eqnarray}
f_{P^{2}}^{d}(X,\mu ,\nu ) &=&-e^{iX}\delta ^{\prime \prime }(\nu )\delta
(\mu ), \\
f_{\frac{1}{2}(PQ+QP)}^{d}(X,\mu ,\nu ) &=&-e^{iX}\delta ^{\prime }(\nu
)\delta ^{\prime }(\mu ).  \notag
\end{eqnarray}
Thus we obtain
\begin{eqnarray}
\left\langle \Delta Q^{2}\right\rangle &=&-\int \frac{e^{-\frac{X^{2}}{\mu
^{2}+\nu ^{2}}}}{\sqrt{2\pi (\mu ^{2}+\nu ^{2})}}e^{iX}\delta (\nu )\delta
^{\prime \prime }(\mu )dXd\mu d\nu =\frac{1}{2}, \\
\left\langle \Delta P^{2}\right\rangle &=&-\int \frac{e^{-\frac{X^{2}}{\mu
^{2}+\nu ^{2}}}}{\sqrt{2\pi (\mu ^{2}+\nu ^{2})}}e^{iX}\delta (\nu )^{\prime
\prime }\delta (\mu )dXd\mu d\nu =\frac{1}{2},  \notag \\
\left\langle \frac{1}{2}\left[ \Delta Q,\Delta P\right] _{+}\right\rangle
&=&-\int \frac{e^{-\frac{X^{2}}{\mu ^{2}+\nu ^{2}}}}{\sqrt{2\pi (\mu
^{2}+\nu ^{2})}}e^{iX}\delta ^{\prime }(\nu )\delta ^{\prime }(\mu )dXd\mu
d\nu =0.  \notag
\end{eqnarray}
so the Schr\"{o}dinger-Robertson uncertainty relation
\begin{equation}
\left\langle \Delta Q^{2}\right\rangle \left\langle \Delta
P^{2}\right\rangle \geq \frac{1}{4}\left\langle \left[ \Delta Q,\Delta P%
\right] _{+}\right\rangle ^{2}+\frac{1}{4}
\end{equation}
is satisfied. In this case it is easy to reconstruct the quantum state $%
\varrho _{\mathcal{T}},$ it is the ground state of a harmonic oscillator.
This can be done or directly, by the use of Eq. (\ref{symprec}) or in two
steps, first by a Radon anti-transform of the tomogram yielding a Wigner
function on phase space, from which the quantum state is readily obtained.

\noindent \textbf{Example 2} An example of a non-quantum, classical
tomographic probability distribution is provided by the following normalized
distribution
\begin{equation*}
\frac{1}{|\mu +\nu |}\chi _{I(0,\mu +\nu )}(X).
\end{equation*}
where $I(0,\mu +\nu )$ is the interval with extrema $0$ and $\mu +\nu .$ As
it is homogeneous, it has the form of a symplectic tomographic distribution.
We have
\begin{eqnarray}
&&\left\langle \Delta Q^{2}\right\rangle =-\int \frac{1}{|\mu +\nu |}\chi
_{I(0,\mu +\nu )}(X)e^{iX}\delta (\nu )\delta ^{\prime \prime }(\mu )dXd\mu
d\nu \\
&=&-\int \frac{1}{|\mu |}\chi _{I(0,\mu )}(X)e^{iX}\delta ^{\prime \prime
}(\mu )dXd\mu =-\int \frac{2\sin \frac{\mu }{2}}{\mu }\exp \left[ i\frac{\mu
}{2}\right] \delta ^{\prime \prime }(\mu )d\mu =\frac{4}{3}  \notag
\end{eqnarray}
Analogously
\begin{equation}
\left\langle \Delta P^{2}\right\rangle =\frac{4}{3}
\end{equation}
While
\begin{eqnarray}
&&\left\langle \frac{1}{2}\left[ \Delta Q,\Delta P\right] _{+}\right\rangle
=-\int \frac{1}{|\mu +\nu |}\chi _{I(0,\mu +\nu )}(X)e^{iX}\delta ^{\prime
}(\nu )\delta ^{\prime }(\mu )dXd\mu d\nu  \notag \\
&=&-\int \frac{2\sin \frac{\mu +\nu }{2}}{\mu +\nu }\exp \left[ i\frac{\mu
+\nu }{2}\right] \delta ^{\prime }(\nu )\delta ^{\prime }(\mu )d\mu d\nu =%
\frac{4}{3}
\end{eqnarray}
so the uncertainty relations are violated:
\begin{equation}
\frac{4}{3}.\frac{4}{3}\ngeq \left( \frac{4}{3}\right) ^{2}+\frac{1}{4}.
\end{equation}

To obtain the classical distribution, we have to evaluate the Radon
anti-transform of the tomographic distribution :
\begin{eqnarray}
&&\frac{1}{\left( 2\pi \right) ^{2}}\int \frac{1}{|\mu +\nu |}\chi _{I(0,\mu
+\nu )}(X)\exp \left[ i(X-\mu q-\nu p)\right] dXd\mu d\nu  \notag \\
&=&\frac{1}{\left( 2\pi \right) ^{2}}\int \frac{2\sin (\frac{\mu +\nu }{2})}{%
\mu +\nu }\exp \left[ i\left(\frac{\mu +\nu }{2}-\mu q-\nu p\right)\right]
d\mu d\nu
\end{eqnarray}
By introducing the new variables
\begin{equation}
\xi =\frac{\mu +\nu }{2};\eta =\frac{\mu -\nu }{2}\Leftrightarrow \mu =\xi
+\eta ;\nu =\xi -\eta
\end{equation}
we then get
\begin{eqnarray}
&&\frac{2}{\left( 2\pi \right) ^{2}}\int \frac{\sin \xi }{\xi }\exp \left[
i\{\xi -\xi (q+p)-\eta (q-p)\}\right] d\xi d\eta \\
&=&\frac{1}{\pi }\delta (q-p)\int \frac{\sin \xi }{\xi }\exp \left[ i\{\xi
-\xi (q+p)\}\right] d\xi  \notag \\
&=&\chi _{\lbrack -1,1]}(1-(q+p))\delta (q-p)=\chi _{\lbrack 0,1]}(q)\delta
(q-p)  \notag
\end{eqnarray}

\noindent \textbf{Example 3} Another non-quantum, classical example is given
by the positive homogeneous normalized distribution
\begin{equation*}
\frac{1}{2|\mu +\nu |}\exp \left[ -\frac{|X|}{|\mu +\nu |}\right]
\end{equation*}
Remembering that
\begin{equation}
\int \frac{1}{2|\mu +\nu |}\exp \left[ -\frac{|X|}{|\mu +\nu |}\right]
e^{iX}dX=\frac{1}{1+|\mu +\nu |^{2}}
\end{equation}
we get
\begin{equation}
\left\langle \Delta Q^{2}\right\rangle =\left\langle \Delta
P^{2}\right\rangle =2;\left\langle \frac{1}{2}\left[ \Delta Q,\Delta P\right]
_{+}\right\rangle =2
\end{equation}
so that again the uncertainty relations are violated, and the tomographic
distribution is not quantum. With the change of variable
\begin{equation}
\xi =\mu +\nu ;\eta =\mu -\nu \Leftrightarrow \mu =\frac{\xi +\eta }{2};\nu =%
\frac{\xi -\eta }{2}
\end{equation}
in the Radon anti-transform, it corresponds to the classical distribution:
\begin{eqnarray}
&&\frac{1}{\left( 2\pi \right) ^{2}}\int \frac{1}{1+|\mu +\nu |^{2}}\exp %
\left[ -i(\mu q+\nu p)\right] d\mu d\nu  \notag \\
&=&\frac{1}{2\left( 2\pi \right) ^{2}}\int \frac{1}{1+\xi ^{2}}\exp \left[ -i%
\frac{\xi }{2}(q+p)-i\frac{\eta }{2}(q-p)\right] d\xi d\eta  \notag \\
&=&\frac{1}{2\pi }\delta (\frac{1}{2}(q-p))\frac{\pi }{2}\exp \left[ -\frac{%
\left\vert q+p\right\vert }{2}\right] =\frac{1}{2}\delta (q-p)\exp \left[
-\left\vert q\right\vert \right]
\end{eqnarray}

\noindent \textbf{Example 4} The analogous approach to spin tomography is
based on the necessary conditions of Eq.s (\ref{spinunc1}, \ref{spinunc2})
and positivity of the determinant of the matrix $S$ of Eq. (\ref{spinunc3}).
All the elements of $S$ can be written in terms of spin tomograms and dual
spin tomographic symbols of the observables $J_{k},J_{h}J_{k}(k,h=1,2,3),$
which can be obtained using the Gram-Schmidt operators for the spin
tomographic star-product scheme given in \cite{spin cubic relation}. For
instance, consider in the qubit tomographic case the vector component
\begin{equation}
\mathcal{T}_{\alpha ,\beta }\left( \Omega \right) =\alpha \cos ^{2}\frac{%
\theta }{2}+\beta \sin ^{2}\frac{\theta }{2}=\frac{1}{2}\left( \alpha +\beta
\right) +\frac{1}{2}\left( \alpha -\beta \right) \cos \theta ,
\end{equation}
the other component being $1-\mathcal{T}_{\alpha ,\beta }\left( \Omega
\right) ,$ corresponding to the operator
\begin{equation}
\rho _{\alpha ,\beta }=\left[
\begin{array}{cc}
\alpha & 0 \\
0 & \beta%
\end{array}
\right] \ .
\end{equation}
The elements
\begin{equation}
\sigma _{J_{h}J_{k}}=\frac{1}{2}\left\langle
J_{h}J_{k}+J_{k}J_{h}\right\rangle -\left\langle J_{h}\right\rangle
\left\langle J_{k}\right\rangle
\end{equation}
of the $S$ matrix, in view of
\begin{equation}
J_{h}J_{k}+J_{k}J_{h}=\frac{1}{2}\delta _{hk}I_{2}\ ,
\end{equation}
read as
\begin{equation*}
\int d\Omega \left\langle \delta _{hk}I_{2}|G_{\Omega }\right\rangle
\mathcal{T}_{\alpha ,\beta }\left( \Omega \right) -\int d\Omega \left\langle
J_{h}|G_{\Omega }\right\rangle \mathcal{T}_{\alpha ,\beta }\left( \Omega
\right) \int d\Omega \left\langle J_{k}|G_{\Omega }\right\rangle \mathcal{T}%
_{\alpha ,\beta }\left( \Omega \right)
\end{equation*}
with
\begin{equation}
\int d\Omega \left\langle \delta _{hk}I_{2}|G_{\Omega }\right\rangle
\mathcal{T}_{\alpha ,\beta }\left( \Omega \right) =\frac{1}{4}\left( \alpha
+\beta \right) \delta _{hk},
\end{equation}
while
\begin{eqnarray}
\left\langle J_{1}|G_{\Omega }\right\rangle &=&\frac{3}{2\pi }\cos \phi \sin
\theta \Rightarrow \left\langle J_{1}\right\rangle =0 \\
\left\langle J_{2}|G_{\Omega }\right\rangle &=&\frac{3}{2\pi }\sin \phi \sin
\theta \Rightarrow \left\langle J_{2}\right\rangle =0 \\
\left\langle J_{3}|G_{\Omega }\right\rangle &=&\frac{3}{4\pi }\cos \theta
\Rightarrow \left\langle J_{3}\right\rangle =\frac{1}{2}\left( \alpha -\beta
\right)
\end{eqnarray}
so that eventually
\begin{equation}
S=\frac{1}{4}\left[
\begin{array}{ccc}
\left( \alpha +\beta \right) & i\left( \alpha -\beta \right) & 0 \\
-i\left( \alpha -\beta \right) & \left( \alpha +\beta \right) & 0 \\
0 & 0 & \left( \alpha +\beta \right) -\left( \alpha -\beta \right) ^{2}%
\end{array}
\right] .
\end{equation}
Now $S$ is non-negative iff $0\leq \alpha ,\beta \leq 1.$ Then, $0\leq
\mathcal{T}_{\alpha ,\beta }\left( \Omega \right) \leq 1$ is the component
of a probability vector corresponding to the (non-normalized) density state $%
\rho _{\alpha ,\beta }.$ When the uncertainty relations are violated, $%
\mathcal{T}_{\alpha ,\beta }\left( \Omega \right) $ is readily recognizable
not to be a probability vector.

\subsection{Equations of motion}

For the tomographic description of quantum mechanics to be equivalent to
other existing ones we have to consider the formulation of dynamical
evolution, along with the integration of the equations of motion usually
done by solving an eigenvalue problem. In the coming two subsections we are
going to address these aspects.

\subsubsection{Time evolution}

The evolution of a state probability distribution in classical hamiltonian
mechanics with potential $U(q)$ is given by the Liouville equation
\begin{equation}
\frac{\partial }{\partial t}f(q,p,t)+p\frac{\partial }{\partial q}f(q,p,t)-%
\frac{\partial }{\partial q}U(q)\frac{\partial }{\partial p}f(q,p,t)=0
\label{liouv}
\end{equation}

We study now evolution in quantum mechanics using Wigner functions and
tomograms. We recall the definition of Wigner function of a density matrix $%
\rho (q,q^{\prime }):$%
\begin{equation}
\mathcal{W}(p,q)=\int \rho (q+\frac{x}{2},q-\frac{x}{2})\exp (-ipx)dx\ .
\end{equation}
The equations of motion for $\rho ,$ derived from the Schr\"{o}dinger
equations, give rise to equations of motion for $\mathcal{W}(p,q).$

The evolution equation for the density matrix $\rho (q,q^{\prime },t)$\ is
obtained from the Schr\"{o}dinger equation for the wave function. In fact,
for the Hamiltonian
\begin{equation}
H(P,Q)=\frac{1}{2}P^{2}+U(Q);(m=1)
\end{equation}
one has in the position representation:
\begin{equation}
i\frac{\partial }{\partial t}\psi (q,t)=\left( -\frac{1}{2}\frac{\partial
^{2}}{\partial q^{2}}+U(q)\right) \psi (q,t)
\end{equation}
and
\begin{equation}
-i\frac{\partial }{\partial t}\psi ^{\ast }(q^{\prime },t)=\left( -\frac{1}{2%
}\frac{\partial ^{2}}{\partial q^{\prime 2}}+U(q^{\prime })\right) \psi
^{\ast }(q^{\prime },t).
\end{equation}
So, upon subtracting the above equations after multiplication by $\psi
^{\ast }(q^{\prime },t)$ and $\psi (q,t)$ respectively, we get the von
Neumann evolution equation for the density matrix $\rho (q,q^{\prime
},t)=\psi (q,t)\psi ^{\ast }(q^{\prime },t)$ of the pure state $\psi :$
\begin{equation}
i\frac{\partial }{\partial t}\rho (q,q^{\prime },t)=\left[ -\frac{1}{2}\frac{%
\partial ^{2}}{\partial q^{2}}+\frac{1}{2}\frac{\partial ^{2}}{\partial
q^{\prime 2}}+U(q)-U(q^{\prime })\right] \rho (q,q^{\prime },t).
\end{equation}

Equations of motion for Wigner function $\mathcal{W}(p,q)$ can be derived by
using the following association among derivatives of the density matrix and
the Wigner function:
\begin{eqnarray}
\frac{\partial }{\partial q}\mathcal{W}(q,p) &\leftrightarrow &\left( \frac{%
\partial }{\partial q}+\frac{\partial }{\partial q^{\prime }}\right) \rho
(q,q^{\prime }) \\
i\frac{\partial }{\partial p}\mathcal{W}(q,p) &\leftrightarrow &\left(
q-q^{\prime }\right) \rho (q,q^{\prime })  \notag \\
q\mathcal{W}(q,p) &\leftrightarrow &\frac{1}{2}\left( q+q^{\prime }\right)
\rho (q,q^{\prime })  \notag \\
ip\mathcal{W}(q,p) &\leftrightarrow &\frac{1}{2}\left( \frac{\partial }{%
\partial q}-\frac{\partial }{\partial q^{\prime }}\right) \rho (q,q^{\prime
})  \notag
\end{eqnarray}
and
\begin{eqnarray}
\frac{\partial }{\partial q}\rho (q,q^{\prime }) &\leftrightarrow &\left(
\frac{1}{2}\frac{\partial }{\partial q}+ip\right) \mathcal{W}(q,p) \\
\frac{\partial }{\partial q^{\prime }}\rho (q,q^{\prime }) &\leftrightarrow
&\left( \frac{1}{2}\frac{\partial }{\partial q}-ip\right) \mathcal{W}(q,p)
\notag \\
q\rho (q,q^{\prime }) &\leftrightarrow &\left( q+\frac{i}{2}\frac{\partial }{%
\partial p}\right) \mathcal{W}(q,p)  \notag \\
q^{\prime }\rho (q,q^{\prime }) &\leftrightarrow &\left( q-\frac{i}{2}\frac{%
\partial }{\partial p}\right) \mathcal{W}(q,p)  \notag
\end{eqnarray}
Then, using the previous correspondence rules, we finally obtain the
evolution equation for the Wigner function $\mathcal{W}(q,p,t),$ which reads
\begin{eqnarray}
i\frac{\partial }{\partial t}\mathcal{W}(q,p,t) &=&-\frac{1}{2}\left[ \left(
\frac{1}{2}\frac{\partial }{\partial q}+ip\right) ^{2}-\left( \frac{1}{2}%
\frac{\partial }{\partial q}-ip\right) ^{2}\right] \mathcal{W}(q,p,t)  \notag
\\
&&+\left[ U\left( q+\frac{i}{2}\frac{\partial }{\partial p}\right) -U\left(
q-\frac{i}{2}\frac{\partial }{\partial p}\right) \right] \mathcal{W}(q,p,t)\
.
\end{eqnarray}
The Wigner function is real. The above equation can be written as
\begin{equation}
\left( \frac{\partial }{\partial t}+p\frac{\partial }{\partial q}\right)
\mathcal{W}(q,p,t)=-i\left[ U\left( q+\frac{i}{2}\frac{\partial }{\partial p}%
\right) -U\left( q-\frac{i}{2}\frac{\partial }{\partial p}\right) \right]
\mathcal{W}(q,p,t)  \label{moyal}
\end{equation}
These equations of motion may be called the Moyal evolution equations. The
expansion of Eq. (\ref{moyal}) up to first order in $\left( \frac{i}{2}%
\partial /\partial p\right) $ yields the classical Liouville equation (\ref%
{liouv})

A similar correspondence table allows us to derive the evolution equations
for tomograms out of the evolution equation for the Wigner function. To get
the evolution equation for the tomogram $\mathcal{T}_{\rho }(X,\mu ,\nu ,t)$
the correspondence table which is obtained from the Radon transform formulae
is
\begin{eqnarray}
\frac{\partial }{\partial q}\mathcal{W}(q,p) &\leftrightarrow &\mu \frac{%
\partial }{\partial X}\mathcal{T}(X,\mu ,\nu )  \label{radon corr form} \\
\frac{\partial }{\partial p}\mathcal{W}(q,p) &\leftrightarrow &\nu \frac{%
\partial }{\partial X}\mathcal{T}(X,\mu ,\nu )  \notag \\
q\mathcal{W}(q,p) &\leftrightarrow &-\frac{\partial }{\partial \mu }\left(
\frac{\partial }{\partial X}\right) ^{-1}\mathcal{T}(X,\mu ,\nu )  \notag \\
p\mathcal{W}(q,p) &\leftrightarrow &-\frac{\partial }{\partial \nu }\left(
\frac{\partial }{\partial X}\right) ^{-1}\mathcal{T}(X,\mu ,\nu )  \notag
\end{eqnarray}
Thus, from the Moyal equation we obtain
\begin{equation}
\left( \frac{\partial }{\partial t}-\mu \frac{\partial }{\partial \nu }%
\right) \mathcal{T}(X,\mu ,\nu ,t)=-i\left[ U\left( -\frac{\partial }{%
\partial \mu }\left( \frac{\partial }{\partial X}\right) ^{-1}+\frac{i}{2}%
\nu \frac{\partial }{\partial X}\right) -c.c.\right] \mathcal{T}(X,\mu ,\nu
,t)\ .
\end{equation}

The "Planck constant expansion" of both Moyal and tomographic evolution
equations is given by the potential energy expansion in powers of $\left( -%
\frac{i}{2}\partial /\partial p\right) $ for Moyal and $\left( -\frac{i}{2}%
\nu \partial /\partial X\right) $ for tomographic equation, respectively.
For example, the Moyal equation for an harmonic oscillator is
\begin{eqnarray}
\left( \frac{\partial }{\partial t}+p\frac{\partial }{\partial q}\right)
\mathcal{W}(q,p,t) &=&-\frac{i}{2}\left[ \left( q+\frac{i}{2}\frac{\partial
}{\partial p}\right) ^{2}-c.c.\right] \mathcal{W}(q,p,t)  \notag \\
&=&q\frac{\partial }{\partial p}\mathcal{W}(q,p,t)\ .
\end{eqnarray}
while the tomographic one reads
\begin{eqnarray}
\left( \frac{\partial }{\partial t}-\mu \frac{\partial }{\partial \nu }%
\right) \mathcal{T}(X,\mu ,\nu ,t) &=&-\frac{i}{2}\left[ \left( -\frac{%
\partial }{\partial \mu }\left( \frac{\partial }{\partial X}\right) ^{-1}+%
\frac{i}{2}\nu \frac{\partial }{\partial X}\right) ^{2}-c.c.\right] \mathcal{%
T}(X,\mu ,\nu ,t)  \notag \\
&=&-\nu \frac{\partial }{\partial \mu }\mathcal{T}(X,\mu ,\nu ,t)\ .
\end{eqnarray}

\subsubsection{Eigenvalue problem}

We may obtain the Moyal and the tomographic form of the eigenvalue equation
if we start from the von Neumann stationary equation for the density matrix.
In the position representation, we have
\begin{equation}
E\rho (q,q^{\prime })=\frac{1}{2}\left[ -\frac{1}{2}\frac{\partial ^{2}}{%
\partial q^{2}}-\frac{1}{2}\frac{\partial ^{2}}{\partial q^{\prime 2}}%
+U(q)+U(q^{\prime })\right] \rho (q,q^{\prime }).
\end{equation}
so that with the previous correspondence rules the Moyal equation for the
Wigner function is
\begin{eqnarray}
E \mathcal{W}(q,p) &=&-\frac{1}{4}\left[ \left( \frac{1}{2}\frac{\partial }{%
\partial q}+ip\right) ^{2}+\left( \frac{1}{2}\frac{\partial }{\partial q}%
-ip\right) ^{2}\right] \mathcal{W}(q,p)  \notag \\
&&+\frac{1}{2}\left[ U\left( q+\frac{i}{2}\frac{\partial }{\partial p}%
\right) +U\left( q-\frac{i}{2}\frac{\partial }{\partial p}\right) \right]
\mathcal{W}(q,p).
\end{eqnarray}
and the corresponding tomographic form of the equation:
\begin{eqnarray}
E\mathcal{T}(X,\mu ,\nu ) &=&-\frac{1}{4}\left[ \left( \frac{1}{2}\mu \frac{%
\partial }{\partial X}-i\frac{\partial }{\partial \nu }\left( \frac{\partial
}{\partial X}\right) ^{-1}\right) ^{2}+c.c.\right] \mathcal{T}(X,\mu ,\nu )
\notag \\
&&+\frac{1}{2}\left[ U\left( -\frac{\partial }{\partial \mu }\left( \frac{%
\partial }{\partial X}\right) ^{-1}+\frac{i}{2}\nu \frac{\partial }{\partial
X}\right) +c.c.\right] \mathcal{T}(X,\mu ,\nu ).
\end{eqnarray}
for the energy levels $E.$ In the usual pictures, the energy spectrum is
obtained by solving an eigenvalue problem of the Hamiltonian operator in the
carrier Hilbert space. Thus, for the $n-$th level, Eq. (\ref{radon corr form}%
) is to give the result
\begin{equation}
\mathcal{T}_{n}(X,\mu ,\nu )=\frac{e^{-\frac{X^{2}}{\mu ^{2}+\nu ^{2}}}}{%
\sqrt{\pi (\mu ^{2}+\nu ^{2})}}\frac{1}{n!2^{n}}H_{n}^{2}\left( \frac{X}{%
\sqrt{\mu ^{2}+\nu ^{2}}}\right) .
\end{equation}

However, neither the Wigner function $\mathcal{W}(q,p)$ nor the tomographic
probability densities $\mathcal{T}(X,\mu ,\nu )$ are vectors of a linear
space. The eigenvalue problem in the Hilbert space of states is mapped onto
the problem of solving integro-differential equations either of Moyal (von
Neumann) form or equations for probability distributions. It is remarkable
that in time evolution ruled by the tomographic equation, the function $%
\mathcal{T}(X,\mu ,\nu ,t)$ at any time $t$ is positive and normalized if
the initial tomogram is. This is obvious due to the correspondence chain
rules substituted in the initial unitary evolution of von Neumann equation.
Nevertheless it would be difficult to see it at a first glance directly in
the tomographic evolution equation, without going back to the von Neumann
equation.\bigskip

\subsection{Composite systems: separability and entanglement}

To define the notion of separability of quantum state of a composite system
let us discuss the example of two qubits. Each of the qubits has in its own
Hilbert space of states the standard basis $\left\{ |m\rangle \right\} $
where $m=\pm 1/2$ is the spin projection on the $z$-axis, that is
\begin{equation}
\left\vert +\frac{1}{2}\right\rangle =\left(
\begin{array}{c}
1 \\
0%
\end{array}
\right) \quad ,\quad \left\vert -\frac{1}{2}\right\rangle =\left(
\begin{array}{c}
0 \\
1%
\end{array}
\right) \quad .
\end{equation}
The spin operator $\vec{S}=(S_{x},S_{y},S_{z})$ is determined by the Pauli
matrices $(\sigma _{x},\sigma _{y},\sigma _{z})$ as
\begin{equation}
\vec{S}=\frac{1}{2}\vec{\sigma}\quad (\hbar=1)
\end{equation}
where
\begin{equation}
\sigma _{x}=\left(
\begin{array}{cc}
0 & 1 \\
1 & 0%
\end{array}
\right) ,\quad \sigma _{y}=\left(
\begin{array}{cc}
0 & -i \\
i & 0%
\end{array}
\right) ,\quad \sigma _{z}=\left(
\begin{array}{cc}
1 & 0 \\
0 & -1%
\end{array}
\right)
\end{equation}
Thus the basis $\left\{ |m\rangle \right\} $ satisfies the eigenvalue
problem
\begin{equation}
\sigma_{z}|m\rangle =m|m\rangle \quad .
\end{equation}

For two qubits the Hilbert space of states is four-dimensional and it is
obtained as tensor product of the two-dimensional Hilbert spaces of each
qubit, so that we have four basis vectors
\begin{equation}
|m_{1}m_{2}\rangle =|m_{1}\rangle |m_{2}\rangle \quad ;\quad m_{1},m_{2}=\pm
\frac{1}{2}
\end{equation}
The previous equation means
\begin{eqnarray}
\left\vert +\frac{1}{2}+\frac{1}{2}\right\rangle &=&\left(
\begin{array}{c}
1 \\
0%
\end{array}
\right) \otimes \left(
\begin{array}{c}
1 \\
0%
\end{array}
\right) =\left(
\begin{array}{c}
1 \\
0 \\
0 \\
0%
\end{array}
\right) \ , \\
\left\vert +\frac{1}{2}-\frac{1}{2}\right\rangle &=&\left(
\begin{array}{c}
1 \\
0%
\end{array}
\right) \otimes \left(
\begin{array}{c}
0 \\
1%
\end{array}
\right) =\left(
\begin{array}{c}
0 \\
1 \\
0 \\
0%
\end{array}
\right) \ , \\
\left\vert -\frac{1}{2}+\frac{1}{2}\right\rangle &=&\left(
\begin{array}{c}
0 \\
1%
\end{array}
\right) \otimes \left(
\begin{array}{c}
1 \\
0%
\end{array}
\right) =\left(
\begin{array}{c}
0 \\
0 \\
1 \\
0%
\end{array}
\right) \ , \\
\left\vert -\frac{1}{2}-\frac{1}{2}\right\rangle &=&\left(
\begin{array}{c}
0 \\
1%
\end{array}
\right) \otimes \left(
\begin{array}{c}
0 \\
1%
\end{array}
\right) =\left(
\begin{array}{c}
0 \\
0 \\
0 \\
1%
\end{array}
\right) \ .
\end{eqnarray}
Any pure state of one qubit $|\psi \rangle $ can be expressed as
superposition of the basis states
\begin{equation}
\left\vert \psi \right\rangle =\psi _{1}\left\vert +\frac{1}{2}\right\rangle
+\psi _{2}\left\vert -\frac{1}{2}\right\rangle =\left(
\begin{array}{c}
\psi _{1} \\
\psi _{2}%
\end{array}
\right) \ .
\end{equation}
The density matrix for one qubit in case of a pure state reads
\begin{equation}
\rho _{\psi }=\left\vert \psi \right\rangle \left\langle \psi \right\vert
=\left(
\begin{array}{cc}
\psi _{1}\psi _{1}^{\ast } & \psi _{1}\psi _{2}^{\ast } \\
\psi _{2}\psi _{1}^{\ast } & \psi _{2}\psi _{2}^{\ast }%
\end{array}
\right) \ .
\end{equation}
The density matrix for a mixed state of one qubit
\begin{equation}
\rho =\left(
\begin{array}{cc}
\rho _{11} & \rho _{12} \\
\rho _{21} & \rho _{22}%
\end{array}
\right) \
\end{equation}
is such that
\begin{equation}
\rho =\rho ^{\dagger }\quad ;\quad \mathrm{Tr}\rho =1
\end{equation}
and
\begin{equation}
\rho \geq 0\Longleftrightarrow \rho _{11}\geq 0\quad ;\quad \rho _{22}\geq
0\quad ;\quad \rho _{11}\rho _{22}-\rho _{12}\rho _{21}\geq 0\ .
\end{equation}
It can can be presented in the form
\begin{equation}
\rho =\lambda _{1}\left\vert \psi _{1}\right\rangle \left\langle \psi
_{1}\right\vert +\lambda _{2}\left\vert \psi _{2}\right\rangle \left\langle
\psi _{2}\right\vert \ ,
\end{equation}
where $\lambda _{1},\lambda _{2}\geq 0$ are its (non-negative) eigenvalues
and $\left\vert \psi _{1}\right\rangle $,$\left\vert \psi _{2}\right\rangle $
the corresponding eigenvectors.

The simply separable state of two qubits is defined as the factorized state
\begin{equation}
|\varphi \rangle =|\varphi _{1}\rangle |\varphi _{2}\rangle =\left(
\begin{array}{c}
\varphi _{11} \\
\varphi _{12}%
\end{array}
\right) \otimes \left(
\begin{array}{c}
\varphi _{21} \\
\varphi _{22}%
\end{array}
\right) =\left(
\begin{array}{c}
\varphi _{11}\varphi _{21} \\
\varphi _{11}\varphi _{22} \\
\varphi _{12}\varphi _{21} \\
\varphi _{12}\varphi _{22}%
\end{array}
\right) \ .
\end{equation}
The density matrix of this state is the tensor product of the density states
of each qubit:
\begin{equation}
\rho _{\varphi }=\left\vert \varphi \right\rangle \left\langle \varphi
\right\vert =\rho _{\varphi _{1}}\otimes \rho _{\varphi _{2}}\ .
\label{eq13}
\end{equation}
A separable state of two qubits is defined as a convex sum of the above
factorized states
\begin{equation}
\rho {(1,2)}=\sum\nolimits_{k}\lambda _{k}\rho _{\varphi _{1}}^{(k)}\otimes
\rho _{\varphi _{2}}^{(k)}\ .  \label{eq14}
\end{equation}
In Eq.(\ref{eq14}) the states $\rho _{\varphi _{1}}^{(k)},\rho _{\varphi
_{2}}^{(k)}$ satisfy the condition of purity
\begin{equation}
\left( \rho _{\varphi _{1,2}}^{(k)}\right) \left( \rho _{\varphi
_{1,2}}^{(k)}\right) =\rho _{\varphi _{1,2}}^{(k)}\ .
\end{equation}

However, this condition can be violated, as the general definition of
separable two qubit state is: a state is separable iff its density matrix
can be given the form
\begin{equation}
\rho {(1,2)}=\sum\nolimits_{k}\lambda _{k}\rho _{1}^{(k)}\otimes \rho
_{2}^{(k)}  \label{eq16}
\end{equation}
where $\sum\nolimits_{k}\lambda _{k}=1$ with $\lambda _{k}\geq 0,$ while $%
\rho _{1}^{(k)},\rho _{2}^{(k)}$ are density matrices of each qubit
respectively.

The above separability condition is written for density states. A condition
which holds for the corresponding tomograms may be obtained by calculating
the tomograms of the two sides of the above equation. The tomogram of the
state $\rho {(1,2)}$ is the joint probability function:
\begin{equation}
\mathcal{T}_{\rho {(1,2)}}(m_{1},m_{2},U)=\left\langle m_{1}m_{2}\left\vert
U\rho (1,2)U^{\dagger }\right\vert m_{1}m_{2}\right\rangle \ .
\end{equation}
As the tomogram of a simple separable state factorizes in the product of
independent probabilities:
\begin{eqnarray}
&&\left\langle m_{1}m_{2}\left\vert U_{1}\otimes U_{2}\left( \rho
_{1}\otimes \rho _{2}\right) U_{1}^{\dagger }\otimes U_{2}^{\dagger
}\right\vert m_{1}m_{2}\right\rangle \\
&&=\left\langle m_{1}\left\vert U_{1}\rho _{1}U_{1}^{\dagger }\right\vert
m_{1}\right\rangle \left\langle m_{2}\left\vert U_{2}\rho _{2}U_{2}^{\dagger
}\right\vert m_{2}\right\rangle \ ,  \notag
\end{eqnarray}
we eventually get from Eq. (\ref{eq16}) the tomographic separability
condition in the form:
\begin{equation}
\mathcal{T}_{\rho {(1,2)}}(m_{1},m_{2},U_{1}\otimes
U_{2})=\sum\nolimits_{k}\lambda _{k}\mathcal{T}_{\rho
_{1}}^{(k)}(m_{1},U_{1})\mathcal{T}_{\rho _{2}}^{(k)}(m_{2},U_{2})\ .
\label{TomSepa}
\end{equation}

The state is called entangled when the $4\times 4$-density matrix $\rho
(1,2) $ cannot be presented in the form (\ref{eq16}), or its tomogram $%
\mathcal{T}_{\rho {(1,2)}}$ in the form of Eq. (\ref{TomSepa}).

Now we reformulate the introduced notions and definitions using the
tomographic representations of the qubit states. This representation is
characterized by a map from density matrices onto the family of the
probability distributions which is invertible. We racall that for one qubit
one has
\begin{equation}
\rho =\left(
\begin{array}{cc}
\rho _{11} & \rho _{12} \\
\rho _{21} & \rho _{22}%
\end{array}
\right) \longleftrightarrow \vec{\mathcal{T}_{\rho }}(\vec{n})=\left(
\begin{array}{c}
\mathcal{T}_{\rho }(+\frac{1}{2},\vec{n}) \\
\mathcal{T}_{\rho }(-\frac{1}{2},\vec{n})%
\end{array}
\right)
\end{equation}
where the unit vector $\vec{n}=(\sin \theta \cos \varphi ,\sin \theta \sin
\varphi ,\cos \theta )$ determines a point on the sphere $S^{2}.$ As it was
observed in section \ref{gen_aspect}, the tomogram here is presented in the
form of a probability vector, whose components $\mathcal{T}_{\rho }(m,\vec{n}%
)=\mathrm{Tr}(U^{\dagger }\left\vert m\right\rangle \left\langle
m\right\vert U\rho )$ are the diagonal elements of the unitarily rotated
density matrix:
\begin{equation}
\mathcal{T}_{\rho }(+\frac{1}{2},\vec{n})=(U\rho U^{\dagger })_{11}\quad
;\quad \mathcal{T}_{\rho }(-\frac{1}{2},\vec{n})=(U\rho U^{\dagger })_{22}
\label{eq18}
\end{equation}
where
\begin{equation}
U=\left(
\begin{array}{cc}
u_{11} & u_{12} \\
u_{21} & u_{22}%
\end{array}
\right) =\left(
\begin{array}{cc}
\cos \frac{\theta }{2}\mathrm{e}^{i(\psi +\varphi )/2} & \sin \frac{\theta }{%
2}\mathrm{e}^{i(\psi -\varphi )/2} \\
-\sin \frac{\theta }{2}\mathrm{e}^{-i(\psi -\varphi )/2} & \cos \frac{\theta
}{2}\mathrm{e}^{-i(\psi +\varphi )/2}%
\end{array}
\right)  \label{eq19}
\end{equation}
is a unitary matrix of the group $SU(2)$ parametrized by the usual Euler
angles $\theta ,\psi ,\varphi $. The matrix $U$ transforms the spinors
according to the rotation labeled by the Euler angles.

For the qubit the tomogram can be considered as a function on the group $%
SU(2)$ (in fact on the homogeneous space $SU(2)/U(1)$). The formulae can be
presented in the form
\begin{eqnarray}
\mathcal{T}_{\rho }(+\frac{1}{2},U) &=&|u_{11}|^{2}\rho
_{11}+|u_{12}|^{2}\rho _{22}+(u_{12}u_{11}^{\ast }\rho _{21}+\mathrm{c.c})
\notag \\
\mathcal{T}_{\rho }(-\frac{1}{2},U) &=&|u_{21}|^{2}\rho
_{11}+|u_{22}|^{2}\rho _{22}+(u_{21}u_{22}^{\ast }\rho _{12}+\mathrm{c.c}).
\label{eq20}
\end{eqnarray}
In terms of Euler angles the tomographic probability reads
\begin{eqnarray}
\mathcal{T}_{\rho }(+\frac{1}{2},U) &=&\cos ^{2}\frac{\theta }{2}\rho
_{11}+\sin ^{2}\frac{\theta }{2}\rho _{22}+\cos \frac{\theta }{2}\sin \frac{%
\theta }{2}(\mathrm{e}^{-i\varphi }\rho _{21}+\mathrm{e}^{i\varphi }\rho
_{12})  \notag \\
\mathcal{T}_{\rho }(-\frac{1}{2},U) &=&\sin ^{2}\frac{\theta }{2}\rho
_{11}+\cos ^{2}\frac{\theta }{2}\rho _{22}-\cos \frac{\theta }{2}\sin \frac{%
\theta }{2}(\mathrm{e}^{i\varphi }\rho _{12}+\mathrm{e}^{-i\varphi }\rho
_{21}).
\end{eqnarray}

One can regard the qubit tomogram in the following manner. The density
matrix $i\rho $ can be considered as a point in the Lie algebra $u(2)$ of
the group $U(2)$; then the formula (\ref{eq18}) defines an orbit of the
unitary group in its Lie algebra.

Assume $\rho _{12}=\rho _{21}=0$: the initial point of the orbit, i.e. $U=I$%
, is determined by the two non-negative numbers $\rho _{11}$ and $\rho _{22}$
which satisfy the simplex condition $\rho _{11}+\rho _{22}=1,$ that is a
segment with extremes in $(1,0),(0,1)$ in $\mathbb{R}^{2}.$ Thus we may
choose in the Lie algebra of the group $U(2)$ a subset, which is a simplex
in an affine subspace modeled on the Cartan subalgebra of $SU(2)$. The
formula for the tomogram (\ref{eq20}) determines the orbit of the group $%
SU(2)$ in the simplex. Further details of this relation between qudit
tomograms and points of a simplex are discussed in \cite{PLASemient}. It is
interesting to note that, under the action of the group, the orbit starting
from any point of the simplex does not go out of the simplex. Thus one can
define the tomographic probability as a map of the pairs $(\vec{\rho},U)$
onto the points of the simplex $\vec{\mathcal{T}}_{\rho }(U).$ Here $\vec{%
\rho}=(\rho _{11},\rho _{22})^{\mathrm{T}}$ is a column vector and $U$ is an
element of the $SU(2)$ group. As we remarked in section \ref{weyl_tom}, we
could also use the full unitary group in Eq. (\ref{eq20}).

The general case $\rho _{12}=\rho _{21}^{\ast }\neq 0$ is analogous due to
the possibility of diagonalizing:
\begin{equation}
U_{0}^{\dagger }\rho U_{0}=
\begin{pmatrix}
\widetilde{\rho }_{11} & 0 \\
0 & \widetilde{\rho }_{22}%
\end{pmatrix}
\ ,  \label{eq22}
\end{equation}
so the generic state $\rho $ is again mapped onto the orbit of the group in
the given simplex, but the initial point of the unitary transformation is
shifted by the diagonalizing unitary matrix $U_{0}$, i.e. now the element $%
UU_{0}$ moves the simplex point $(\widetilde{\rho }_{11},\widetilde{\rho }%
_{22})$.

We stress that the initial points of the orbits belonging to the simplex can
be considered as elements of a Cartan subalgebra of the unitary group
belonging to non-negative weights (the Weyl chamber from which all other
weights can be obtained by discrete reflections (Weyl group)). Thus the
formula for the tomogram (\ref{eq20}) can be written as:
\begin{eqnarray}
\mathcal{T}_{\rho }(+\frac{1}{2},U) &=&|(UU_{0})_{11}|^{2}\widetilde{\rho }%
_{11}+|(UU_{0})_{12}|^{2}\widetilde{\rho }_{22}  \notag \\
\mathcal{T}_{\rho }(-\frac{1}{2},U) &=&|(UU_{0})_{21}|^{2}\widetilde{\rho }%
_{11}+|(UU_{0})_{22}|^{2}\widetilde{\rho }_{22}\ ,  \label{eq23}
\end{eqnarray}
in the form of a bi-stochastic map acting on the simplex where the shift
matrix $U_{0}$ and simplex point coordinates $(\widetilde{\rho }_{11},%
\widetilde{\rho }_{22})$ are connected with eigenvalues and eigenvectors of
the density matrix $\rho $ by Eq.(\ref{eq22}) (for further details, see \cite%
{PLASemient}). The equation (\ref{eq23}) can be reinterpreted in the
following way. The ortho-stochastic $2\times 2-$matrices
\begin{equation}
M(U)=
\begin{pmatrix}
|u_{11}|^{2} & |u_{12}|^{2} \\
|u_{21}|^{2} & |u_{22}|^{2}%
\end{pmatrix}%
\end{equation}
belong to the semigroup of bistochastic $2\times 2-$matrices.

We recall that, in the $n-$dimensional case, if $e\in \mathbb{R}^{n}$
denotes the column vector with all components $+1$ and $e^{\mathrm{T}}$ its
transpose, an $n\times n$-matrix $M$ is called (column) stochastic iff $e^{%
\mathrm{T}}M=e^{\mathrm{T}}$ and bistochastic iff
\begin{equation}
Me=e\ \text{and}\ e^{\mathrm{T}}M=e^{\mathrm{T}}\ .
\end{equation}
All such $M$'s are non-negative matrices. For invertible matrices $M,$ if we
consider not only bistochastic matrices but also their inverse, which need
not be non-negative any more, we get an open dense subset of $GL(n-1,\mathbb{%
R}),$ the general linear group of $(n-1)\times (n-1)$-matrices \cite%
{PLASemient}. Then formula (\ref{eq23}), when $U$ varies on all the elements
of the unitary group, provides an orbit of the semigroup of matrices $M(U)$
in the simplex. Thus the tomogram of the quantum qubit state is orbit of the
semigroup in the simplex, which is a subset of the first positive chamber of
the Cartan subalgebra of $SU(2)$ group. The statement is correct for unitary
tomogram of any qudit state. Since a tomogram in the tomographic
representation is identified with a quantum state, for a qudit one can say
that the quantum state is the orbit of the semigroup of bistochastic maps $%
M(U),$ parametrized by the pairs $\{(U,\vec{\mathcal{T}}_{\rho }(U))\}$
which are the graph of the tomogram, in the simplex which is a subset of a
chosen Cartan subalgebra of $SU(n)$ group. Due to this picture we write the
qudit analog of Eq.(\ref{eq23}) in matrix form as
\begin{equation}
\vec{\mathcal{T}}_{\rho }(U)=M(UU_{0})\overrightarrow{\widetilde{\rho }}.
\end{equation}
Here $\overrightarrow{\widetilde{\rho }}=(\widetilde{\rho }_{11},\widetilde{%
\rho }_{22},\ldots ,\widetilde{\rho }_{nn})^{\mathrm{T}}$ is the probability
column vector consisting of the eigenvalues of $\rho $ or, equivalently, a
point in the simplex in the given subalgebra. The ortho-stochastic matrix $%
M(UU_{0})$ has matrix elements
\begin{equation}
\left( M(UU_{0})\right) _{js}=|(UU_{0})_{js}|^{2}.
\end{equation}
The columns of the matrix $U_{0}$ are normalized eigenvectors of $\rho ,$
and a `gauge' has been chosen by fixing the phase factors of $U_{0}$ and an
ordering of both the components of $\overrightarrow{\widetilde{\rho }}$ and
the columns of $U_{0}$ so that $U_{0}^{\dagger }\rho U_{0}=\mathrm{diag}%
\left[ \widetilde{\rho }_{11},\widetilde{\rho }_{22},\ldots ,\widetilde{\rho
}_{nn}\right] .$

The component of the vector $\vec{\mathcal{T}}_{\rho }(U)$ are tomographic
probabilities. They are defining a spin tomogram if one takes as $n\times n$%
-matrix $U\in SU(n)$ the matrix of a $\left( 2j+1=n\right) -$dimensional
irreducible representation of the group $SU(2).$

For two qubits the condition determining a separable state of a composite
(bipartite) system can be rewritten in the form
\begin{equation}
\vec{\mathcal{T}}_{\rho (1,2)}(U_{1}\otimes U_{2})=\sum\limits_{k}\lambda
_{k}\left( |U_{1}U_{10}^{(k)}|^{2}\otimes |U_{2}U_{20}^{(k)}|^{2}\right)
\vec{\rho}_{1}^{(k)}\otimes \vec{\rho}_{2}^{(k)}  \label{eq27}
\end{equation}
or
\begin{equation}
|(U_{1}\otimes U_{2})U_{0}(1,2)|^{2}\vec{\rho}(1,2)=\sum\limits_{k}\lambda
_{k}\left( |U_{1}U_{10}^{(k)}|^{2}\otimes |U_{2}U_{20}^{(k)}|^{2}\right)
\vec{\rho}_{1}^{(k)}\otimes \vec{\rho}_{2}^{(k)}\,,  \label{eq28}
\end{equation}
where the notation $|A|^{2}$ means $(|A|^{2})_{js}=|A_{js}|^{2}$.

Thus, given the (non-negative) eigenvalues and eigenvectors of the density
matrix of two qubit systems $\rho (1,2)$, which are the components of the
vector $\vec{\rho}(1,2)$ and the corresponding columns of the unitary matrix
$U_{0}(1,2)$ respectively, the state is separable if the vector in the left
hand side of Eq.(\ref{eq28}) is a convex sum of vectors, which may be
written as tensor products of the tomographic probability vectors of each
qubit, i.e., with eigenvalues given by the components of $\vec{\rho}_{1},%
\vec{\rho}_{2}$ and eigenvectors given by the columns of the unitary
matrices $U_{10}^{(k)}$ and $U_{20}^{(k)}$.

In other words, the state is separable if the semigroup orbit determined by
semigroup $|(U_{1}\otimes U_{2})U_{0}(1,2)|^{2}$ acting on the simplex can
be presented as a convex set of orbits of semigroups determined by
semigroups $|U_{1}U_{10}^{(k)}|^{2}$ and $|U_{2}U_{20}^{(k)}|^{2}$ acting on
their own simplex.

Now we illustrate the previous theory by discussing some examples.

\noindent \textbf{Example 1 }Let us consider the simplest example of simply
separable state of two qubits: $\left\vert \uparrow \uparrow \right\rangle
=\left\vert \uparrow \right\rangle _{1}\left\vert \uparrow \right\rangle
_{2}.$ The density matrix of this state is diagonal:
\begin{equation}
\rho (1,2)=
\begin{pmatrix}
1 & 0 & 0 & 0 \\
0 & 0 & 0 & 0 \\
0 & 0 & 0 & 0 \\
0 & 0 & 0 & 0%
\end{pmatrix}
.  \label{eq69}
\end{equation}
The eigenvectors of this matrix may be chosen as
\begin{equation}
\vec{u}_{01}=
\begin{pmatrix}
1 \\
0 \\
0 \\
0%
\end{pmatrix}
;\vec{u}_{02}=
\begin{pmatrix}
0 \\
1 \\
0 \\
0%
\end{pmatrix}
;\vec{u}_{03}=
\begin{pmatrix}
0 \\
0 \\
1 \\
0%
\end{pmatrix}
;\vec{u}_{04}=
\begin{pmatrix}
0 \\
0 \\
0 \\
1%
\end{pmatrix}
;
\end{equation}
and the eigenvalues are $(1,0,0,0),$.that is:
\begin{equation}
\overrightarrow{\widetilde{\rho }}(1,2)=
\begin{pmatrix}
1 \\
0 \\
0 \\
0%
\end{pmatrix}
\label{eq74}
\end{equation}
The unitary matrix $U_{0}$ constructed from the eigenvectors is, of course,
the identity matrix:
\begin{equation}
U_{0}(1,2)=
\begin{pmatrix}
1 & 0 & 0 & 0 \\
0 & 1 & 0 & 0 \\
0 & 0 & 1 & 0 \\
0 & 0 & 0 & 1%
\end{pmatrix}
=I_{4}.
\end{equation}
The tomogram Eq.(\ref{eq28}) of the state reads
\begin{equation}
\vec{\mathcal{T}}_{\rho (1,2)}(U_{1}\otimes U_{2})=|U_{1}\otimes U_{2}|^{2}%
\overrightarrow{\widetilde{\rho }}(1,2)  \label{eq72}
\end{equation}
because
\begin{equation}
(U_{1}\otimes U_{2})U_{0}=(U_{1}\otimes U_{2})I_{4}=U_{1}\otimes U_{2}
\end{equation}
and thus Eq.(\ref{eq72}) yields
\begin{equation}
\vec{\mathcal{T}}_{\rho (1,2)}(U_{1}\otimes U_{2})=
\begin{pmatrix}
\cos ^{2}\frac{\theta _{1}}{2}\cos ^{2}\frac{\theta _{2}}{2} \\
\cos ^{2}\frac{\theta _{1}}{2}\sin ^{2}\frac{\theta _{2}}{2} \\
\sin ^{2}\frac{\theta _{1}}{2}\cos ^{2}\frac{\theta _{2}}{2} \\
\sin ^{2}\frac{\theta _{1}}{2}\sin ^{2}\frac{\theta _{2}}{2}%
\end{pmatrix}
=|U_{1}U_{10}|^{2}\overrightarrow{\widetilde{\rho }}_{1}\otimes
|U_{2}U_{20}|^{2}\overrightarrow{\widetilde{\rho }}_{2}  \label{eq75}
\end{equation}
Here: $U_{10}=U_{20}=I_{2}$ \ ; $\overrightarrow{\widetilde{\rho }}_{1}=%
\overrightarrow{\widetilde{\rho }}_{2}=
\begin{pmatrix}
1 \\
0%
\end{pmatrix}
.$

The formula (\ref{eq27}) contains only one term, with $\lambda _{k}=1$ times
the tensor product of the $2-$vectors
\begin{equation}
|U_{1}U_{10}|^{2}\overrightarrow{\widetilde{\rho }}_{1}=
\begin{pmatrix}
\cos ^{2}\frac{\theta _{1}}{2} \\
\sin ^{2}\frac{\theta _{1}}{2}%
\end{pmatrix}
\ \ ;\ \ |U_{2}U_{20}|^{2}\overrightarrow{\widetilde{\rho }}_{2}=
\begin{pmatrix}
\cos ^{2}\frac{\theta _{2}}{2} \\
\sin ^{2}\frac{\theta _{2}}{2}%
\end{pmatrix}
\label{eq76}
\end{equation}

\noindent \textbf{Example 2} Another example of simply separable state of
two qubits is the state with both spin-projections oriented along the $x$%
-axis. The state vector reads
\begin{equation}
\frac{1}{\sqrt{2}}
\begin{pmatrix}
1 & 1 \\
1 & -1%
\end{pmatrix}
\begin{pmatrix}
1 \\
0%
\end{pmatrix}
_{1}\otimes \frac{1}{\sqrt{2}}
\begin{pmatrix}
1 & 1 \\
1 & -1%
\end{pmatrix}
\begin{pmatrix}
1 \\
0%
\end{pmatrix}
_{2}=\frac{1}{4}
\begin{pmatrix}
1 \\
1 \\
1 \\
1%
\end{pmatrix}%
\end{equation}
Thus one has the density matrix of the two qubits state, which is a pure
simple separable state:
\begin{equation}
\rho (1,2)=\frac{1}{4}
\begin{pmatrix}
1 & 1 & 1 & 1 \\
1 & 1 & 1 & 1 \\
1 & 1 & 1 & 1 \\
1 & 1 & 1 & 1%
\end{pmatrix}
\label{eq78}
\end{equation}
The eigenvalues of this matrix yield the vector:
\begin{equation}
\overrightarrow{\widetilde{\rho }}(1,2)=
\begin{pmatrix}
1 \\
0 \\
0 \\
0%
\end{pmatrix}
\label{eq79}
\end{equation}
The corresponding eigenvectors of the matrix $\rho (1,2)$ may be fixed as
\begin{equation}
\vec{u}_{01}=\frac{1}{2}
\begin{pmatrix}
1 \\
1 \\
1 \\
1%
\end{pmatrix}
;\vec{u}_{02}=\frac{1}{2}
\begin{pmatrix}
1 \\
-1 \\
1 \\
-1%
\end{pmatrix}
;\vec{u}_{03}=\frac{1}{2}
\begin{pmatrix}
1 \\
1 \\
-1 \\
-1%
\end{pmatrix}
;\vec{u}_{04}=\frac{1}{2}
\begin{pmatrix}
1 \\
-1 \\
-1 \\
1%
\end{pmatrix}
;  \label{eq80}
\end{equation}
The above vectors are the columns of the unitary matrix $U_{0}$:
\begin{equation}
U_{0}=\frac{1}{2}
\begin{pmatrix}
1 & 1 & 1 & 1 \\
1 & -1 & 1 & -1 \\
1 & 1 & -1 & -1 \\
1 & -1 & -1 & 1%
\end{pmatrix}
=\frac{1}{\sqrt{2}}
\begin{pmatrix}
1 & 1 \\
1 & -1%
\end{pmatrix}
\otimes \frac{1}{\sqrt{2}}
\begin{pmatrix}
1 & 1 \\
1 & -1%
\end{pmatrix}
.
\end{equation}

The matrix $(U_{1}\otimes U_{2})U_{0}$ is a product:
\begin{equation}
(U_{1}\otimes U_{2})U_{0}=\frac{1}{\sqrt{2}}U_{1}
\begin{pmatrix}
1 & 1 \\
1 & -1%
\end{pmatrix}
\otimes \frac{1}{\sqrt{2}}U_{2}
\begin{pmatrix}
1 & 1 \\
1 & -1%
\end{pmatrix}%
\end{equation}
Here the matrices $U_{1}$ and $U_{2}$ have the form of Eq.(\ref{eq19}).

The corresponding orthostochastic matrix is
\begin{equation}
|(U_{1}\otimes U_{2})U_{0}|^{2}=\left\vert \frac{1}{\sqrt{2}}U_{1}
\begin{pmatrix}
1 & 1 \\
1 & -1%
\end{pmatrix}
\right\vert ^{2}\otimes \left\vert \frac{1}{\sqrt{2}}U_{2}
\begin{pmatrix}
1 & 1 \\
1 & -1%
\end{pmatrix}
\right\vert ^{2}
\end{equation}
due to the obvious property
\begin{equation}
\left\vert a\otimes b\right\vert ^{2}=\left\vert a\right\vert ^{2}\otimes
\left\vert b\right\vert ^{2}.
\end{equation}
>From this orthostochastic matrix and the vector (\ref{eq79}) we get the
tomogram probability vector as:
\begin{eqnarray}
&&\vec{\mathcal{T}}_{\rho (1,2)}(U_{1}\otimes U_{2})=|(U_{1}\otimes
U_{2})U_{0}|^{2}\overrightarrow{\widetilde{\rho }}(1,2)  \label{eq85} \\
&=&\frac{1}{4}
\begin{pmatrix}
\left\vert \cos \frac{\theta _{1}}{2}e^{i\varphi _{1}}+\sin \frac{\theta _{1}%
}{2}e^{-i\varphi _{1}}\right\vert ^{2}\left\vert \cos \frac{\theta _{2}}{2}%
e^{i\varphi _{2}}+\sin \frac{\theta _{2}}{2}e^{-i\varphi _{2}}\right\vert
^{2} \\
\left\vert \cos \frac{\theta _{1}}{2}e^{i\varphi _{1}}+\sin \frac{\theta _{1}%
}{2}e^{-i\varphi _{1}}\right\vert ^{2}\left\vert -\sin \frac{\theta _{2}}{2}%
e^{i\varphi _{2}}+\cos \frac{\theta _{2}}{2}e^{-i\varphi _{2}}\right\vert
^{2} \\
\left\vert -\sin \frac{\theta _{1}}{2}e^{i\varphi _{1}}+\cos \frac{\theta
_{1}}{2}e^{-i\varphi _{1}}\right\vert ^{2}\left\vert \cos \frac{\theta _{2}}{%
2}e^{i\varphi _{2}}+\sin \frac{\theta _{2}}{2}e^{-i\varphi _{2}}\right\vert
^{2} \\
\left\vert -\sin \frac{\theta _{1}}{2}e^{i\varphi _{1}}+\cos \frac{\theta
_{1}}{2}e^{-i\varphi _{1}}\right\vert ^{2}\left\vert -\sin \frac{\theta _{2}%
}{2}e^{i\varphi _{2}}+\cos \frac{\theta _{2}}{2}e^{-i\varphi
_{2}}\right\vert ^{2}%
\end{pmatrix}
.  \notag
\end{eqnarray}
The tomogram has the form of tensor product, corresponding to only one term
with $\lambda _{k}=1$ in Eq.(\ref{eq27}):
\begin{eqnarray}
&&\vec{\mathcal{T}}_{\rho (1,2)}(U_{1}\otimes U_{2})  \label{eq86} \\
&=&\frac{1}{2}
\begin{pmatrix}
\left\vert \cos \frac{\theta _{1}}{2}e^{i\varphi _{1}}+\sin \frac{\theta _{1}%
}{2}e^{-i\varphi _{1}}\right\vert ^{2} \\
\left\vert -\sin \frac{\theta _{1}}{2}e^{i\varphi _{1}}+\cos \frac{\theta
_{1}}{2}e^{-i\varphi _{1}}\right\vert ^{2}%
\end{pmatrix}
\otimes \frac{1}{2}
\begin{pmatrix}
\left\vert \cos \frac{\theta _{2}}{2}e^{i\varphi _{2}}+\sin \frac{\theta _{2}%
}{2}e^{-i\varphi _{2}}\right\vert ^{2} \\
\left\vert -\sin \frac{\theta _{2}}{2}e^{i\varphi _{2}}+\cos \frac{\theta
_{2}}{2}e^{-i\varphi _{2}}\right\vert ^{2}%
\end{pmatrix}
\notag
\end{eqnarray}

Finally, taking a convex sum of the density matrices (\ref{eq69}) and (\ref%
{eq78}):
\begin{equation}
\rho (1,2)=\cos ^{2}\delta
\begin{pmatrix}
1 & 0 & 0 & 0 \\
0 & 0 & 0 & 0 \\
0 & 0 & 0 & 0 \\
0 & 0 & 0 & 0%
\end{pmatrix}
+\frac{1}{4}\sin ^{2}\delta
\begin{pmatrix}
1 & 1 & 1 & 1 \\
1 & 1 & 1 & 1 \\
1 & 1 & 1 & 1 \\
1 & 1 & 1 & 1%
\end{pmatrix}
,  \label{eq87}
\end{equation}
we get a separable state whose tomogram, by construction, has the form of
the convex sum of Eq.(\ref{eq27}), with $\lambda _{1}=\cos ^{2}\delta
,\lambda _{2}=\sin ^{2}\delta :$
\begin{eqnarray}
&&\vec{\mathcal{T}}_{\rho (1,2)}(U_{1}\otimes U_{2})=\cos ^{2}\delta
\begin{pmatrix}
\cos ^{2}\frac{\theta _{1}}{2}\cos ^{2}\frac{\theta _{2}}{2} \\
\cos ^{2}\frac{\theta _{1}}{2}\sin ^{2}\frac{\theta _{2}}{2} \\
\sin ^{2}\frac{\theta _{1}}{2}\cos ^{2}\frac{\theta _{2}}{2} \\
\sin ^{2}\frac{\theta _{1}}{2}\sin ^{2}\frac{\theta _{2}}{2}%
\end{pmatrix}
+  \label{eq88} \\
&&+\frac{\sin ^{2}\delta }{4}
\begin{pmatrix}
\left\vert \cos \frac{\theta _{1}}{2}e^{i\varphi _{1}}+\sin \frac{\theta _{1}%
}{2}e^{-i\varphi _{1}}\right\vert ^{2}\cdot \left\vert \cos \frac{\theta _{2}%
}{2}e^{i\varphi _{2}}+\sin \frac{\theta _{2}}{2}e^{-i\varphi
_{2}}\right\vert ^{2} \\
\left\vert \cos \frac{\theta _{1}}{2}e^{i\varphi _{1}}+\sin \frac{\theta _{1}%
}{2}e^{-i\varphi _{1}}\right\vert ^{2}\cdot \left\vert -\sin \frac{\theta
_{2}}{2}e^{i\varphi _{2}}+\cos \frac{\theta _{2}}{2}e^{-i\varphi
_{2}}\right\vert ^{2} \\
\left\vert -\sin \frac{\theta _{1}}{2}e^{i\varphi _{1}}+\cos \frac{\theta
_{1}}{2}e^{-i\varphi _{1}}\right\vert ^{2}\cdot \left\vert \cos \frac{\theta
_{2}}{2}e^{i\varphi _{2}}+\sin \frac{\theta _{2}}{2}e^{-i\varphi
_{2}}\right\vert ^{2} \\
\left\vert -\sin \frac{\theta _{1}}{2}e^{i\varphi _{1}}+\cos \frac{\theta
_{1}}{2}e^{-i\varphi _{1}}\right\vert ^{2}\cdot \left\vert -\sin \frac{%
\theta _{2}}{2}e^{i\varphi _{2}}+\cos \frac{\theta _{2}}{2}e^{-i\varphi
_{2}}\right\vert ^{2}%
\end{pmatrix}
.  \notag
\end{eqnarray}

\noindent \textbf{Example 3} Let us now consider an example of an entangled
two qubit state
\begin{equation}
\left\vert \psi \right\rangle =\frac{1}{\sqrt{2}}\left( \left\vert +\frac{1}{%
2}\right\rangle \left\vert -\frac{1}{2}\right\rangle +\left\vert -\frac{1}{2}%
\right\rangle \left\vert +\frac{1}{2}\right\rangle \right) .  \label{eq30}
\end{equation}
The density matrix reads
\begin{equation}
\left\vert \psi \right\rangle \left\langle \psi \right\vert =\frac{1}{2}
\begin{pmatrix}
0 & 0 & 0 & 0 \\
0 & 1 & 1 & 0 \\
0 & 1 & 1 & 0 \\
0 & 0 & 0 & 0%
\end{pmatrix}
;  \label{eq31}
\end{equation}
the four eigenvalues yield the probability vector
\begin{equation}
\overrightarrow{\widetilde{\rho }}=
\begin{pmatrix}
0 \\
1 \\
0 \\
0%
\end{pmatrix}
.
\end{equation}
The corresponding eigenvectors may be chosen as
\begin{equation}
\vec{u}_{01}=
\begin{pmatrix}
1 \\
0 \\
0 \\
0%
\end{pmatrix}
,\vec{u}_{02}=\frac{1}{\sqrt{2}}
\begin{pmatrix}
0 \\
1 \\
1 \\
0%
\end{pmatrix}
,\vec{u}_{03}=\frac{1}{\sqrt{2}}
\begin{pmatrix}
0 \\
1 \\
-1 \\
0%
\end{pmatrix}
,\vec{u}_{04}=
\begin{pmatrix}
0 \\
0 \\
0 \\
1%
\end{pmatrix}%
\end{equation}
and the unitary matrix $U_{0}$\ diagonalizing the density matrix is
\begin{equation}
U_{0}=||\vec{u}_{01},\vec{u}_{02},\vec{u}_{03},\vec{u}_{04}||=
\begin{pmatrix}
1 & 0 & 0 & 0 \\
0 & \frac{1}{\sqrt{2}} & \frac{1}{\sqrt{2}} & 0 \\
0 & \frac{1}{\sqrt{2}} & -\frac{1}{\sqrt{2}} & 0 \\
0 & 0 & 0 & 1%
\end{pmatrix}
.
\end{equation}
The second column of the matrix $UU_{0}$ has the form:
\begin{equation}
\frac{1}{\sqrt{2}}\left(
\begin{array}{c}
u_{12}+u_{13} \\
u_{22}+u_{23} \\
u_{32}+u_{33} \\
u_{42}+u_{43}%
\end{array}
\right)
\end{equation}
Then the second column of the corresponding orthostochastic matrix $\left(
M(UU_{0})\right) _{js}=|\left( UU_{0}\right) _{js}|^{2}$ is the vector
\begin{equation}
\vec{\rho}=\frac{1}{2}\left(
\begin{array}{c}
|u_{12}+u_{13}|^{2} \\
|u_{22}+u_{23}|^{2} \\
|u_{32}+u_{33}|^{2} \\
|u_{42}+u_{43}|^{2}%
\end{array}
\right) .
\end{equation}
Thus, the tomogram of the entangled two qubit state (\ref{eq30}) is the
probability vector
\begin{equation*}
\vec{\mathcal{T}}_{\rho }(U)=M(UU_{0})\overrightarrow{\widetilde{\rho }}=%
\frac{1}{2}\left(
\begin{array}{c}
|u_{12}+u_{13}|^{2} \\
|u_{22}+u_{23}|^{2} \\
|u_{32}+u_{33}|^{2} \\
|u_{42}+u_{43}|^{2}%
\end{array}
\right)
\end{equation*}

For the subgroup $U=U_{1}\otimes U_{2},$ where
\begin{equation}
U_{1}=\left(
\begin{array}{cc}
a_{11} & a_{12} \\
a_{121} & a_{22}%
\end{array}
\right) \quad ;\quad U_{2}=\left(
\begin{array}{cc}
b_{11} & b_{12} \\
b_{21} & b_{22}%
\end{array}
\right) \quad ,
\end{equation}
the tomogram of the state reads:
\begin{equation}
\vec{\mathcal{T}}_{\rho }(U=U_{1}\otimes U_{2})=\frac{1}{2}\left(
\begin{array}{c}
|a_{11}b_{12}+a_{12}b_{11}|^{2} \\
|a_{11}b_{22}+a_{12}b_{21}|^{2} \\
|a_{21}b_{12}+a_{22}b_{11}|^{2} \\
|a_{21}b_{22}+a_{22}b_{21}|^{2}%
\end{array}
\right) .
\end{equation}
By using the Euler angles as parameters:
\begin{equation}
U_{1}=\left(
\begin{array}{cc}
a_{11} & a_{12} \\
a_{121} & a_{22}%
\end{array}
\right) =\left(
\begin{array}{cc}
\cos \frac{\theta _{1}}{2}e^{i(\psi _{1}+\varphi _{1})/2} & \sin \frac{%
\theta _{1}}{2}e^{i(\psi _{1}-\varphi _{1})/2} \\
-\sin \frac{\theta _{1}}{2}e^{-i(\psi _{1}-\varphi _{1})/2} & \cos \frac{%
\theta _{1}}{2}e^{-i(\psi _{1}+\varphi _{1})/2}%
\end{array}
\right) ,
\end{equation}
\begin{equation}
U_{2}=\left(
\begin{array}{cc}
b_{11} & b_{12} \\
b_{121} & b_{22}%
\end{array}
\right) =\left(
\begin{array}{cc}
\cos \frac{\theta _{2}}{2}e^{i(\psi _{2}+\varphi _{2})/2} & \sin \frac{%
\theta _{2}}{2}e^{i(\psi _{2}-\varphi _{2})/2} \\
-\sin \frac{\theta _{2}}{2}e^{-i(\psi _{2}-\varphi _{2})/2} & \cos \frac{%
\theta _{2}}{2}e^{-i(\psi _{2}+\varphi _{2})/2}%
\end{array}
\right) ,
\end{equation}
the components of the above tomogram are the following explicit functions of
$\left( \vec{n}_{1},\vec{n}_{2}\right) \in S^{2}\otimes S^{2}:$%
\begin{eqnarray}
\mathcal{T}_{\rho }(+\frac{1}{2},+\frac{1}{2},\vec{n}_{1},\vec{n}_{2}) &=&%
\frac{1}{2}[\cos ^{2}\frac{\theta _{1}}{2}\sin ^{2}\frac{\theta _{2}}{2}%
+\sin ^{2}\frac{\theta _{1}}{2}\cos ^{2}\frac{\theta _{2}}{2} \\
&+&\cos \frac{\theta _{1}}{2}\sin \frac{\theta _{1}}{2}\cos \frac{\theta _{2}%
}{2}\sin \frac{\theta _{2}}{2}(e^{i(\varphi _{1}-\varphi
_{2})}+e^{-i(\varphi _{1}-\varphi _{2})})]  \notag \\
\mathcal{T}_{\rho }(+\frac{1}{2},-\frac{1}{2},\vec{n}_{1},\vec{n}_{2}) &=&%
\frac{1}{2}[\cos ^{2}\frac{\theta _{1}}{2}\cos ^{2}\frac{\theta _{2}}{2}%
+\sin ^{2}\frac{\theta _{1}}{2}\sin ^{2}\frac{\theta _{2}}{2}  \notag \\
&-&\cos \frac{\theta _{1}}{2}\sin \frac{\theta _{1}}{2}\cos \frac{\theta _{2}%
}{2}\sin \frac{\theta _{2}}{2}(e^{i(\varphi _{1}-\varphi
_{2})}-e^{-i(\varphi _{1}-\varphi _{2})})]  \notag \\
\mathcal{T}_{\rho }(-\frac{1}{2},+\frac{1}{2},\vec{n}_{1},\vec{n}_{2}) &=&%
\mathcal{T}_{\rho }(+\frac{1}{2},-\frac{1}{2},\vec{n}_{1},\vec{n}_{2})
\notag \\
\mathcal{T}_{\rho }(-\frac{1}{2},-\frac{1}{2},\vec{n}_{1},\vec{n}_{2}) &=&%
\mathcal{T}_{\rho }(+\frac{1}{2},+\frac{1}{2},\vec{n}_{1},\vec{n}_{2})
\notag  \label{eq42}
\end{eqnarray}
The problem of separability amounts to the existence of a decomposition of
the above probability vector as a convex sum, according to Eq.(\ref{eq27}):
\begin{eqnarray}
&&\vec{\mathcal{T}}_{\rho }(U_{1}\otimes U_{2})  \label{eq61} \\
&=&\sum\limits_{k}\lambda _{k}\left[
\begin{pmatrix}
\cos ^{2}\frac{\theta _{k}}{2} & \sin ^{2}\frac{\theta _{k}}{2} \\
\sin ^{2}\frac{\theta _{k}}{2} & \cos ^{2}\frac{\theta _{k}}{2}%
\end{pmatrix}
\otimes
\begin{pmatrix}
\cos ^{2}\frac{\chi _{k}}{2} & \sin ^{2}\frac{\chi _{k}}{2} \\
\sin ^{2}\frac{\chi _{k}}{2} & \cos ^{2}\frac{\chi _{k}}{2}%
\end{pmatrix}
\right] \widetilde{\rho }_{1}^{(k)}\otimes \widetilde{\rho }_{2}^{(k)}
\notag \\
&=&\sum\limits_{k}\lambda _{k}
\begin{pmatrix}
\cos ^{2}\frac{\theta _{k}}{2}\widetilde{\rho }_{11}^{(k)}+\sin ^{2}\frac{%
\theta _{k}}{2}\widetilde{\rho }_{12}^{(k)} \\
\sin ^{2}\frac{\theta _{k}}{2}\widetilde{\rho }_{11}^{(k)}+\cos \frac{\theta
_{k}}{2}\widetilde{\rho }_{12}^{(k)}%
\end{pmatrix}
\otimes
\begin{pmatrix}
\cos ^{2}\frac{\chi _{k}}{2}\widetilde{\rho }_{21}^{(k)}+\sin ^{2}\frac{\chi
_{k}}{2}\widetilde{\rho }_{22}^{(k)} \\
\sin ^{2}\frac{\chi _{k}}{2}\widetilde{\rho }_{21}^{(k)}+\cos \frac{\chi _{k}%
}{2}\widetilde{\rho }_{22}^{(k)}%
\end{pmatrix}
\notag
\end{eqnarray}
which eventually leads to the following equation
\begin{eqnarray}
&&\frac{1}{2}\left( \cos ^{2}\frac{\theta _{1}}{2}\sin ^{2}\frac{\theta _{2}%
}{2}+\sin ^{2}\frac{\theta _{1}}{2}\cos ^{2}\frac{\theta _{2}}{2}+\sin
\theta _{1}\sin \theta _{2}\cos (\varphi _{1}-\varphi _{2})\right)  \notag \\
&=&\sum\limits_{k}\lambda _{k}\left( \cos ^{2}\frac{\theta _{k}}{2}%
\widetilde{\rho }_{11}^{(k)}+\sin ^{2}\frac{\theta _{k}}{2}\widetilde{\rho }%
_{12}^{(k)}\right) \left( \cos ^{2}\frac{\chi _{k}}{2}\widetilde{\rho }%
_{21}^{(k)}+\sin ^{2}\frac{\chi _{k}}{2}\widetilde{\rho }_{22}^{(k)}\right)
\label{eq44}
\end{eqnarray}
with an infinite number of unknown variables $\lambda _{k},\theta _{k},\chi
_{k},\widetilde{\rho }_{11}^{(k)},\widetilde{\rho }_{12}^{(k)},\widetilde{%
\rho }_{21}^{(k)},\widetilde{\rho }_{22}^{(k)}$. This equation has no
solutions, so the decomposition of Eq.(\ref{eq61}) is impossible for this
tomographic probability vector: the state is entangled. The proof is given
in the next subsection.

To resume, we have presented examples of tomograms for two pure simply
separable states. One state corresponds to both spins directed along $z$%
-axes. The density matrix of this state in the natural basis has the form of
Eq.(\ref{eq69}) and the tomogram is given in the form of probability vector (%
\ref{eq75}) with the tensor product form of probability vectors of Eq.(\ref%
{eq76}). Another state corresponds to both spins directed along the $x$%
-axis. The density matrix of this state has the form (\ref{eq78}) and the
tomogram is given by (\ref{eq85}). The tensor product form of the tomogram
is given by Eq.(\ref{eq86}). The separable, but not simply separable, mixed
state with density matrix (\ref{eq87}) has the tomogram given by Eq.(\ref%
{eq88}) in the form of convex series (\ref{eq27}), with only two nonzero
terms in the series.

An example of entangled state is given by the pure state (\ref{eq30}), with
density matrix (\ref{eq31}) and tomogram (\ref{eq42}).

\subsection{Inequalities}

To prove that equation (\ref{eq44}) has no solution let us use the following
inequality valid for stochastic $4\times 4$-matrices which are a tensor
product of two stochastic $2\times 2$-matrices, i.e.
\begin{equation}
M=M_{1}\otimes M_{2}.  \label{ineq69}
\end{equation}
Let us introduce the matrix
\begin{equation}
I_{0}=
\begin{pmatrix}
1 & -1 & -1 & 1 \\
1 & -1 & -1 & 1 \\
1 & -1 & -1 & 1 \\
-1 & 1 & 1 & -1%
\end{pmatrix}
\label{ineq70}
\end{equation}
Then we prove that
\begin{equation}
\left\vert \mathrm{Tr}(MI_{0})\right\vert \leq 2  \label{ineq71}
\end{equation}
is a necessary condition for the validity of Eq.(\ref{ineq69}).

In fact, the stochastic $4\times 4$-matrix has the form:
\begin{equation}
M=
\begin{pmatrix}
p_{1}M_{2} & q_{1}M_{2} \\
p_{2}M_{2} & q_{2}M_{2}%
\end{pmatrix}
,
\end{equation}
where the stochastic matrices $M_{1}$ and $M_{2}$\ read
\begin{equation}
M_{1}=
\begin{pmatrix}
p_{1} & q_{1} \\
p_{2} & q_{2}%
\end{pmatrix}
;M_{2}=
\begin{pmatrix}
s_{1} & t_{1} \\
s_{2} & t_{2}%
\end{pmatrix}
.
\end{equation}
The columns of $M_{1}$ and $M_{2}$ are probability vectors: $%
p_{1}+p_{2}=q_{1}+q_{2}=s_{1}+s_{2}=t_{1}+t_{2}=1$ and all the matrix
elements are non-negative. The trace in Eq. (\ref{ineq71}) reads:
\begin{eqnarray}
\mathrm{Tr}(MI_{0})
&=&p_{1}(s_{1}-s_{2})-p_{2}(s_{1}-s_{2})+p_{1}(t_{1}-t_{2})-p_{2}(t_{1}-t_{2})
\label{ineq75} \\
&+&q_{1}(s_{1}-s_{2})-q_{2}(s_{1}-s_{2})-q_{1}(t_{1}-t_{2})+q_{2}(t_{1}-t_{2})
\notag \\
&=&(p_{1}-p_{2})[(s_{1}-s_{2})+(t_{1}-t_{2})]+(q_{1}-q_{2})[(s_{1}-s_{2})-(t_{1}-t_{2})]
\notag \\
=: &&p(s+t)+q(s-t)  \notag
\end{eqnarray}
The differences $p,q,s,t$ of the probabilistic distributions satisfy
respectively the inequalities
\begin{equation}
|p_{1}-p_{2}|\leq 1\ ,|q_{1}-q_{2}|\leq 1\ ,|s_{1}-s_{2}|\leq 1\
,|t_{1}-t_{2}|\leq 1\   \label{ineq76}
\end{equation}

To prove that the modulus of the sum in Eq.(\ref{ineq75}) does not exceed
the number 2, consider the function $f$:
\begin{equation}
f(p,q,s,t):=p(s+t)+q(s-t)
\end{equation}
that is a harmonic function of the four variables $p,q,s,t$, which are
constrained to belong to the hypercube $K_{4}=\left\{ |p|\leq 1,|q|\leq
1,|s|\leq 1,|t|\leq 1\right\} .$ Then maximum and minimum values of $f$ on $%
K_{4}$ are reached on the boundary of the hypercube. Note that $f$ has the
special properties that, when any number of its variables is taken constant,
$f$ is still harmonic in the remaining variables. Therefore $f$ is harmonic
when restricted on each face of the boundary of the hypercube, which is an
hypercube $K_{3}$ of one less dimension. In each face $K_{3}$\ the max and
min of $f$ lie on its boundary. By repeating this argument, eventually the
max and min of $f$ are found to lie on the vertices of the initial hypercube
$K_{4}$, which are $2^{4}=16$ points. For the given function it is then a
trivial matter to check that $\max f=2$ and $\min f=-2.$

The argument has an immediate generalization to any number of dimensions:

\noindent\textbf{Proposition}: Let $f$ be a function of $n$ variables $%
(x_{1},...,x_{n})$ harmonic on the hypercube $K_{n}=$ $\{|x_{i}|\leq
1;i=1,..,n\}$, such that $f$ is still harmonic when restricted to any face $%
K_{s},0<s\leq n-1,$ belonging to the boundary of $K_{n}$. Then $f$ has its
extrema on the $2^{n}$ vertices\ of $K_{n}.$

>From inequality (\ref{ineq71}) it follows that
\begin{equation}
\left\vert \mathrm{Tr}\left( I_{0}\sum\nolimits_{k}\lambda _{k}M_{k}\right)
\right\vert \leq 2  \label{ineq80}
\end{equation}
where $\lambda _{k}\geq 0,\sum\nolimits_{k}\lambda _{k}=1$ and the matrices $%
M_{k}$ have the form of Eq.(\ref{ineq69}). Then, the proof of non-existence
of solution to Eq.(\ref{eq44}) is reduced to check the violation of
inequality Eq.(\ref{ineq80}).

Let us construct the $4\times 4$-matrix $M$ using the probability vector (%
\ref{eq42}). We take this vector for four pairs of argument matrices,
namely:
\begin{equation}
M=||\vec{\mathcal{T}}_{\rho }(U_{1}^{(1)}\otimes U_{2}^{(1)}),\vec{\mathcal{T%
}}_{\rho }(U_{1}^{(1)}\otimes U_{2}^{(2)}),\vec{\mathcal{T}}_{\rho
}(U_{1}^{(2)}\otimes U_{2}^{(1)}),\vec{\mathcal{T}}_{\rho
}(U_{1}^{(2)}\otimes U_{2}^{(2)})||.  \label{ineq81}
\end{equation}
Thus, the constructed matrix $M$ is a function of eight angles: the angles $%
\theta _{a},\varphi _{a}$ which are the Euler angles determining the $%
2\times 2$-matrix $U_{1}^{(1)}$, those $\theta _{d},\varphi _{d}$
determining the $2\times 2$-matrix $U_{1}^{(2)}$, while $\theta _{b},\varphi
_{b}$ determine the $2\times 2$-matrix $U_{2}^{(1)}$ and $\theta
_{c},\varphi _{c}$ the $2\times 2$-matrix $U_{2}^{(2)}$.

If one assumes that the equality (\ref{eq61}) is valid, then the constructed
matrix (\ref{ineq81}) has the form of a convex sum $\sum_{k}$ $\lambda
_{k}M_{k}$ satisfying the inequality (\ref{ineq80}) for all values of the
eight angles $\theta _{a},\varphi _{a},\theta _{d},\varphi _{d},\theta
_{b},\varphi _{b},\theta _{c},\varphi _{c}.$ On the other hand we know the
explicit form of such a matrix. In fact, the first column of this matrix is
\begin{equation}
\vec{\mathcal{T}}_{\rho }(U_{1}^{(1)}\otimes U_{2}^{(1)})=\left(
\begin{array}{c}
x_{ab} \\
\frac{1}{2}-x_{ab} \\
\frac{1}{2}-x_{ab} \\
x_{ab}%
\end{array}
\right)  \label{eq95}
\end{equation}
where
\begin{equation}
x_{ab}=\frac{1}{2}\left( \cos ^{2}\frac{\theta _{a}}{2}\sin ^{2}\frac{\theta
_{b}}{2}+\sin ^{2}\frac{\theta _{a}}{2}\cos ^{2}\frac{\theta _{b}}{2}+\sin
\theta _{a}\sin \theta _{b}\cos (\varphi _{a}-\varphi _{b})\right) ,
\label{eq96}
\end{equation}
the second column is obtained from the first one (\ref{eq95}) by the
following replacement:
\begin{equation}
a\rightarrow a\quad ,\quad b\rightarrow c\quad ,
\end{equation}
the third column by:
\begin{equation}
a\rightarrow d\quad ,\quad b\rightarrow b\quad ,
\end{equation}
and the fourth column by:
\begin{equation}
a\rightarrow d\quad ,\quad b\rightarrow c\quad .
\end{equation}

Taking the trace of the obtained matrix $M$ with the matrix $I_{0}$ of (\ref%
{ineq70}),\ we get the following function of the eight angles:
\begin{equation}
B=4(x_{ab}+x_{ac}+x_{db}-x_{dc})-2  \label{ineq86}
\end{equation}
Now we look for values of the angles $\theta _{a},\varphi _{a},$ $\theta
_{d},\varphi _{d},$ $\theta _{b},\varphi _{b},$ $\theta _{c},\varphi _{c}$
for which $|B|$ exceeds $2$ (in fact \cite{cirel} the maximum of this
function is equal to $2\sqrt{2}$). Such values do exist and this implies
that the hypothesis (\ref{eq61}) for the given probability vector is false.
In particular, the maximum of $|B|$ is achieved when $%
x_{ab}=x_{ac}=x_{db}=1-x_{dc},$ and the corresponding stochastic matrix
reads
\begin{equation}
M=
\begin{pmatrix}
\frac{2+\sqrt{2}}{8} & \frac{2+\sqrt{2}}{8} & \frac{2+\sqrt{2}}{8} & \frac{2-%
\sqrt{2}}{8} \\
\frac{2-\sqrt{2}}{8} & \frac{2-\sqrt{2}}{8} & \frac{2-\sqrt{2}}{8} & \frac{2+%
\sqrt{2}}{8} \\
\frac{2-\sqrt{2}}{8} & \frac{2-\sqrt{2}}{8} & \frac{2-\sqrt{2}}{8} & \frac{2+%
\sqrt{2}}{8} \\
\frac{2+\sqrt{2}}{8} & \frac{2+\sqrt{2}}{8} & \frac{2+\sqrt{2}}{8} & \frac{2-%
\sqrt{2}}{8}%
\end{pmatrix}
\ .
\end{equation}
It means that Eq. (\ref{eq44}) has no solution. The inequality
\begin{equation}
|B|\leq 2
\end{equation}
is called Bell inequality \cite{Bell64} (or CHSH inequality \cite{CHSH}).

\newpage

\section{Conclusions}

To conclude we point out the main aspects discussed in the paper. We
reviewed the probability representation of quantum states, in which wave
functions or density states are replaced by tomographic probability
distributions containing a complete information on quantum states. The
mathematical mechanism to construct all the possible probability
descriptions of quantum states was clarified. It amounts to constructing
complete sets of rank-one projectors in the Hilbert space of operators
acting on the underlying Hilbert space of state vectors. These sets are
complete or overcomplete. The tomograms depending on continuous variables
were shown to be used both in the classical and quantum domains. The set of
all tomographic functions describing the quantum states, the classical
states or no states at all was characterized. The characterization is
expressed in terms of inequalities, given by Eq.s ( \ref{Class}), (\ref%
{Quantum}). We suggested a method of direct experimental checking the
Heisenberg uncertainty relations \cite{Ad.Sci.Lett.}. In experiments \cite%
{Raymer, Mlynek, Lvovski, Solimeno} the optical tomogram of a photon state $%
\mathcal{T}(X,\theta )$ was measured in order to find the Wigner function $%
\mathcal{W}(q,p)$ by means of the Radon anti-transform of the tomogram. We
pointed out that the new inequality (\ref{gamma}) for the dispersion of the
homodyne photon quadrature $X=\cos \theta Q+\sin \theta P,$ expressed in
terms of integrals containing the directly measured optical tomogram $%
\mathcal{T}(X,\theta ),$ provides the method of checking Heisenberg
uncertainty relations. The accuracy of the experiment on checking the
uncertainty relations is in one-to-one correspondence with measuring optical
tomogram in the above experiments. The inequality (\ref{gamma}) can be
violated if one uses tomograms of classical states. The Heisenberg
uncertainty relation checking can be possibly accompanied by a checking of
the inequality for tomographic entropy discussed in \cite{DN}.

The entanglement of spin system states was given in terms of properties of
the spin tomograms. The Bell inequalities were shown to reflect the
properties of entanglement in terms of properties of joint probability
distributions (spin tomograms) to be expressed (or not expressed) as convex
sums of joint probability distributions without correlations. The quantum
spin states provide some bounds for the correlations. The Cirelson \cite%
{cirel} bound $2\sqrt{2}$ corresponds to the Bell number characterizing a
maximally entangled state of two qubits expressed in terms of the system
tomograms properties. The tomographic-like joint probability distributions
for which the Bell number is greater than that bound (the maximum can be
equal to $4,$ see \cite{pop}) do not correspond neither to states with
quantum spin correlations nor to classical correlations, characterized by a
bound equal to $2$. The stochastic matrix which has the Bell number $4$ is,
for example, of the form:
\begin{equation}
M=
\begin{pmatrix}
1 & \frac{1}{3} & \frac{1}{2} & 0 \\
0 & 0 & 0 & \frac{1}{4} \\
0 & 0 & 0 & \frac{3}{4} \\
0 & \frac{2}{3} & \frac{1}{2} & 0%
\end{pmatrix}
\ .
\end{equation}
Suppose, in an experiment on measuring the spin state tomogram of two
qubits, to get the probability vector (\ref{eq72}) and the stochastic
matrix, whose columns are the values of this vector for 4 pairs of Euler
angles, to be just the above matrix $M.$ This result contradicts quantum
mechanics. Moreover, the `classical'\ states of two qubits have the bound $%
2. $ Thus, the value $4$ of the Bell number corresponds to a tomogram which
is neither quantum nor classical, and this is similar to the example with
the scaled Wigner function for continuous variables discussed in section \ref%
{Distrib_quasi}. We mean `classical' in the following sense: in classical
probability theory, given a distribution function of two random classical
variables, there are two possibilities. One is that the random variables are
uncorrelated, i.e, the joint probability distribution has a factorized form
as a product of two probability distributions, describing statistical
properties of each of the random variables. Another possibility is that the
random variables are correlated, so the joint probability distribution is a
convex sum of factorized probability distributions. The correlations
described by such sums we call `classical'. In a sense, it is a terminology
that refers to a class of situations in which the joint probability
distribution of two correlated random variables is thought of as a convex
mixture of distributione without correlations.

This picture for discrete spin variables is analogous to the picture with
continuous variables, where tomograms violating both inequalities (\ref%
{Class}), (\ref{Quantum}) do not correspond neither to classical nor to
quantum states.

We have shown that in tomographic approach to quantum mechanics an important
role is played by semigroups and their orbits in simplexes, since from a
geometrical point of view the tomograms, being probability distributions,
are points of simplexes and their dependence on extra parameters provides
some domains in the simplexes as orbits of semigroups. For qubit systems the
semigroups are obtained from unitary matrices taking the map of their
elements onto their square moduli. There exist generalizations of symplectic
tomographic maps both for classical mechanics of a top \cite{Aso1} and
quantum mechanics with using maps with curvilinear coordinate lines Radon
transforms \cite{Aso2}. The analysis presented in this paper may be extended
to those cases of generalized Radon transform. We hope to study these maps
and their properties in future papers.

Among the variety of open problems, we should mention few of them which are
more crucial for the full equivalence of the tomographic picture of quantum
mechanics with the existing ones. We do not have yet a complete and
autonomous characterization of tomograms whose inverse Radon transform gives
rise to a Wigner function or to a classical probability distribution.
Similarly we do not have a complete characterization of the continuity
property of the Radon transform and its inverse. Some of those aspects play
a relevant role to establish the topological properties of the star-product
among the observable functions and the action of the observables on the
states. We shall consider these interesting problems in a forthcoming paper.

\noindent \textbf{Acknowledgements} V. I. Man'ko thanks I. N. F. N. and
University `Federico II'\ of Naples for their hospitality and R. F. F. I.
for partial support.

\end{document}